\documentclass[twoside, a4paper, reqno, english, 11pt]{article}

\usepackage{amssymb,amsmath,amsthm}
\usepackage{graphicx}
\usepackage[margin=2cm]{geometry}
\usepackage[toc,page]{appendix}
\usepackage{float}
\usepackage{caption}
\DeclareCaptionFont{10pt}{\fontsize{10pt}{12pt}\selectfont}
\usepackage{afterpage}
\usepackage{listings}
\usepackage{fancyhdr}
\usepackage{color}
\usepackage{hyperref}
\usepackage{babel,blindtext}
\usepackage[normalem]{ulem}
\usepackage{multirow}
\usepackage{lscape}
\usepackage{graphicx}
\usepackage{subfig}
\usepackage{mathtools}
\usepackage{indentfirst}
\usepackage{algorithm}
\usepackage[noend]{algpseudocode}
\usepackage{bm}
\usepackage{booktabs}
\usepackage{natbib}
\usepackage{eurosym}
\usepackage[table]{xcolor}
\usepackage{authblk}
\usepackage[symbol]{footmisc}
\usepackage{abstract}

\DeclareMathOperator{\BG}{BG}

\begin{document}

\title{Copula Averaging for Tail Dependence in Insurance Claims Data}

\author[1,2]{Sen Hu}
\author[1,2]{Adrian O'Hagan\footnote{E-mail for correspondence: \texttt{adrian.ohagan@ucd.ie}}}

\affil[1]{School of Mathematics and Statistics, University College Dublin, Dublin 4, Ireland}
\affil[2]{Insight Centre for Data Analytics, University College Dublin, Dublin 4, Ireland}

\date{}

\maketitle

\begin{abstract}
Analysing dependent risks is an important task for insurance companies.
A dependency is reflected in the fact that information about one random variable provides information about the likely distribution of values of another random variable. 
Insurance companies in particular must investigate such dependencies between different lines of business and the effects that an extreme loss event, such as an earthquake or hurricane, has across multiple lines of business simultaneously. 
Copulas provide a popular model-based approach to analysing the dependency between risks, and the coefficient of tail dependence is a measure of dependence for extreme losses.
Besides commonly used empirical estimators for estimating the tail dependence coefficient, copula fitting can lead to estimation of such coefficients directly or can verify their existence.   
Generally, a range of copula models is available to fit a data set well, leading to multiple different tail dependence results; a method based on Bayesian model averaging is designed to obtain a unified estimate of tail dependence. 
In this article, this model-based coefficient estimation method is illustrated through a variety of copula fitting approaches and results are presented for several simulated data sets and also a real general insurance loss data set.
\end{abstract}



\section{Introduction}

Using copulas to calculate joint distributions of two or more random variables, to model extreme losses and to capture tail dependency has become increasingly popular in actuarial science and finance (\citealp{Frees1998}; \citealp{Embrechts2003}; \citealp{Cherubini2004}).
One key merit of the copula approach is that it separates the joint distribution into two segments: a set of marginal distributions for each random variable and a copula function that combines the marginals to form a joint distribution that solely accounts for the dependence structure.
Although often it is preferred to model joint risks using multivariate distributions directly, sometimes this may be conceptually challenging due to the complex specification and implementation of multivariate distributions compared to their univariate counterparts, or the limited choices of applications available using multivariate distributions. 
In turn, copulas allow the focus to remain on efficiently modelling each marginal risk separately before bonding them using a copula framework, especially given that there exists a rich set of resources for univariate modelling in the literature. 
For example, it is common that the historical behaviour of some types of insurance losses have already been well modeled and understood by actuaries and it can then be desirable for these models to be incorporated directly into modelling joint distributions of losses, which the copula approach facilitates.

Because the copula function completely specifies the dependence structure between random variables, it provides an efficient approach to computing dependence among risks.
Although there are different correlation measures in the broad statistical or copula literature, such as Pearson's correlation coefficient, Kendall's tau or Pearson's rho, sometimes the dependence is not linear. One particular quantity of interest is correlation in the tail of a joint distribution, i.e. tail dependence, assessed via tail dependence coefficient (TDC). For example, upper tail dependence coefficient can be viewed as the probability of realising an extremely large value from one random variable or line of business given that an extremely large value has occurred in a (potentially) related variable or line of business. 
In most real-world data sets, the true underlying TDC is unknown and the TDC cannot be calculated directly due to its definition requiring a limiting measure (see Equation~\ref{chap3:eq:TDC}). 
Commonly, TDCs can be nonparametrically estimated using empirical copulas. 
In the literature various methods have been proposed to empirically estimate the TDCs (mostly nonparametrically) -- see \cite{Sweeting2013} for a summary. 
However, the performance of these estimations is not stable and varies largely. It can also be shown that fitting copulas to data sets is highly sensitive to small changes in the data, primarily because the data in the copula space is strictly within $[0,1]^{d}$.
More importantly, they often over-estimate the TDCs when the coefficients are small or zero, i.e. they cannot verify whether the TDCs actually exist.

For both insurance and reinsurance, understanding tail dependence between different lines of business and the effects that an extreme loss event has across multiple lines of business simultaneously is vital, especially under the regulatory requirements of Solvency II to calculate the solvency capital requirement (\citealp{SolvencyII}). Such dependence exists both in underwriting risks (e.g. catastrophic events including hurricanes and earthquakes) and in market and credit risks. The 2007--2010 financial crisis is a good example of this (\citealp{Salmon2012}). With the current climate change environment, insurers are facing more emerging risks and challenges from extreme climate conditions, which will further adversely impact many areas of business; the report of \cite{CROForum2019} provides further detail.
Hence, insurers must investigate and be able to estimate the tail dependence coefficient effectively and accurately when pricing multiple products together in the broad pricing process.


Compared with copulas, multivariate distributions are often difficult to use to produce a simplified summary of tail dependence, where as a wide range of dependency structures can be applied using different parameter settings for many types of copulas, especially there exists a considerable number of copulas in the literature. 
Some Archimedean copulas (e.g. Gumbel copula (\citealp{Gumbel1960}) and Joe copula (\citealp{Joe1997})) focus on modelling the right tail of the data, i.e. data with joint large extreme values; some (e.g. Clayton copula (\citealp{Clayton1978})) focus on the left tail, i.e. joint low extreme values; and symmetric copulas (e.g. Frank copula (\citealp{Frank1979}), Gaussian copula and t copula within the elliptical copula family) give equal importance to both tails. 
Furthermore, for more complex dependence structures, there are nested (Archimedean) copulas for nested correlations of high dimensional data (\citealp{Joe1997}); vine copulas for more flexible modelling of high dimensional data (\citealp{Bedford2002}), and rotated copulas for accommodating dependence (\citealp{Joe1997}). 
Because copulas are essentially distribution functions, they could further be mixed to form mixtures of copulas for more complex dependencies, see for example \cite{Kosmidis2016} and \cite{Arakelian2014}. 

Faced with the vast literature on copulas, it is often unclear how to choose between competing copulas and their associated estimates of tail dependence when more than one copula provides reasonably good fit to the data. 
Typically, copula models can be selected using either standard likelihood-based approaches such as Akaike information criterion (AIC) (\citealp{Akaike1974}) and Bayesian information criterion (BIC) (\citealp{Schwarz1978}), or minimum distance approach, which compares the distance between an empirically fitted copula and each copula function determined by quantities such as Kendall's tau or the tail dependence coefficient $\lambda$ (TDC) (\citealp{Genest1993}; \citealp{Frees1998}; \citealp{Caillault2005}). In a recent development in the copula literature, the focused information criteria (FIC) (\citealp{Ko2019fic}) and copula information criteria (CIC) (\citealp{Gronneberg2014}; \citealp{Ko2019cic}) are proposed as selection criteria for copulas. 
However, any selection criteria still leads to the problem that choosing only one ``best" copula ignores model uncertainty in inference. 


Bayesian model averaging (BMA) has been shown to be useful in terms of model selection and model predictions and accounting for model uncertainty and bias, which provides a convenient and statistically robust solution to this problem. A detailed review of model uncertainty and BMA is given by \cite{Draper1995} and \cite{Hoeting1999}. 
BMA has been proposed to use in combination with copulas in various modelling settings:
in hydrology and meteorology \cite{Moller2013} and \cite{Madadgar2014} proposed using BMA to average different model forecasts while a chosen copula takes the dependence among forecasts of different variables into consideration; 
\cite{Ramsey2018} proposed to use BMA and copulas for modelling dependence between crop yields and futures prices in spatial analysis, with a focus on averaging different locations and copulas simultaneously. Similar ideas are also presented in \cite{Pelster2016}. 
In all present literature, BMA is typically used in prediction or forecast ensembles to address uncertainties and biases and averages multiple (typically regression-based) models, which may have copulas as elements in the model or for post-processing, instead of aiming at directly combining quantities of interest from multiple copulas for model selection purposes.
Furthermore, when interest lies in better estimating the TDC, it allows a weighted average of the TDCs from different copulas to be combined to form a single, blended estimate of tail dependence.   
From the perspective of modelling maxima and estimating the TDCs, \cite{Sabourin2013} and \cite{Apputhurai2011} proposed to use BMA to model multivariate extremes, but using multivariate extreme distributions instead of copulas or extreme copulas.

One goal of this paper is to assemble results for empirical estimation of the TDCs using existing methodologies and to contrast them with the proposed applications of BMA, to show that a copula model-based method provides a better and more stable estimate, or at least as equally good estimate as the empirical one, via various model fitting strands.
To the authors' knowledge, the proposed method of using BMA to combine multiple copula models and acquire a better understanding and estimation of the TDCs is a novel contribution. 

The remainder of the article proceeds as follows: 
Section~\ref{sec:copula} reviews and highlights the characteristics of copulas in general and provides more refined details of the specific copulas employed. It also illustrates different copula model fitting approaches including mixtures of copulas, as well as how the empirical copula can be implemented to provide a nonparametric estimate of the TDCs.
Section~\ref{sec:BMA} briefly reviews Bayesian model averaging.
Section~\ref{sec:methods} details how the conventional copula framework can be adapted to employ Bayesian model averaging. 
Sections~\ref{sec:simstudy3} and~\ref{sec:simstudy2} detail two simulation studies where artificial data sets are generated to illustrate the methods. 
Section~\ref{sec:irish} provides the results obtained from analysing a real world data set from an Irish general insurance company. 
The article concludes in Section~\ref{sec:conclusion}, summarising the main research findings and highlighting potential further advances.
The package \textsf{copula} (\citealp{Hofert2018}) is extensively used in R (\citealp{R2018}), and software implementation in R for empirical TDC estimators, copula regression and mixture of copulas is available in \cite{mvClaim2019}. 

\section{Copula review and model fitting methods}
\label{sec:copula}

This section considers bivariate cases, but the results are readily extendible to higher dimensional cases.
Suppose there are two random variables, $Y_1$ and $Y_2$, with joint distribution function $F(y_1,y_2)$ and two marginal distributions $F_1(y_1), F_2(y_2)$ respectively.
A $2$-dimensional copula is a distribution function on $[0,1]\times [0,1]$ with standard uniform marginal distributions, such as
\begin{equation*}
C(u,v)=C(F_1(y_1),F_2(y_2))=F(F_1^{-1}(F_1(y_1)),F_2^{-1}(F_2(y_2)))=F(y_1,y_2) .
\end{equation*}
The function $C(u,v)$ is called a copula (\citealp{Joe1997}; \citealp{Nelsen2007}; \citealp{Joe2014}). 
The copula contains all information on the dependence structure between the margins, whereas the marginal distribution functions $F_1, F_2$ contain all information on the marginals. 
If $\bar{F}(Y_1,Y_2)$ is joint survival function for $Y_1, Y_2$ and $\bar{F}_1, \bar{F}_2$ are univariate survival marginals, then $\bar{C}(\bar{F}_1(y_1),\bar{F}_2(y_2))=\bar{F}(y_1,y_2)$ is a survival copula (\citealp{Nelsen2007}). Note that the survival copula can be regarded as a $180^{\circ}$ rotated copula.    
Furthermore, Sklar's theorem (\citealp{Sklar1959}) states that any multivariate joint distribution function can be written in terms of univariate marginal distribution functions and a copula. 
This leads to a prime advantage of using copulas -- allowing modelling of the marginals and the dependence structure separately. Following Sklar's theorem, if $C$, $F$, $F_1$ and $F_2$ are absolutely continuous and differentiable with respective to density $c, f, f_1$ and $f_2$, then the joint density of $(Y_1, Y_2)$ and the copula density are (\citealp{Denuit2005}, \citealp{Hofert2018book}):
\begin{equation}
\begin{split}
&f(y_1, y_2) = f_1(y_1)f_2(y_2)c(F_1(y_1),F_2(y_2)) , \\
&c(F_1(y_1),F_2(y_2)) = \frac{f(y_1, y_2)}{f_1(y_1)f_2(y_2)} .
\end{split}
\label{eq:copula_density}
\end{equation}

One important copula class is the Archimedean copula. If a bivariate copula is of the form 
$C(\boldsymbol{u};\psi)=\psi^{(-1)}(\psi (u_1)+\psi(u_2))$
for some generator function $\psi$ that is completely monotone, where $\boldsymbol{u} \in [0,1]^{2}$, then the copula is said to be an (one-parameter) Archimedean copula. This can be extended to two-parameter Archimedean copulas -- see \cite{Joe1997} and \cite{Nelsen2007} for more details. 
Another important copula class is the elliptical copula. These are copulas fitted to elliptical, and therefore symmetric, distributions. 
The most common copulas in this family are the Gaussian and Student-t copulas. 
These copulas became popular due to the tractable nature of the multivariate normal and multivariate t distributions. 
Table~\ref{tab:copula_property} presents the properties, Kendall's tau and the TDCs of some popular copulas, which will be used later in this article. 

\begin{table}[H]
	\centering
	\caption{Selected parametric copulas from the one-parameter Archimedean copula family and elliptical copula family, showing their parameter ($\theta$) ranges, Kendall's tau and tail dependence coefficients (TDCs) $\lambda_L, \lambda_U$, see \cite{Nelsen2007}. Note that $D_1$ is the Debye function. }
	\label{tab:copula_property}
	\resizebox{\textwidth}{!}{
		\begin{tabular}{lcccc}
			\toprule[0.15 em]
			Name & Generator & Parameter ($\theta$) & Kendall's tau & TDCs ($\lambda_L, \lambda_U$) \\
			\midrule
			Joe & $-\log(1-(1-t)^{\theta})$ & $ [1, \infty) $ & $1+\frac{4}{\theta^2}\int_{0}^{1} t \log(t) (1-t)^{\frac{2(1-\theta)}{\theta}}dt$ & $(0, 2-2^{1/\theta})$ \\
			Gumbel & $(-\log t)^{\theta}$ & $ [1, \infty) $ & $1-\frac{1}{\theta}$ & $(0, 2-2^{1/\theta})$ \\
			Clayton & $\frac{1}{\theta}(t^{-\theta}-1)$ & $ [-1, \infty] \backslash \{0\} $ & $\frac{\theta}{\theta+2}$ & $(2^{-1/\theta}, 0)$ \\
			Frank & $-\log(\frac{e^{-\theta t}-1}{e^{-\theta}-1})$ & $ \mathbb{R} \backslash \{0\}$ & $1-\frac{4}{\theta}+4\frac{D_1(\theta)}{\theta}$ & $(0,0)$ \\
			\midrule
			Gaussian & - & [-1, 1] & $\frac{2}{\pi} \arcsin \theta$ & (0,0) \\
			Student t & - & [-1, 1] & $\frac{2}{\pi} \arcsin \theta$ & $2t_{v+1}( -\sqrt{\frac{(v+1)(1-\theta)}{(1+\theta)}})$\\
			\bottomrule[0.15 em]
	\end{tabular} }
\end{table}

There are a number of ways to measure dependence among risks, notably Kendall's tau or Spearman's rho, which focus on global dependence. 
The concept of tail dependence is a sufficient and more specialised measure of dependence in the tails for extreme co-movements, which makes it a key concept in many financial and insurance applications (\citealp{Sweeting2013}). 
The extent to which this happens is measured by the tail dependence coefficient (TDC), and the analytic form for the lower and upper TDCs, $\lambda_L$ and $\lambda_U$ respectively, are:
\begin{equation}
\label{eq:TDC}
\begin{split}
\lambda_U &=\lim_{u\rightarrow 1^{-}} P\left(Y_2>F_2^{-1}(u)|Y_1>F_1^{-1}(u)\right) \\
&= \lim_{u\rightarrow 1^{-}} \frac{ P\left( F_2(Y_2) > u, F_1(Y_1) > u \right) }{ P(F_1(Y_1) > u) }
= \lim_{u\rightarrow 1^{-}} \frac{1-2u+C(u,u)}{1-u} \in [0,1] \\
\lambda_L &= \lim_{u \rightarrow 0^{+}} P\left(Y_2 \leq F_1^{-1}(u) | Y_1 \leq F_1^{-1}(u) \right) 
= \lim_{u \rightarrow 0^{+}} \frac{C(u, u)}{u} \in [0,1] ,
\end{split}
\end{equation}
as defined in \cite{Sibuya1960}, \cite{Nelsen2007}, and \cite{Sweeting2013},
where the limit in $\lambda_U$ means 1 is approached from below, and the limit in $\lambda_L$ means 0 is approached from above.
If the limit exists and the result is between zero and one, then tail dependence exists; if the result is zero then there is no tail dependence. 
It is worth noting that in extreme value theory, if some copulas have $\lambda=0$, they can still be discriminated from one another based on rate of convergence of the above conditional probability to zero (\citealp{Ledford1996}; \citealp{Engelke2019}). However, given the actuarial application focus of this article, it is beyond the scope of this work and will not be considered.

There are various ways to estimate the parameter $\theta$ of a given parametric copula $C_{\theta}$, which also depends on estimation of the marginal distributions. 
A popular method is rank-based estimations, through which the marginals are estimated nonparametrically based on the ranked observations. This includes estimation based on Kendall's tau or Spearman's rho, or more commonly maximum pseudo-likelihood estimation (MPLE) (we note that it has been argued in the literature that the use of AIC or BIC-like formulae based on pseudo-likelihood is not strictly theoretically correct, despite its popularity (\citealp{Gronneberg2014})).
Alternatively, if the marginals are not rank-based, i.e. using full parametric estimation, the popular maximum likelihood estimation (MLE) can be used: either maximising the joint likelihood of copula and marginal distributions, or a method called Inference Function for Margins (IFM) (\citealp{Joe1996}; \citealp{Joe2005}). The latter first fits each marginal by maximum likelihood, then fits the copula conditional on the fitted univariate parameters. For a good review, see \cite{Genest2007}. 
If both the copula and marginals are estimated nonparametrically, it leads to empirical copula. This approach has the benefit that when estimating a copula it does not require any assumptions with respect to the distributional nature of the copula or the underlying data, provided that there is a sufficient quantity of data to facilitate its fitting. 
In the bivariate setting, suppose $(Y_{1i},Y_{2i})$ for $i \in {1,…,N}$ is a random data sample of length $N$, 
constructed similarly to the empirical distribution function. The empirical copula is defined as (\citealp{Frahm2005}) 
\begin{equation}
\label{eq:empirical_copula}
\widehat{C}(u,v) = \frac{1}{N} \sum_{i=1}^{N} \mathbb{I}( \frac{R_{1i}}{N} \leq u, \frac{R_{2i}}{N} \leq v),
\end{equation}
where for example $R_{1i}$ is the rank of observation $Y_{1i}$, i.e. $R_{1i} = \sum_{m=1}^{N} \mathbb{I}(Y_{1m} \leq Y_{1i})$.
Equivalently, let $Y_{1(1)} \leq \cdots \leq Y_{1(N)}$ and $Y_{2(1)} \leq \cdots \leq Y_{2(N)}$ be the ordered sample values. The empirical copula can then be defined as (\citealp{Dobric2005})
\begin{equation*}
\widehat{C}(\frac{i}{N}, \frac{j}{N}) = \frac{1}{N} \sum_{m}^{N} \mathbb{I}(Y_{1m} \leq Y_{1(i)}, Y_{2m} \leq Y_{2(j)}).
\end{equation*}
Table~\ref{tab:copula_estimation_methods} shows a summary of the estimation methods -- more details can be found in \cite{Hofert2018book}. 
In this article, MPLE, MLE and empirical copula all are used for analysis due to their popularity: the empirical copula is used for empirically estimating the TDCs, while MPLE and MLE are used for copula fitting in simulation studies and Irish GI data study respectively. 

\begin{table}[h!]
	\centering
	\caption{Different estimation methods for copula models.}
	\label{tab:copula_estimation_methods}
	\begin{tabular}{lcc}
		\toprule[.15 em]
		\ & \multicolumn{2}{c}{ \textbf{Marginals} } \\ 
		\textbf{Copula} & nonparametric &  parametric \\
		nonparametric & \cellcolor{lightgray!60}\textit{empirical copula} &\cellcolor{lightgray!60} \textit{plug-in} \\
		parametric & \cellcolor{lightgray!60}\textit{pseudo-likelihood} &\cellcolor{lightgray!60} \textit{likelihood} \\
		\bottomrule[.15 em]
	\end{tabular} 
\end{table}


When the marginal distributions are estimated parametrically, covariates can be incorporated in the marginals following a GLM framework as extra information to assist in marginal and further parametric copula estimations. This is sometimes called copula regression (\citealp{Yan2007}; \citealp{Masarotto2017}; \citealp{Kramer2013}). 
Furthermore, copulas can also be mixed to form finite mixtures of copulas for complex and heterogeneous dependence structures, see \cite{Kosmidis2016} and \cite{Arakelian2014}. 
Therefore, copula estimation can be performed via such finite mixtures of copula regressions, following the standard mixture models construction and inference such as that found in \cite{Gormley2011}. 
Suppose there are bivariate identical and independently distributed data $\bm{y}_1, \ldots, \bm{y}_N$ of size $N$, which consist of $G$ components.  
There also exist vectors of covariates $\bm{x}_1, \ldots, \bm{x}_N$, based on which let $\bm{X}_j$ be the covariate design matrix for margin $j$, for $j = 1, 2$; i.e. covariates are not necessarily the same for both marginals. 
Each component $g$ can be specified by a copula function $C_g$ and two marginal distributions of CDF $F_{jg}$ and PDF $f_{jg}$, for $g = 1, \ldots, G$: now let $\bm{\alpha}_g$ be the copula parameter, $\bm{\beta}_{jg}$ be the regression coefficients of margin $j$, $\gamma_{jg}$ be any distribution specific parameters of margin $j$ (e.g. shape parameters of gamma distributions), so that each margin follows a specified distribution within a GLM framework such as $\mu_{ji} = \delta^{-1}(\bm{x}_{ijg}^{\top} \bm{\beta}_{jg})$ where $\delta$ is the link function.
Then each component can be regarded as a bivariate distribution (or component-specific copula regression), denoted as $h_g(\bm{y}_i;\bm{\theta}_g)$, where $\bm{\theta}_g = \left\{\bm{\alpha}_g, \bm{\beta}_{jg}, \bm{\gamma}_{jg} \right\} $. 
Based on Equation~\ref{chap3:eq:copula_density}, the observed data likelihood of the mixture copula model is
\begin{equation}
\begin{split}
\mathcal{L}(\bm{\theta}) &= \prod_{i=1}^{N} \sum_{g=1}^{G} \tau_g h_g(\bm{y}_i;\bm{\theta}_g) \\
&= \prod_{i=1}^{N} \sum_{g=1}^{G} \tau_g c_g\left(F_1(y_{1i};\bm{\beta}_{1g},\bm{\gamma}_{1g}), F_2(y_{2i};\bm{\beta}_{2g},\bm{\gamma}_{2g});\bm{\alpha}_g\right) f_1(y_{1i};\bm{\beta}_{1g},\bm{\gamma}_{1g}) f_2(y_{2i};\bm{\beta}_{2g},\bm{\gamma}_{2g})
\end{split}
\label{eq:mixture_copulas_model}
\end{equation}
where $\tau_g$ is the mixing proportion, i.e. $\sum_{g=1}^{G} \tau_g = 1$, and $c_g$ is the density of a specified copula.
This model can be estimated via the EM algorithm (\citealp{Dempster1977}). Further details can be found in Appendix~\ref{app:EM_mixture_copula_regression}. 

It is also worth noting that in this work copula models are fitted to the whole data set for modelling dependence throughout the distribution, rather than to only extreme data points via threshold-based approaches and censored likelihood (\citealp{Huser2016}). While both are popular methods in estimating the extent of extreme dependence, availing of all data to understand the overall dependence and copulas' goodness-of-fit to the data is of interest, particularly through the second method of employing BMA in Section~\ref{sec:method2} where the target is that the BMA simulated data is the most similar to the original data, and through the mixture of copulas with covariates approach in Section~\ref{sec:irish} where the method is used for clustering and predictions of the overall data.

\subsection{Empirical TDC estimation}
\label{sec:utdc}

The TDCs are defined on a limiting measure and are functions of the copula, so they cannot be calculated exactly unless the true data-generating copula is known. 
For non-standard copulas or distributions the analytical form of the TDCs may not be available. Hence, estimation of the tail dependence is difficult. 
In the literature there are many methods proposed to estimate the upper and lower TDCs nonparametrically and empirically using methods such as the empirical copula, tail copula, Kendall's tau, and copula approximation (via maximum and independence copulas). For a detailed review, see \cite{Frahm2005}, \cite{Fischer2006}, and \cite{Sweeting2013}. 
Note that not all proposed estimators in the literature work well from the authors' experience through various simulation studies. 
A range of estimators is selected for consideration, not to review and compare which of the empirical or nonparametric estimators work best, but to use them as a benchmark for Bayesian model averaging approach:
\vskip .2 cm
\begin{itemize}
	\item \textbf{Estimator 1} (\citealp{Dobric2005}; \citealp{Fischer2006}): \\
	\begin{equation*}
	\widehat{\lambda}_U^{(1)} = \frac{\widehat{C} \left( (1-\frac{k}{N},1] \times (1-\frac{k}{N},1] \right) }{ 1-(1-\frac{k}{N})}, \hskip .3cm
	\widehat{\lambda}_L^{(1)} = \frac{\widehat{C}(\frac{k}{N},\frac{k}{N})}{\frac{k}{N}} .
	\end{equation*}
	This is a direct result combining the empirical copula definition in Equation~\ref{eq:empirical_copula} with the definitions of TDCs in Equation~\ref{eq:TDC}. The empirical copula with sets in $\widehat{\lambda}_U^{(1)}$ can be regarded as an empirical survival copula. The value (i.e. the cutoff value) in this proposed method $k\approx\sqrt{N}$ has been shown to be appropriate in \cite{Dobric2005}. 
	\vskip .2cm
	\item \textbf{Estimator 2} (\citealp{Dobric2005}; \citealp{Fischer2006}): \\
	\begin{equation*}
	\widehat{\lambda}_U^{(1)} = 2 - \frac{\log \widehat{C}(1-\frac{k}{N}, 1-\frac{k}{N})}{\log (1-\frac{k}{N})} , \hskip .3cm
	\widehat{\lambda}_L^{(1)} = 2 - \frac{\log \widehat{C} \left( (\frac{k}{N},1] \times (\frac{k}{N},1] \right) }{\log (1-\frac{k}{N})} ,  	
	\end{equation*}
	This result is based on \cite{Coles1999}, which provides asymptotically equivalent versions of Equation~\ref{eq:TDC}:
	\begin{equation*}
	\lambda_U = 2 - \lim_{u\rightarrow 1^{-}} \frac{\log C(u,u)}{\log(u)}, \hskip .3cm \lambda_L = 2 - \lim_{u\rightarrow 0^{+}} \frac{\log(1-2u+C(u,u))}{\log(1-u)} .
	\end{equation*}
	Similarly, the value $k\approx\sqrt{N}$ has been shown to be appropriate.
	\vskip 0.2cm	
	\item \textbf{Estimator 3} (\citealp{Dobric2005}; \cite{Fischer2006}):  \\ 
	$\widehat{\lambda}_U^{(3)}$ and $\widehat{\lambda}_L^{(3)}$ are least squares (OLS) solutions to to the ad-hoc regression models respectively 
	\begin{equation*}
	\begin{split}
	\widehat{C}((1-\frac{i}{N},1] \times (1-\frac{i}{N},1] ) &= \lambda_U \frac{i}{N}+\epsilon_i, \\
	\widehat{C}(\frac{i}{N}, \frac{i}{N}) &= \lambda_L \frac{i}{N}+\epsilon_i,
	\end{split}
	\end{equation*}
	for $ i=1,\ldots, k$ and error term $\epsilon_i$. Again $k$ is determined as a function of $N$, and $k \approx \sqrt{N}$ has been shown to be appropriate. 
	\vskip 0.2cm	
	\item \textbf{Estimator 4} (\citealp{Caillault2005}): \\
	For $i \in \{1,\ldots,N-1 \}$, let the two statistics (defined similarly to the TDC definition in Equation~\ref{eq:TDC})
	\begin{equation}
	\label{chap3:eq:tdc_trajectory}
	\widehat{\lambda}_U(\frac{i}{N}) = \frac{1-(\frac{2i}{N})+\widehat{C}(\frac{i}{N}, \frac{i}{N})}{1-\frac{i}{N}} , \hskip .4cm
	\widehat{\lambda}_L(\frac{i}{N}) = \frac{\widehat{C}(\frac{i}{N}, \frac{i}{N})}{\frac{i}{N}}
	\end{equation} 
	represent the empirical trajectories of $\lambda_U$ and $\lambda_L$ on $[0,1]$ with respect to $\frac{i}{N}$, where $\widehat{C}$ is the empirical copula. 
	On $[0,1]$, $\lambda_U$ and $\lambda_L$ are decreasing and increasing functions of $u$ respectively. Therefore, graphically choose the last value $i=i_0$ such that the decreasing (or increasing) property of $\lambda_U$ (or $\lambda_L$) is preserved, which leads to 
	\begin{equation*}
	\widehat{\lambda}_U^{(4)} = \widehat{\lambda}_U(\frac{i_0}{N}), \hskip .4cm \widehat{\lambda}_L^{(4)} = \widehat{\lambda}_L(\frac{i_0}{N}).
	\end{equation*}
	A detailed discussion of how to best choose $i_0$ using a bootstrap method can be found in \cite{Caillault2005}.    
	\vskip .2cm	
	\item \textbf{Estimator 5} (\citealp{Schmidt2006}): \\ 
	\begin{equation*}
	\begin{split}
	\widehat{\lambda}_U^{(5)} &= \widehat{\Lambda}_{U}(1,1) \hskip .2cm \text{or} \hskip .2cm \widehat{\Lambda}_{U}^{EVT}(1,1) , \\
	\widehat{\lambda}_L^{(5)} &= \widehat{\Lambda}_{L}(1,1) \hskip .2cm \text{or} \hskip .2cm \widehat{\Lambda}_{L}^{EVT}(1,1) ,
	\end{split}
	\end{equation*}
	where $\widehat{\Lambda}_{U}(x,y)$ is the empirical upper tail copula and $\widehat{\Lambda}_{U}^{EVT}(x,y)$ is an upper tail copula estimated by a stable tail-dependence function. 
	The definitions of (empirical) tail copula and stable tail-dependence functions and details on choosing optimal parameters can be found in \cite{Schmidt2006}.
	Generally, both methods return similar results, with the difference depending on sample size and choice of parameters used in the estimation.  This nonparametric estimator can be readily implemented by the R package \textsf{FRAPO} (\citealp{Pfaff2016}).
\end{itemize}

These estimations have different accuracies, depending on the data sets used.
Although generally they have worked well in simulation studies, from the authors' experience the best estimator is not consistent and varies from case by case. 
Moreover, when dealing with very small or zero TDCs, the nonparametric estimators generally do not work well - either over-estimating or not able to distinguish the cases where tail dependence does not exist.
More details on their performance can be found in Sections~\ref{sec:simstudy3} and~\ref{sec:simstudy2}.
The authors acknowledge that, parallel to the nonparametric and empirical TDC estimation literature, in extreme value theory the TDCs of many non-standard distributions or copulas can sometimes still be derived theoretically under some regularity conditions, for example see \cite{Engelke2019}, which is an active research field of mathematical probability. However, given the application focus of this paper, this direction is not pursued.

\section{Bayesian model averaging}
\label{sec:BMA}

When there are multiple competing copulas that all show relatively similar goodness of fit, typically only a single best copula is selected and treated as the true copula underlying the data generating process. 
This approach ignores model uncertainty. As a result, the quantity of interest may be over- or under-estimated. 
A popular alternative method is to select and combine a group of models with relatively similar goodness of fits with respect to the quantity of interest in order to calculate a more statistically robust unified estimate. 
Combining models has been investigated widely in the statistical literature, and BMA is a common method that involves selecting models using posterior model probabilities. For a more detailed review and its application examples see \cite{Draper1995}, \cite{Hoeting1999} and \cite{Hu2018}. 
Suppose $\Delta$ is the quantity of interest, such as the TDC of the copula, and 
$C_1, \ldots, C_K$ are selected copulas being considered.
The posterior distribution of $\Delta$ given data $\mathcal{D}$ is
\begin{equation*}
P(\Delta|\mathcal{D}) = \sum_{k=1}^{K} P(\Delta | C_k, \mathcal{D}) P(C_k | \mathcal{D}) ,
\end{equation*}
This is an average of the posterior distributions $P(\Delta | C_k, \mathcal{D})$ under each of the models $C_k$ considered, weighted by their posterior model probabilities $P(C_k | \mathcal{D})$.
Due to the intractable integral required in calculating the model weights, the Bayesian Information Criterion (BIC) (\citealp{Schwarz1978}) can be employed to approximate the posterior model probability for BMA (\citealp{Madigan1994}; \citealp{Hoeting1999}). 
If all the copulas are equally likely \textit{a priori}, the weight for copula $C_k$ is given by: 
\begin{equation}
\label{eq:BMA_weights}
W_{C_k} = P(C_k | \mathcal{D}) \simeq \frac{\exp(-\frac{1}{2} BIC_k)}{\sum_{k^{\prime}=1}^{K} \exp(-\frac{1}{2} BIC_{k^{\prime}})} . 
\end{equation}
$BIC_k$ is the BIC value of the $k^{th}$ copula, calculated as:
$$BIC_k = -2 \log \hat{\ell}_k + p_k \log N$$
with $N$ being the total number of observations, $\hat{\ell}_k$ the fitted copula (pseudo-) log-likelihood and $p_k$ the number of parameters estimated for copula $C_k$.
This means that the weights are being determined in the context of the goodness of fit of the proposed copulas and their respective levels of parsimony, hence the method can be extended to accommodate more complex nested copula structures. 
Note that if all copulas to be selected are one-parameter Archimedean, then $p_k$ and $N$ are the same among all copulas and the BICs are simply proportional to the fitted log-likelihood $\hat{\ell}$.

\section{Copula averaging for TDC estimation}
\label{sec:methods}

In this article, the tail dependence coefficient (TDC) is the main target quantity of interest; more specifically, the objective of the method is to improve the estimation of the upper and lower TDCs using BMA, compared to TDCs from individual fitted copula models (including finite mixtures of copula models).
More importantly, although nonparametric empirical estimation has been popular in the literature, our review has shown that their results can be unstable and sensitive to specific data, or sometimes even need subjective judgement, especially for weak tail dependence or non-existent tail dependence.   
Therefore, model-based estimation, especially using BMA to account for model uncertainty and to combine a wider variety of models, provides a more robust modelling approach.
Nevertheless, nonparametric estimation can still play an important role in TDC estimation. 
When the true upper TDC is unknown, an estimate of this value empirically may reflect the approximate nature of dependence in the data.
This becomes important when assessing the improvement or merit of BMA estimation compared with using individual copula models.
It is also worth noting that, when implementing BMA, the quantity of interest must be consistently defined across all models.
Although the overall target of the proposed method is the TDCs $\lambda_L$ and $\lambda_U$, the quantity of interest when implementing BMA could vary. Depending on this, two methods of using BMA are proposed and detailed below. 

\subsection{Method 1}
\label{sec:method1}

In the context of treating TDC as the quantity of interest within BMA, this method directly blends the calculated $\widehat{\lambda}$ values from multiple fitted copula models to produce a single unified estimate of the TDC.
Suppose $\widehat{\lambda}_{C_k}$ is the TDC estimated from a fitted copula $C_k$ to the data. 
The BMA weighted estimate of TDC, $\widehat{\lambda}_{BMA}$, across multiple fitted copulas $C_k$, for $k=1,...,K$ can then be calculated as
\begin{equation}
\widehat{\lambda}_{BMA} = \sum_{k=1}^{K} W_{C_k}\widehat{\lambda}_{C_k} ,
\end{equation} 
where $W_{C_k}$ is calculated using Equation~\ref{eq:BMA_weights}. 
Calculation of $\lambda$ for the copulas used in this article can be found in Table~\ref{tab:copula_property}.   
Note that not all copulas have non-zero TDCs, such as the Gaussian ($\lambda_L = \lambda_U = 0$) or the Clayton ($\lambda_U =0$). Therefore, when the quantity of interest is $\lambda$, caution needs to be exercised - for example the choice of whether to fit all available copulas or use only the ones where $\lambda$ exists. But at the same time it leads to an advantage of such a method -- it is able to distinguish from the cases when the TDCs are zero, unlike the empirical estimation.
It is worth remembering that when implementing BMA, the quantity of interest must be consistently defined, and TDC is a consistent definition based on Equation~\ref{eq:TDC}.
By contrast, because each copula is defined differently (e.g. different Archimedean copulas use a different generator function and different parameter support), copula parameters cannot be directly blended by BMA.

\subsection{Method 2}
\label{sec:method2}

This method treats the proportion of the sample size $N$ arising from each of the available copulas as the quantity of interest, while the modelling target is still the TDCs. 
After calculating the BMA posterior model weights $W_{C_k}$, a BMA simulated data set of the same sample size $N$ as the original data is simulated proportionally from each fitted copula, with the simulation size $n_k$ from copula $C_k$ being proportional to its BMA weight $W_{C_k}$, i.e. $N=\sum_{k=1}^{K} n_k$ where $n_k=N W_{C_k}$. 
The purpose of this mixture simulation is to average the ``similarity-to-the-given-data" of each fitted copula so that (1) this simulation process may better reflect the underlying true and often complex data-generating process; (2) the new BMA set combines characteristics of the tail behaviour from each fitted copula; (3) the new BMA set is the most similar set to the given original sample data compared to simulations from individual fitted copulas. 
This method is similar to the ``minimum distance" approach of copula selection mentioned in \cite{Genest1993}, \cite{Frees1998}, and \cite{Caillault2005}. 
One of the targets is to achieve good general fit to the data, but more importantly, because this is a simulation-based method that could reflect clearer dependence characteristics of the given data based on the fitted copulas, the empirical TDCs that are estimated from the BMA set can be considered an average value of individually fitted copulas and be closer to the true value. If repeated many times and each time the TDCs are estimated empirically, then the BMA-induced TDCs can be calculated by taking the average (similar to bagging or bootstrapping).
Note that this method works best provided that a good nonparametric TDC estimation method is employed.
Because the motivation of such a method is that the BMA simulated data set is the most similar set to the data, data or distribution similarity comparison measures such as Wasserstein distance (\citealp{Vallender1974}; \citealp{Schuhmacher2018}) or the distance measure used in the ``minimum distance" approach in \cite{Frees1998} or \cite{Caillault2005} can be employed for verification of such motivations. The $L^2$ distance is defined in \cite{Caillault2005} as  
\begin{equation*}
D_2 = \sum_{t=0}^{N} \left| C_{\widehat{\theta}}(\widehat{F}_{1}(x_t), \widehat{F}_{2}(y_t)) - \widehat{C}(\frac{t}{N}, \frac{t}{N}) \right| ^2 ,
\end{equation*}
where $\widehat{F}_{1}(x_t)$ is empirical distribution function, $C_{\widehat{\theta}}$ is a fitted copula and $\widehat{C}$ is an empirical copula.
In this paper, a slightly modified $L^2$ distance is used (since the purpose is not copula selection), which is defined as:
\begin{equation*}
D_{2}^{\star} = \left( \sum_{t=0}^{N} \left| \widehat{C}_1 (\frac{t}{N}, \frac{t}{N}) - \widehat{C}_2 (\frac{t}{N}, \frac{t}{N}) \right|^2 \right)^{1/2} ,
\end{equation*}  
where $\widehat{C}_1$ and $\widehat{C}_2$ are empirical copulas for two (pseudo-) data sets.

\section{Simulation study I}
\label{sec:simstudy3}

To examine the merits of using BMA and to compensate for the fact that the real Irish insurer data set in Section~\ref{sec:irish} is confidential and not publicly available, an artificial data set is simulated and studied.
The main issue with artificial simulation is that, if the data are simply simulated from one copula, for example the t-copula, when fitting different copulas to the data the best fit will naturally be t, most likely with a posterior model probability of $1$ (see Appendix~\ref{app:simstudy4}). 
Hence, simulating a data set with a more complex dependence structure is required. 
Furthermore, the authors note that the proposed model-based BMA estimation of TDC works best when the underlying true copula or data distribution is non-standard. When the true copula or distribution is of a standard form (such as within the Archimedean family) with analytical TDC expressions available, naturally the true model can be found if the pool of potential copula options is large enough. 
In real-world insurance data sets, the dependence structures are typically more complex than that provided by any single bivariate copula.  
Therefore, ideally the data used in a simulation study should have the following conditions for demonstrating the validity of the BMA approach:
\begin{enumerate}
	\item Its dependence structure is complex so that multiple copulas have competing goodness-of-fit. 
	\item The simulating distribution or copula is non-standard, and does not have an analytical form. 
	\item The true TDC values are known to use as benchmarks for the BMA approach. 
\end{enumerate}
However, conditions (2) and (3) are hardly satisfied simultaneously. Therefore, this first simulation study considers a case where only conditions (1) and (2) are met, and the second simulation study in Section~\ref{sec:simstudy2} considers a case where conditions (1) and (3) are met.

In the continuous case of general insurance claims data, a gamma-type distribution is commonly used for claim severity. 
For modelling dependence between joint risks, an appropriate distribution is the bivariate gamma, which has been shown to model insurance claims data well, for example see \cite{Hu2019}. 
In this simulation study the bivariate gamma definition provided by \cite{Cheriyan1941} and \cite{Ramabhadran1951} is considered: 
let $X_1, X_2, X_3$ be independent gamma random variables, where $X_i \sim Gamma(\alpha_i, \beta)$ for $i=1, 2, 3$, with different shape parameters $\alpha_i > 0$ and a common rate parameter $\beta > 0$. Then vector $\boldsymbol{Y}$ is defined as:
\begin{equation*}
\boldsymbol{Y} = \begin{bmatrix} Y_1 \\ Y_2  \end{bmatrix}  =  \begin{bmatrix} X_1+X_3 \\ X_2+ X_3  \end{bmatrix} \sim \BG(\alpha_1, \alpha_2, \alpha_3, \beta) \ ,
\end{equation*} 
with distribution density
\begin{equation*} 
\begin{split}
f_{Y_1, Y_2}(y_1,y_2) = \frac{\beta^{\alpha_1+\alpha_2+\alpha_3}e^{-\beta(y_1+y_2)}}{\Gamma(\alpha_1)\Gamma(\alpha_2)\Gamma(\alpha_3)} \int_{x_3=0}^{\min(y_1, y_2)} e^{\beta x_3} x_3^{\alpha_3 - 1} (y_1 - x_3)^{\alpha_1 -1} (y_2 - x_3)^{\alpha_2 -1} dx_3 \  .
\end{split}
\end{equation*} 
Due to the fact that the distribution function does not have an analytical form, there are no explicit copula families defined for this type of distribution (unlike some similarly constructed distributions such as the Raftery copula (\citealp{Raftery1984}; \citealp{Nelsen2007}). 
Therefore, the true TDCs are unknown (if they exist).
This gives a complex dependence structure with a non-standard copula while being similar to real-world insurance claims data.
More details of this type of distribution can be found in \cite{Hu2019}.

\begin{figure}[h!]
	\centering
	\begin{minipage}{.6\textwidth}
		\includegraphics[width=\linewidth]{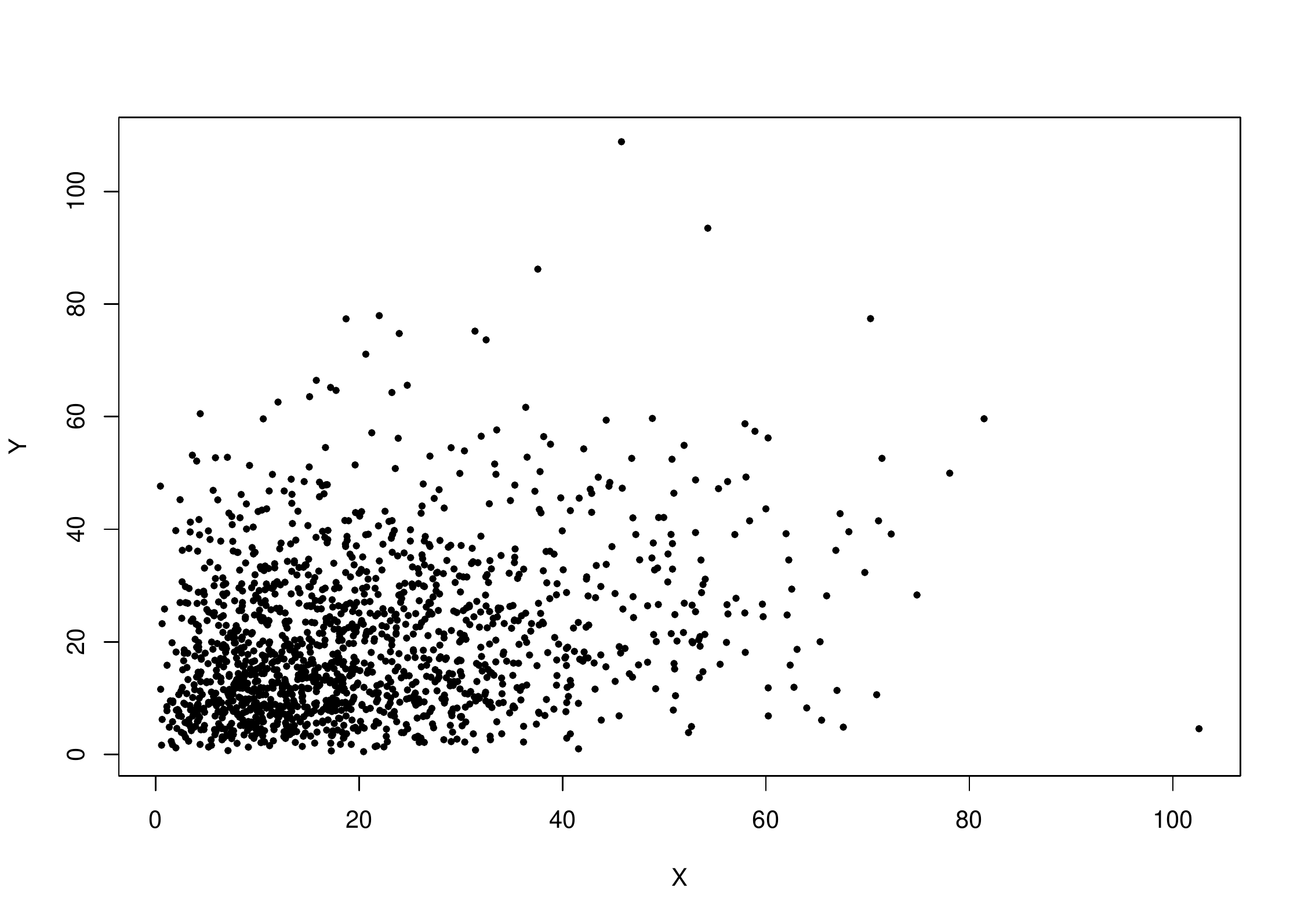}
		\caption*{(a)}	
	\end{minipage}
	\begin{minipage}{.6\linewidth}
		\includegraphics[width=\linewidth]{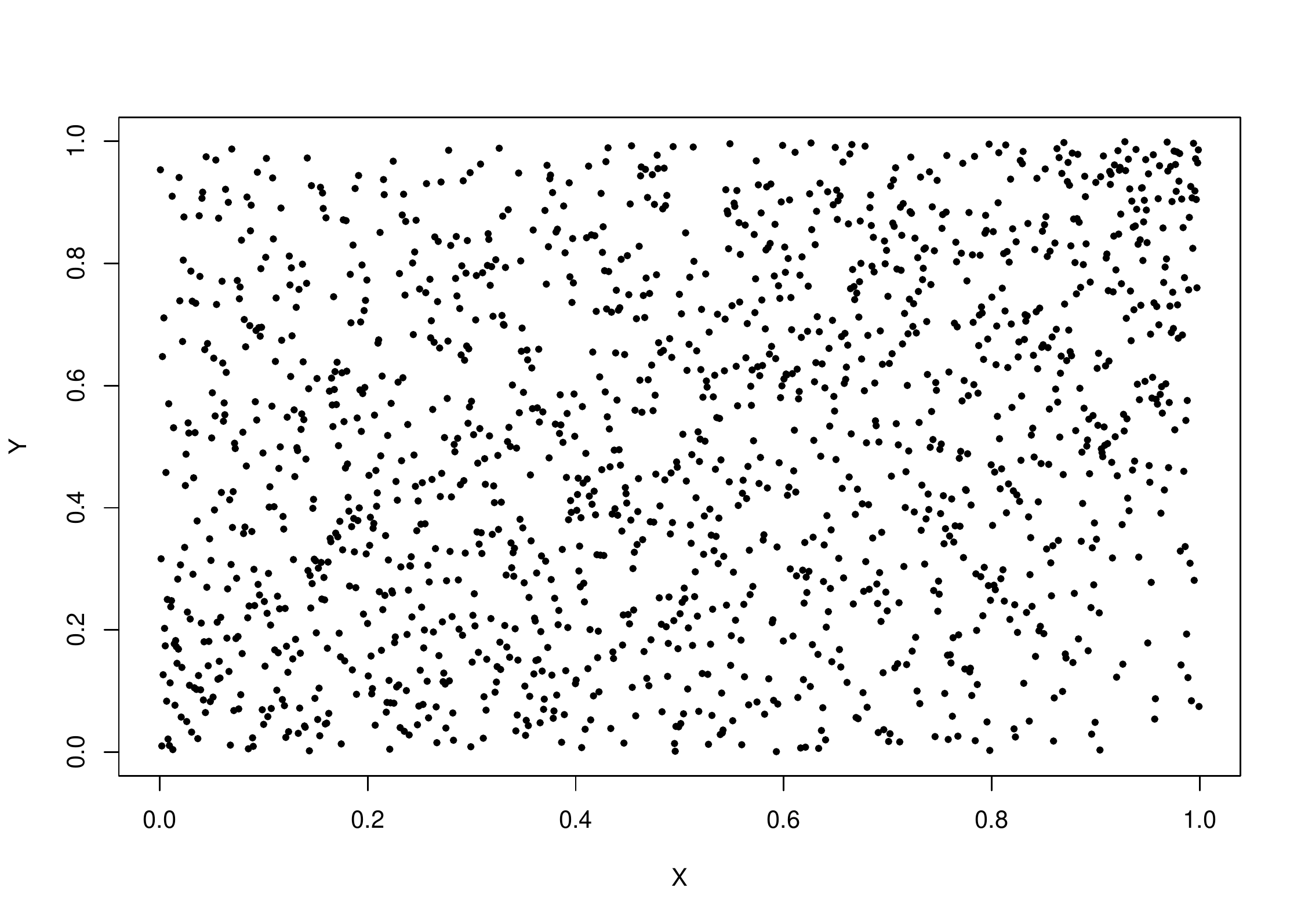}		
		\caption*{(b)}
	\end{minipage}
	\caption{Plot of the simulation data: (a) shows the simulated data in the data space; (b) shows the pseudo-data according to their rank order in the copula $[0,1]\times [0,1]$ space.}
	\label{fig:simstudy3_plot}
\end{figure}

$N=1500$ data points are generated from a bivariate gamma distribution $BG(\alpha_1=1.5, \alpha_2=1.5, \alpha_3=0.5, \beta=0.1)$. 
Therefore, the two marginals $X$ and $Y$ are both gamma$(2, 0.1)$. Figure~\ref{fig:simstudy3_plot} shows the simulated data in both the data domain and the copula domain. 
From Figure~\ref{fig:simstudy3_plot}(b) it is obvious that there is some concentration of points in the top right corner, indicating a degree of upper tail dependence. By contrast, there is no obvious point concentration in the lower left corner, which indicates that there is no or at most very weak lower tail dependence. This is expected since the given bivariate gamma distribution is designed to have heavy joint right tails.

Since there is no explicit form of copula defined for this type of bivariate gamma distribution, the true TDCs need to be estimated using the empirical estimators from Section~\ref{sec:utdc}, with the results given in Table~\ref{tab:simstudy3_TDC_estimation}.  
This verifies that the upper tail dependence is stronger than the lower tail dependence. Because of the unstable empirical estimates of the TDCs, and the fact that these nonparametric estimators perform poorly for small TDCs, and as suggested by Figure~\ref{fig:simstudy3_plot}(b), it could be argued that both TDCs are potentially over-estimated, especially the lower TDC.

For illustration simplicity, the MPLE method is used for copula fitting in this example. When fitting different copulas to the pseudo-data, the summaries of the fitted copulas are shown in Table~\ref{tab:simstudy3_copula_summary}, including their BICs, BMA weights and the estimated $\widehat{\lambda}_U$ and $\widehat{\lambda}_L$. Note that the BMA weights are spread across multiple copulas including the t, Gaussian, Gumbel, survival Clayton and Frank. The weight of the Joe is only 0.01\%, which is negligible. All copulas except the t do not have a lower TDC while the t accounts for only 5\% model weight. This suggests that lower tail dependence is very weak or close to zero in the data, as expected. \\

\begin{table}[h!]
	\centering
	\caption{Empirically estimated upper and lower TDCs $\widehat{\lambda}_U$ and $\widehat{\lambda}_L$ for the simulation study I data based on estimators in Section~\ref{sec:utdc}.}
	\label{tab:simstudy3_TDC_estimation}
	\begin{tabular}{lrrrrrrc}
		\toprule[.15 em]
		\ & $\widehat{\lambda}^{(1)}$ & $\widehat{\lambda}^{(2)}$ & $\widehat{\lambda}^{(3)}$ & $\widehat{\lambda}^{(4)}$ & $\widehat{\lambda}^{(5)}$ & mean & range \\
		\midrule
		$\widehat{\lambda}_{U}$ & 0.167 & 0.146 & 0.099 & 0.111 & 0.132 & 0.131 & [0.099, 0.167] \\
		$\widehat{\lambda}_{L}$ & 0.103 & 0.081 & 0.109 & 0.085 & 0.105 & 0.097 & [0.081, 0.109] \\
		\bottomrule[.15 em]
	\end{tabular} 
\end{table}

\begin{table}[h!]
	\centering
	\caption{Summary of the fitted copulas, including fitted BIC values, BMA weights ($W_j$) calculated based on BICs and the estimated $\widehat{\lambda}_U$ and $\widehat{\lambda}_L$ based on the estimated copula parameters.}
	\label{tab:simstudy3_copula_summary}
		\begin{tabular}{lrrrrrrrrrrrr}
			\toprule[.15 em]
			\ & t & Gaussian & Joe & survival Joe& Gumbel \\
			\midrule
			BIC & -89.35 & -91.94 & -76.95 & -36.00 & -94.26  \\
			BMA $W_j$ & 0.052 & 0.190 & 0 & 0 & 0.607 \\
			$\widehat{\lambda}_{U}$ & 0.004 & 0 & 0.250 & 0 & 0.201 \\
			$\widehat{\lambda}_{L}$ & 0.004 & 0 & 0 & 0.192 & 0 \\
			\midrule[.15 em]
			\ & survival Gumbel & Clayton & survival Clayton & Frank  \\
			\midrule
			BIC & -67.88 & -48.98 & -90.94 & -88.54 \\
			BMA $W_j$ & 0 & 0 & 0.116 & 0.035 \\
			$\widehat{\lambda}_{U}$ & 0 & 0 & 0.117 & 0 \\
			$\widehat{\lambda}_{L}$ & 0.183 & 0.061 & 0 & 0  \\
			\bottomrule[.15 em]
	\end{tabular} 
\end{table}

\newpage
\noindent\textbf{Method 1} \\

Using method 1 the fitted TDCs are directly averaged to calculate a unified TDC based on the fitted copulas' weights, with the result and comparison to results of other estimations shown in Table~\ref{tab:simstudy3_method1_result}. The BMA-induced TDCs are the closest to the potential true values, which can be regarded as the (mean of) empirical estimations, compared to the best fitting Gumbel copula which has the highest individual model weight and will be used as a benchmark best individual model. 

\begin{table}[H]
	\centering
	\caption{Method 1 results for the simulation study I: comparison of TDCs estimated by the empirical estimators, the best fitting Gumbel copula and BMA.}
	\label{tab:simstudy3_method1_result}
	\begin{tabular}{l|ccc}
		\toprule[.15 em]
		\ & Empirical & Gumbel & BMA \\
		\midrule
		$\widehat{\lambda}_U$ & [0.099, 0.167] & 0.201 & 0.1359 \\
		$\widehat{\lambda}_L$ & [0.081, 0.109] & 0 & 0.0002 \\ 
		\bottomrule[.15 em]
	\end{tabular}
\end{table}

\noindent\textbf{Method 2} \\

Using method 2, a new BMA set of simulated observations of the same size $1500$ is generated combining simulated data from each of the fitted copulas, with the number of observations from each proportional to its BMA weight. The motivation is that simulation using BMA can best reflect the characteristics of multiple well-fitting copulas, and hence also the characteristics of the given data. In turn the empirical TDCs that can be estimated from the BMA set can best represent and be closest to the true value. Such motivations can be verified by measuring the distance between the BMA set and the observed data, using either Wasserstein or $L^2$ distance. 
Table~\ref{tab:simstudy3_method2_result} shows the empirical estimates of the TDCs from simulated data from of the best fitting individual copula (the Gumbel) and from the BMA set. Regardless of the choice of empirical estimator, BMA always provides superior estimations. The distance measures in Table~\ref{tab:simstudy3_method2_distance} also show the BMA set is much closer to the original data set than the best fitting Gumbel copula simulated data.

\begin{table}[!h]
	\centering
	\caption{Mean of the empirical nonparametric estimates across 1000 repetitions, using proposed method 2. In each iteration empirical TDCs are estimated using all five estimators on simulations from the best fitting individual copula and the BMA simulated data set. Standard deviations are shown in brackets.}
	\label{tab:simstudy3_method2_result}
	\begin{tabular}{lcc||ccc}
		\toprule[.15 em]
		\ & Gumbel & BMA & \ & Gumbel & BMA \\
		\midrule
		$\overline{\lambda}_U^{(1)}$ & 0.249 (0.06) & 0.211 (0.06) & $\overline{\lambda}_L^{(1)}$ & 0.053 (0.04) & 0.058 (0.04) \\
		$\overline{\lambda}_U^{(2)}$ & 0.229 (0.06) & 0.192 (0.06) & $\overline{\lambda}_L^{(2)}$ & 0.030 (0.04) & 0.034 (0.04) \\
		$\overline{\lambda}_U^{(3)}$ & 0.213 (0.06) & 0.173 (0.06) & $\overline{\lambda}_L^{(3)}$ & 0.042 (0.03) & 0.048 (0.03) \\
		$\overline{\lambda}_U^{(4)}$ & 0.174 (0.05) & 0.143 (0.05) & $\overline{\lambda}_L^{(4)}$ & 0.033 (0.03) & 0.037 (0.03) \\ 
		$\overline{\lambda}_U^{(5)}$ & 0.215 (0.06) & 0.177 (0.06) & $\overline{\lambda}_L^{(5)}$ & 0.053 (0.04) & 0.057 (0.04) \\
		\bottomrule[.15 em]
	\end{tabular}
\end{table}

\begin{table}[!h]
	\centering
	\caption{Comparison of distances between simulations from the best fitting individual copula and the BMA simulated data set to the simulation study I data set. The underlined values highlight that BMA based approach provides the best fit to the given data.}
	\label{tab:simstudy3_method2_distance}
	\begin{tabular}{l|rrrrr}
		\toprule[.15 em]
		Distance & Gumbel & BMA \\
		\midrule 
		Wasserstein & 35.59 & \underline{34.18} \\
		$L^2$  & 0.28 & \underline{0.25} \\
		\bottomrule[.15 em]
	\end{tabular}
\end{table}


\newpage
\noindent\textbf{Sensitivity study for simulation study I} \\

To date in this simulation study, given the distribution and its parameters, the merits of employing BMA in copula fitting and TDC estimation have been shown for one simulated data set.
However, it must be investigated whether this result is consistent in multiple different simulations using the same distribution and parameter values.
From the authors' experience the fitting of copulas is very sensitive to the data set simulated - given the total sample size of 1500 is not very large, every run of the simulation and BMA process may return different results. Therefore, interest also lies in the generalization of this particular case to check that BMA improves the estimation in most cases or at least matches the ``best" individual copula. 

\begin{figure}[!h]
	\centering
	\includegraphics[width=.9\linewidth, height=.68\linewidth]{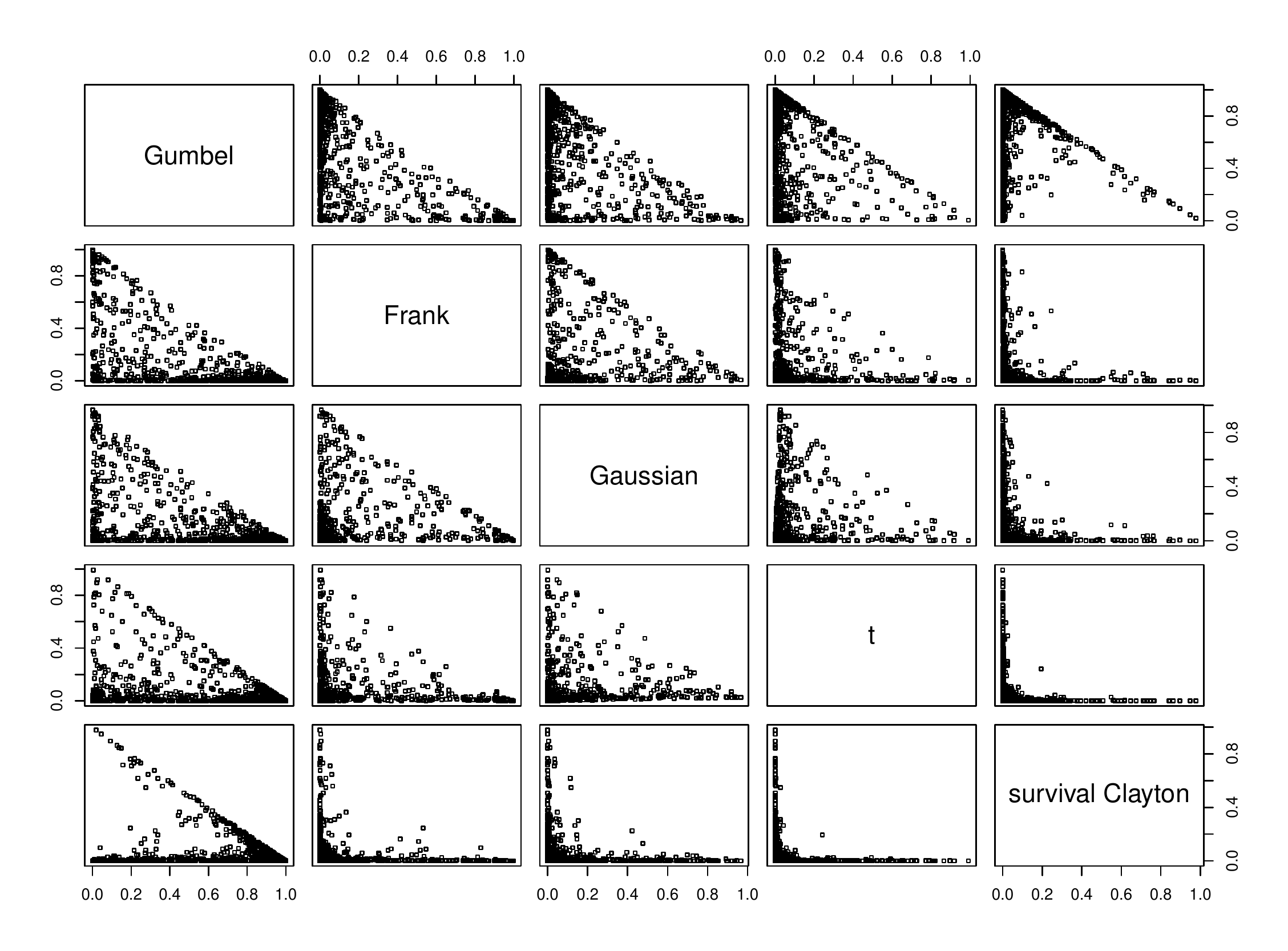}
	\caption{The BMA weights for non-zero weighted fitted copulas (Gumbel, Frank, Gaussian, t, survival Clayton), when repeating the simulation 1000 times. }
	\label{fig:simstudy3_sensitivity_weights_plots}
\end{figure}

\begin{figure}[!h]
	\centering
	\begin{minipage}{.7\textwidth}	\includegraphics[width=\linewidth]{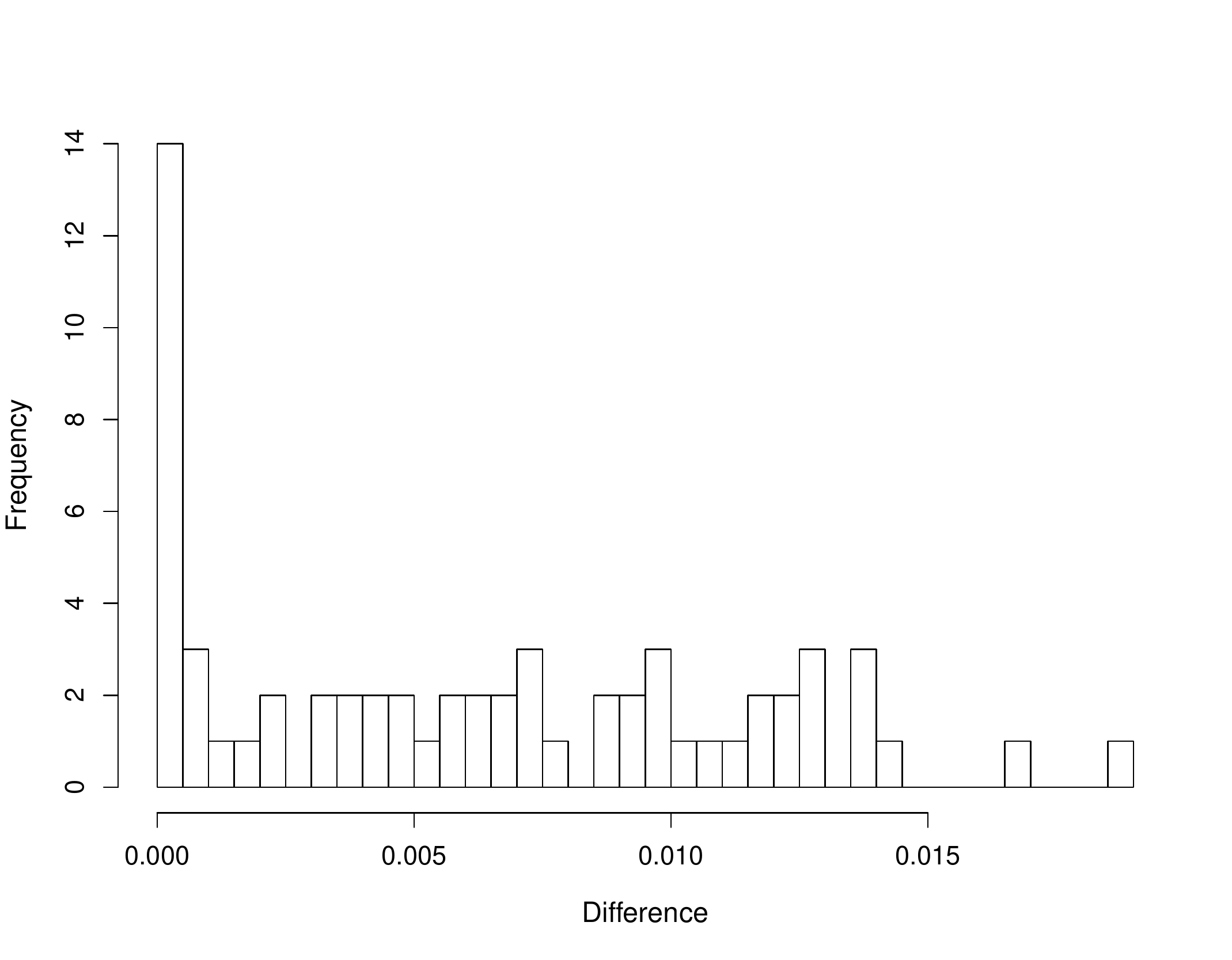}
		\caption*{(a) $\lambda_L$}
	\end{minipage}
	\begin{minipage}{.7\textwidth}	\includegraphics[width=\linewidth]{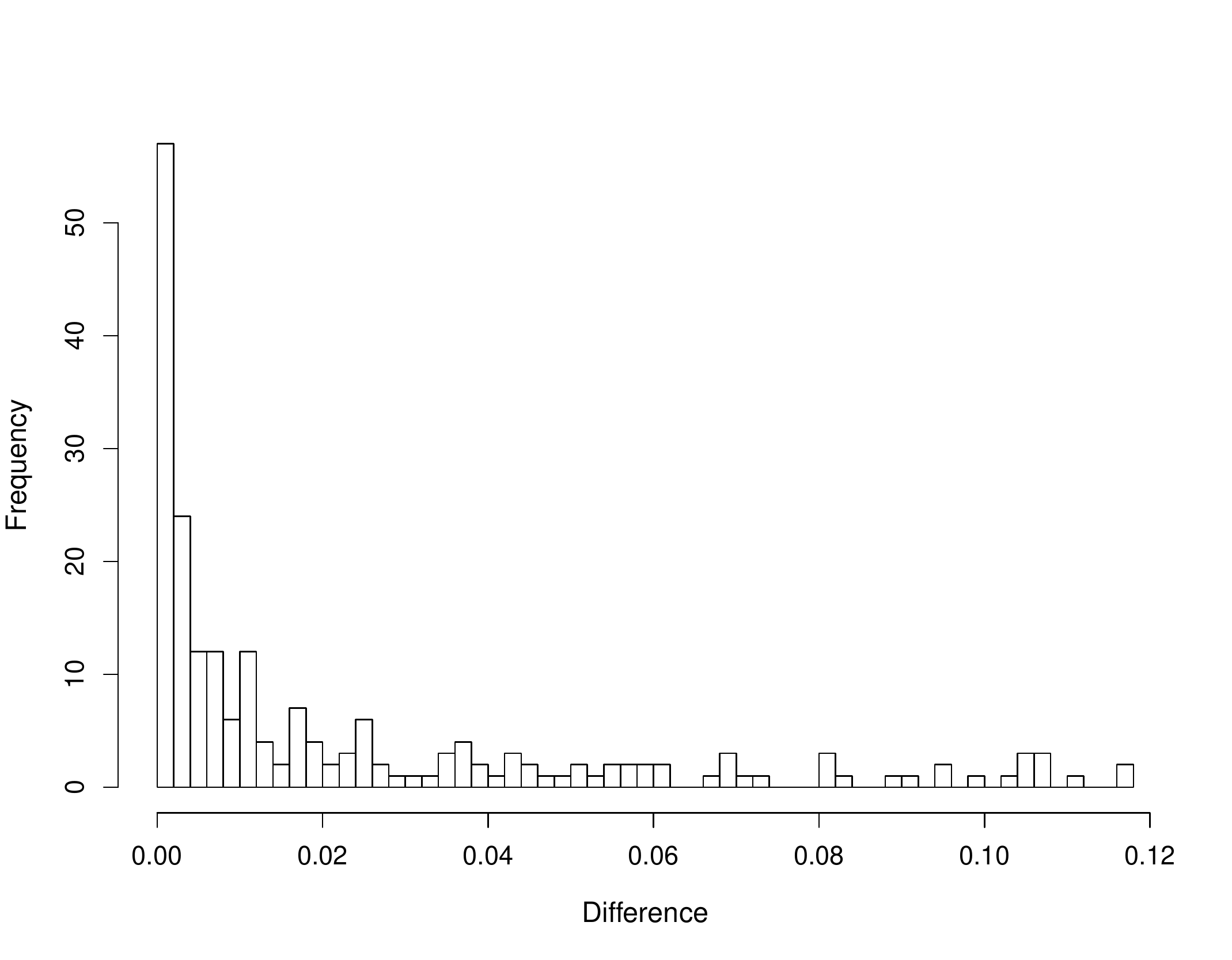}
		\caption*{(b) $\lambda_U$}
	\end{minipage}	
	\caption{Histogram of the difference in estimated TDCs between the single ``best" fitted copula and the BMA method when BMA TDC estimates are slightly worse: (a) shows the result for lower TDC; (b) shows the results for upper TDC. In both cases, the differences are very small.}
	\label{fig:simstudy3_sensitivity_diff}
\end{figure}

Repeating the simulation and BMA fitting process 1000 times leads to the results in Table~\ref{tab:simstudy3_sensitivity}, which includes the averaged model weights for each copula and the averaged estimated upper and lower TDCs of each copula. It is interesting to observe that most copulas have non-zero mean weights, except the survival Joe. The Gumbel has dominating goodness-of-fit while the Gaussian, Frank, survival Clayton and t have relatively large weights, similar to the individual case presented in Table~\ref{tab:simstudy3_copula_summary}.
The distributions of the weights for the non-zero weighted copulas - Gumbel, Frank, Gaussian, t and survival Clayton - are shown in Figure~\ref{fig:simstudy3_sensitivity_weights_plots}. It shows that the weights distribution among repetitions is very volatile across different simulations, ranging from 0 to 1.  
Table~\ref{tab:simstudy3_sensitivity} also shows that the five best fitting copulas all under-estimate the TDCs.
Note that because the standard deviations are very small (when they exist), the fitted TDCs are mostly very consistent across iterations. 
By focusing only on using method 1 in each repetition (since it does not involve further simulation), for $\lambda_L$ BMA improved the estimation in 93.7\% of cases (i.e. the BMA estimated $\lambda_L$ is the closest to the mean of the empirical estimates). For the remaining 6.3\% of cases, the dominating best copula estimates the TDC very well but the small weights from the other slightly worse fitting copulas pull the BMA TDC estimates slightly off course.
However, the differences between the best fitting individual copula's estimates and the BMA estimates are very small (within only about 0.015) -- see Figure~\ref{fig:simstudy3_sensitivity_diff}.    
For $\lambda_U$, in 79.1\% of cases BMA improved the estimation. For the remaining 20.9\% of cases, the differences between the best fitting copula's estimates and the BMA estimates are again very small (mostly within about 0.1). Therefore, it concludes that BMA is mostly either comparable to the one ``best" copula or considerably improves the TDCs estimation beyond using a single copula. Alternatively, it can be said that BMA ``often helps and rarely hurts". 

\begin{table}[!h]
	\caption{Summary of the BMA weights and the estimated upper TDCs for fitted copulas from 1000 iterations of simulation study I.}
	\label{tab:simstudy3_sensitivity}
	\centering
		\begin{tabular}{l|rrrrr}
			\toprule[.15 em]
			\ & t & Gaussian & Joe & survival Joe & Gumbel \\
			\midrule
			$\bar{W}_{C}$ & 0.069 (0.15) & 0.098 (0.19) & 0.008 (0.04) & 0 (0) & 0.660 (0.35) \\
			$\bar{\lambda}_{U}$ & 0.014 (0.01) & 0 (0) & 0.241 (0.03) & 0 (0) & 0.193 (0.02) \\
			$\bar{\lambda}_{L}$ & 0.014 (0.01) & 0 (0) & 0 (0) & 0.183 (0.03) & 0 (0) \\
			\midrule[.15 em]
			\ & survival Gumbel & Clayton & survival Clayton & Frank \\
			\midrule
			$\bar{W}_{C}$ & 0.001 (0.01) & 0 (0) & 0.054 (0.13) & 0.111 (0.24) \\
			$\bar{\lambda}_{U}$ & 0 (0) & 0 (0) & 0.103 (0.03) & 0 (0) \\
			$\bar{\lambda}_{L}$ & 0.173 (0.02) & 0.052 (0.02) & 0 (0) & 0 (0) \\
			\bottomrule[.15 em]
	\end{tabular} 
\end{table}	

\clearpage
\section{Simulation study II}
\label{sec:simstudy2}

In this simulation study, following the same reasoning for constructing complex copula dependency as in Section~\ref{sec:simstudy3}, 
a finite mixture of copulas is used to simulate an artificial data set. 
In this way, the mixture has an analytical form and its TDC can be written explicitly: if $C(u,v;\theta)=p_1 C_1(u,v;\theta_1) + p_2 C_2(u,v;\theta_2) + p_3 C_3(u,v;\theta_3)$ is a convex linear combination of copulas, then
\begin{equation}
\label{eq:mixture_utdc}
\begin{split}
\lambda_U &= \lim_{u\rightarrow 1} \frac{1-2u+C(u,u;\theta)}{1-u} \\
&= \lim_{u\rightarrow 1} \frac{(p_1+p_2+p_3)(1-2u) + p_1 C_1(u,u;\theta_1) + p_2 C_2(u,u;\theta_2) + p_3 C_3(u,u;\theta_3)}{(p_1+p_2+p_3)(1-u)} \\
&= p_1\lambda_{U, C_1} + p_2 \lambda_{U, C_2} + p_3 \lambda_{U, C_3} .
\end{split}
\end{equation} 
Therefore, the simulation conditions (1) and (3) are met: there is a complex dependence structure that individual copulas cannot capture entirely, and the true TDC values are known. 

Note that the use of mixtures of copulas is solely for creating complex dependence structure with known TDC values in this simulation study in order to demonstrate the merits of BMA. 
This simulation method naturally suggests fitting finite mixtures of copulas as an alternative, rather than using an individual copula approach. 
However, mixture models still result in one ``best" mixture model with a single estimate of upper (or lower) TDC, as in Equation~\ref{eq:mixture_utdc}, and this best single model approach still suffers from model uncertainty.
Furthermore, in this simulation study the mixture only occurs in the copula space, i.e. the marginal distributions are not mixtures. So the data do not show clear heterogeneity characteristics if the simulation process is unknown.  
Incorporating BMA with mixtures of copulas will lead to combining multiple mixtures, similar to its use in model-based clustering (\citealp{Wei2015}; \citealp{Russell2015}); and the TDCs can be averaged over multiple mixtures in a similar way to the method proposed here -- this option is illustrated in the GI insurer data in Section~\ref{sec:irish}. 
The authors emphasize that the purpose of this work is not contrasting BMA against the mixture of copulas approach. 

\begin{figure}[ht!]
	\centering
	\begin{minipage}{.65\linewidth}
		\includegraphics[width=\textwidth]{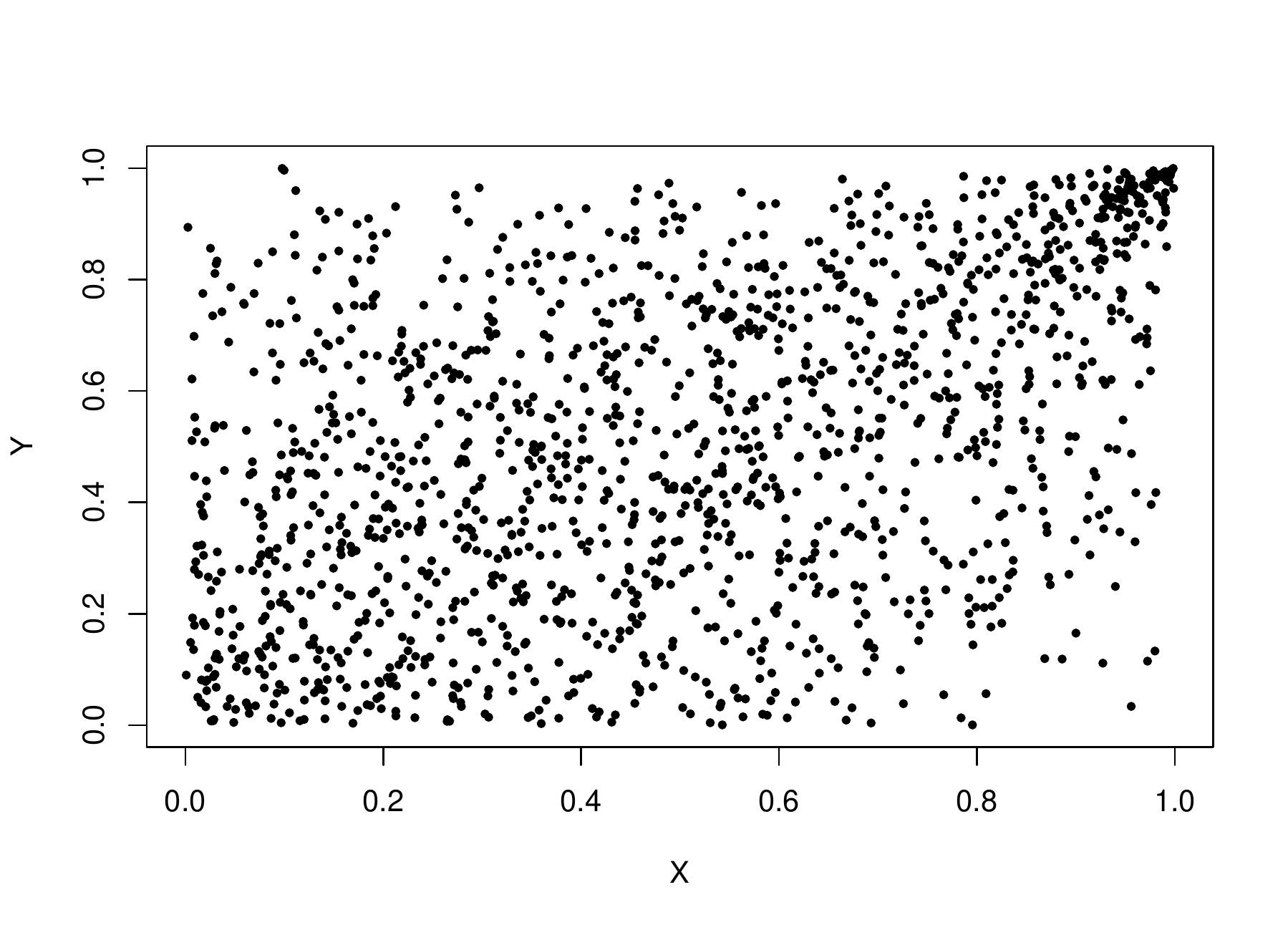}
		\caption*{(a)}	
	\end{minipage}
	\begin{minipage}{.65\linewidth}
		\includegraphics[width=\textwidth]{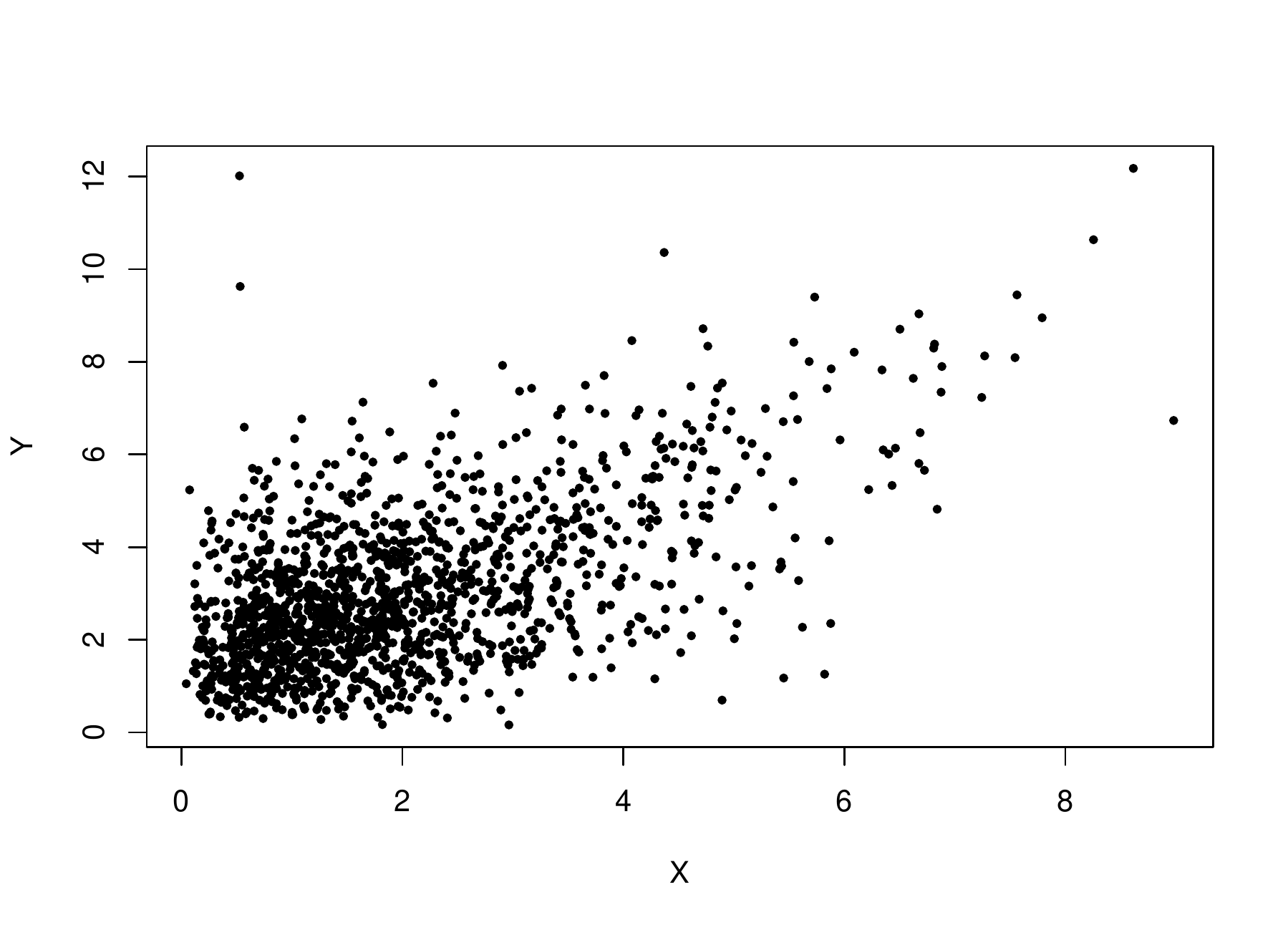}		
		\caption*{(b)}
	\end{minipage}
	\caption{Plot of the simulation data: (a) shows the data in the copula $[0,1]\times [0,1]$ space; (b) shows the data in the data space after taking marginals into account.}
	\label{fig:simstudy2_plot}
\end{figure}

An artificial data set is simulated from a mixture of three copulas, each contributing 500 data points: Joe ($\theta=1.7$, very heavy right tail), Gumbel ($\theta=1.5$, heavy right tail) and survival Clayton ($\theta=1.3$, i.e. $180^{\circ}$ rotated Clayton, heavy right tail), for which the upper TDC exits in all three cases but not the lower TDC. The total sample size is $N=1500$.  
The first marginal ($``X"$) is set as gamma(2,1), while the second marginal ($``Y"$) is set as gamma(3,1), due to the fact that the gamma distribution is a common choice for insurance claim severity modelling. Hence the simulated data can be regarded as a variation of bivariate gamma distributions. 
Figure~\ref{fig:simstudy2_plot} shows the simulated data in both copula space and the data space.

Not all copulas used in the simulation have a lower TDC, so $\lambda_L$ is not of interest in this example.
The theoretical true upper TDCs of the Joe($\theta=1.7$), the Gumbel($\theta=1.5$) and the survival Clayton($\theta=1.3$) are $0.497$, $0.413$ and $0.587$ respectively and are equally mixed, so the overall true upper TDC $\lambda_U = 0.499$. 
First the upper TDC is estimated nonparametrically using an empirical copula as in Section~\ref{sec:utdc},  which leads to: 
$\widehat{\lambda}_{U}^{(1)} = 0.546, 
\widehat{\lambda}_{U}^{(2)}= 0.554, 
\widehat{\lambda}_{U}^{(3)}= 0.431, 
\widehat{\lambda}_{U}^{(4)}= 0.483, 
\widehat{\lambda}_{U}^{(5)}= 0.526$. Hence estimator $\lambda_U^{(4)}$ works best in this case, while the others approximate closely to the true value. 
If the true $\lambda_U$ is unknown, it might be reasonable to estimate the value as lying in the range $[0.431, 0.554]$ or via the mean value of $0.508$. 

\begin{table}[h!]
	\caption{Summary of the fitted copulas for the simulation study II data, including fitted BIC values, BMA weights ($W_j$) based on the BICs and the estimated upper TDC $\widehat{\lambda}_U$ based on the estimated copula parameter.}
	\label{tab:simstudy2_bic_weights}
	\centering
	\resizebox{.9\linewidth}{!}{
		\begin{tabular}{l|rrrrrrrrrrrr}
			\toprule[.15 em]
			\ & t & Gaussian & Joe & Gumbel & Clayton & Frank & survival Clayton \\
			\midrule
			BIC & -415.71 & -386.54 & -506.92 & -503.38 & -171.30 & -359.51 & -503.76 \\
			BMA $W_j$ & 0 & 0 & 0.73 & 0.12 & 0 & 0 & 0.15 \\
			$\widehat{\lambda}_U$ & 0.16 & 0 & 0.53 & 0.42 & 0 & 0 & 0.48\\ 
			\bottomrule[.15 em]
	\end{tabular} }
\end{table}

When different copulas are fitted using pseudo-likelihood-based methods, based on the fitted BICs the posterior model probabilities are calculated as in Table~\ref{tab:simstudy2_bic_weights}, together with their BICs and estimated $\widehat{\lambda}_U$. 
Note that positive BMA weights are only allocated to the true generating copulas of the data, which is reassuring. \\

\noindent\textbf{Method 1}\\

The fitted upper TDCs are directly averaged based on the copulas' weights, leading to the results in Table~\ref{tab:simstudy2_method1_result}.
It shows that the BMA-weighted upper TDC is the closest to the true value and the mean of the empirical estimations, compared to the best fitting individual copula (the Joe copula which has the highest model weight). 

\begin{table}[H]
	\centering
	\caption{Results of using method 1 for $\lambda_U$ estimation: the BMA-estimated $\widehat{\lambda}_U$ is the closest to the truth, compared with the best fitting Joe copula. It also falls within the middle of the range of values estimated by the empirical copula.}
	\label{tab:simstudy2_method1_result}		\begin{tabular}{l|cccc}
		\toprule[.15 em]
		\ & True & Empirical & Joe & BMA \\
		\midrule
		$\lambda_U$ & 0.499 & [0.431, 0.554] & 0.526 & 0.505 \\
		\bottomrule[.15 em]
	\end{tabular}
\end{table}

\noindent\textbf{Method 2} \\

When treating the upper TDC no longer as the parameter of interest for BMA, but still the overall target parameter, 
a new BMA data set is simulated from the fitted copulas proportional to their BMA weights. 
The motivation is that using BMA can improve model fit and data similarity to the given data when simulating multiple copulas simultaneously. 
The empirical upper TDC is calculated based on the BMA simulated set and the simulated data from the best fitting Joe copula.
For robustness, the process is repeated 1000 times and the results are shown in Table~\ref{tab:simstudy2_method2_result} using the estimation methods from Section~\ref{sec:utdc}.
Noticeably for simulations from the Joe copula, all estimators give very close estimation to the true value of $0.53$ (i.e. the fitted upper TDC by the Joe, and this might potentially be used to choose which estimators work best if the true TDCs are unknown). 
The empirically estimated TDCs on the BMA set are better than those from any individual copula.
To further reinforce the conclusion, using Wasserstein and $L^2$ distances in each simulation repetition it is shown the BMA method outperforms the best fitting Joe copula (and in fact any other individual copula) in Table~\ref{tab:simstudy2_distance}. \\

\begin{table}[h!]
	\centering
	\caption{Results of using method 2 for simulation study II data: mean of the empirical nonparametric estimates across 1000 repetitions. In each iteration empirical TDCs are estimated using all five estimators on simulations from the best fitting individual copula and the BMA simulated set. Standard deviations are shown in brackets.}
	\label{tab:simstudy2_method2_result}
	\begin{tabular}{l|cc}
		\toprule[.15 em]
		\ & Joe & BMA \\
		\midrule
		$\overline{\lambda}_U^{(1)}$ & 0.548 (0.07) & 0.535 (0.07) \\
		$\overline{\lambda}_U^{(2)}$ & 0.542 (0.07) & 0.521 (0.07) \\
		$\overline{\lambda}_U^{(3)}$ & 0.516 (0.07) & 0.505 (0.07) \\
		$\overline{\lambda}_U^{(4)}$ & 0.487 (0.04) & 0.476 (0.04) \\ 
		$\overline{\lambda}_U^{(5)}$ & 0.521 (0.07) & 0.505 (0.07) \\
		\bottomrule[.15 em]
	\end{tabular}
\end{table}

\begin{table}[h!]
	\centering
	\caption{Comparison of distances between individual copula simulations and the BMA simulated sets versus the original data set. The underlined value indicates that the BMA approach provides the best fit to the data.}
	\label{tab:simstudy2_distance}
	\begin{tabular}{l|rrrrrrrr}
		\toprule[.15 em]
		Distance & Joe & BMA \\ 
		\midrule
		Wasserstein & 35.35 & \underline{33.98} \\
		$L^2$ & 0.29 & \underline{0.25} \\
		\bottomrule[.15 em]
	\end{tabular}
\end{table}



\noindent\textbf{Sensitivity study for simulation study II}  \\

\begin{figure}[h!]
	\centering
	\begin{minipage}{\textwidth}
		\centering
		\begin{minipage}{.3\linewidth}
			\includegraphics[width=\linewidth, height=6cm]{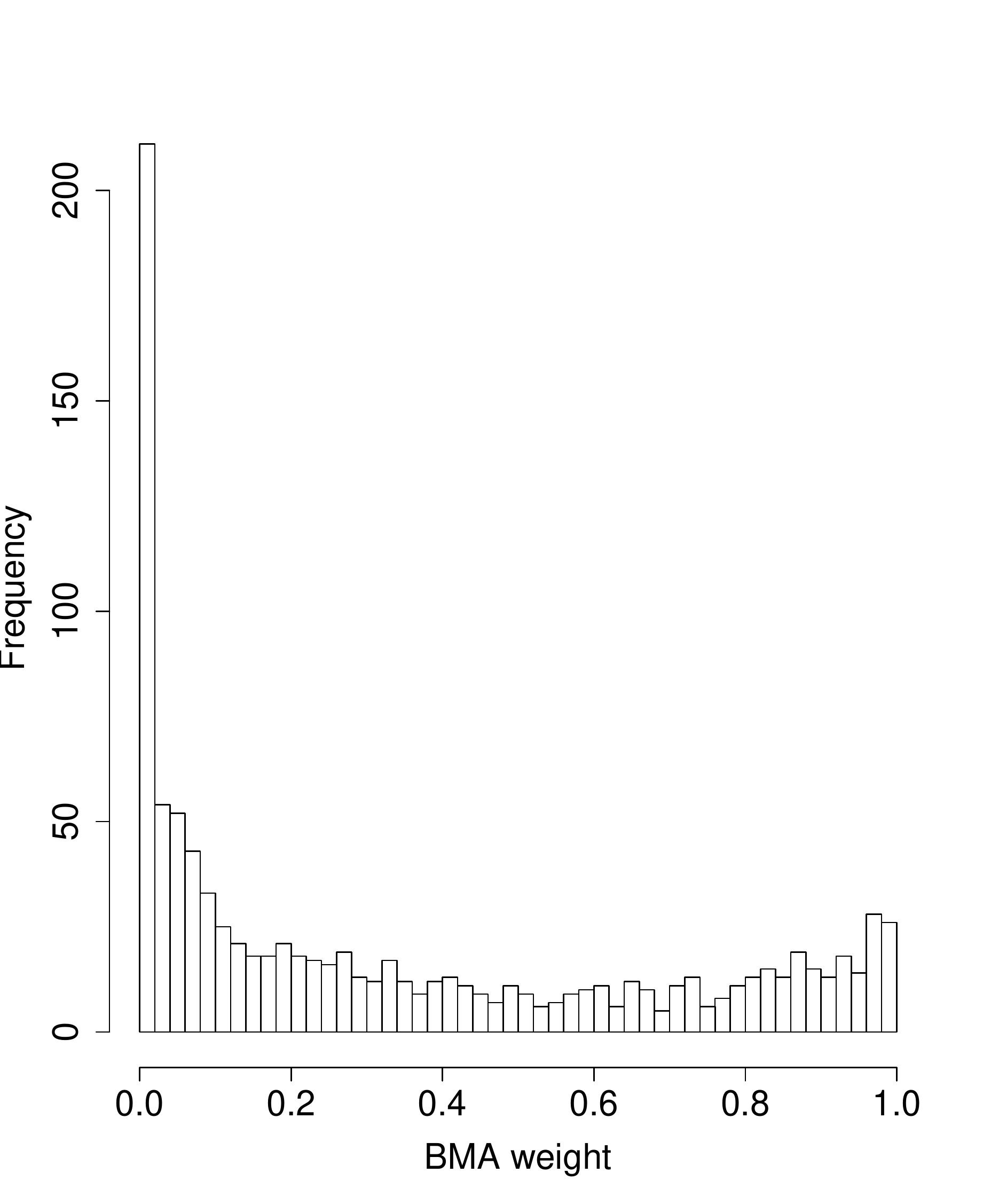}
			\caption*{Joe}
		\end{minipage}
		\begin{minipage}{.3\linewidth}
			\includegraphics[width=\linewidth, height=6cm]{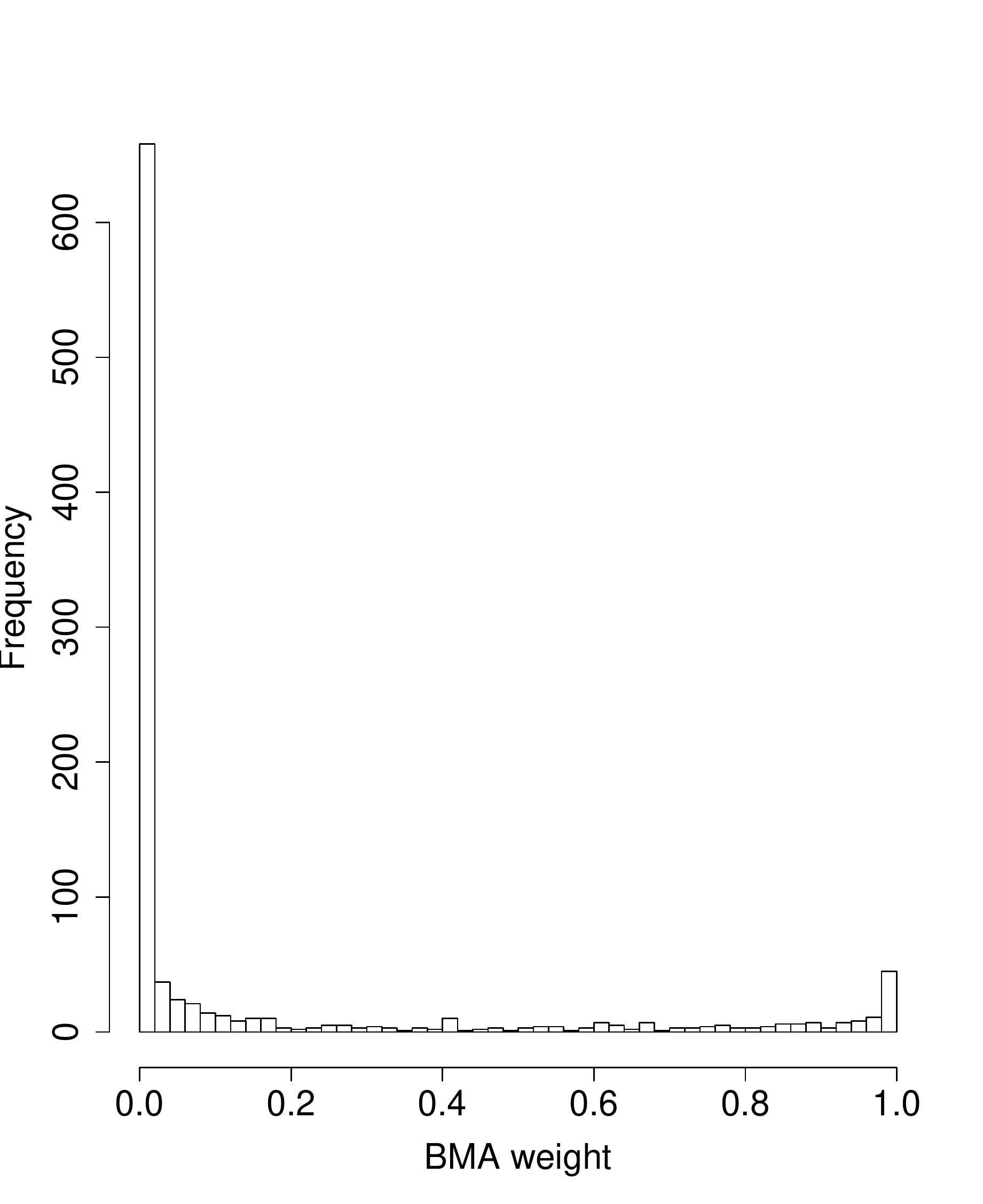}
			\caption*{Gumbel}
		\end{minipage}
		\begin{minipage}{.3\linewidth}
			\includegraphics[width=\linewidth, height=6cm]{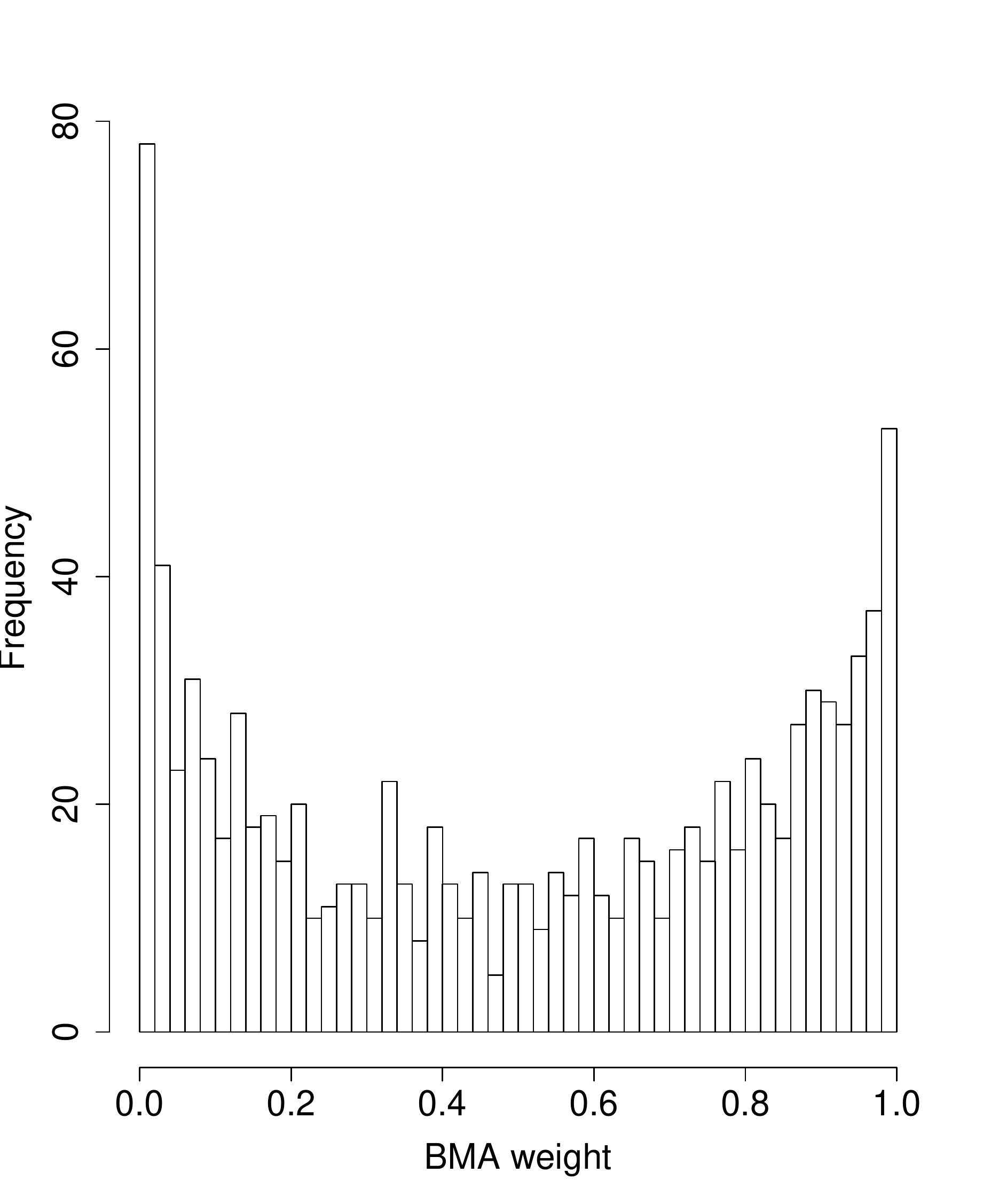}
			\caption*{survival Clayton}
		\end{minipage}		
		\caption*{(a)}
	\end{minipage} 
	\begin{minipage}{\textwidth}
		\centering
		\includegraphics[width=10cm, height=9cm]{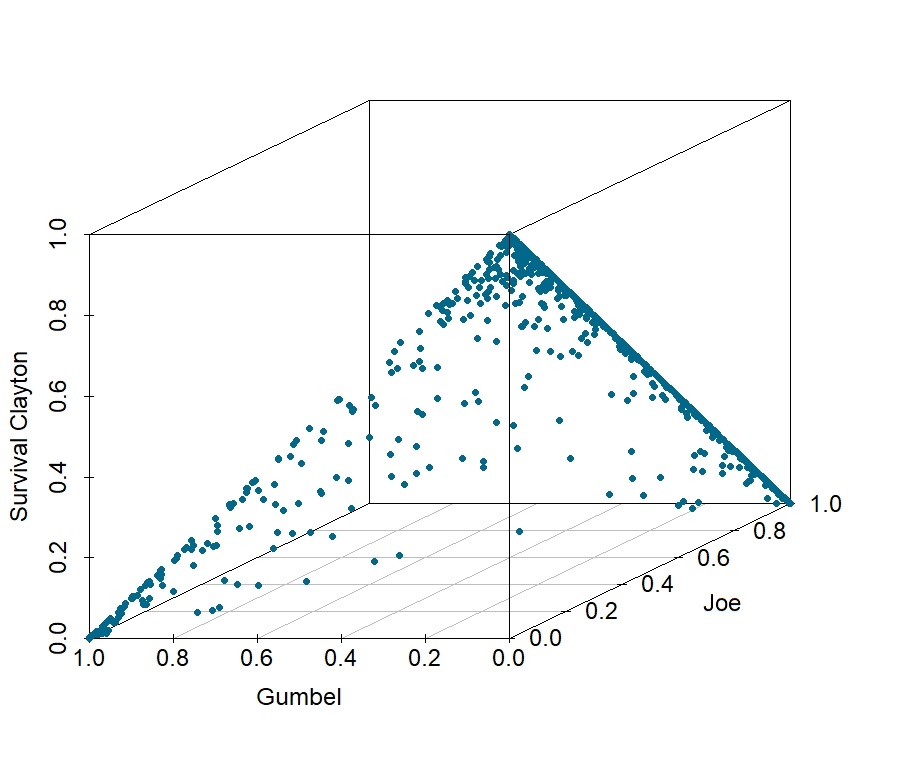}
		\caption*{(b)}
	\end{minipage}
	\caption{The BMA weights for non-zero weighted fitted copulas (Joe, Gumbel and survival Clayton), when repeating the simulation and fitting process 1000 times, for the simulation study II: (a) shows the weights distributions as histograms; (b) shows the weights in a three dimensional plot. }
	\label{fig:simstudy2_sensitivity_weights_plots}
\end{figure}

As in Section~\ref{sec:simstudy3},
although the simulation study shows the usefulness of employing BMA in copula fitting and TDC estimation, interest also lies in the generalization of this particular simulated data and how often BMA works well. 
Hence it must be verified that this approach is consistent beyond this individual simulated data set.  
Repeating the same data-simulation process 1000 times using the same parameters and sample size, all copula options were fitted in each iteration and BMA was implemented to combine the upper TDCs from the fitted copulas. In each run, the BMA weights and fitted upper TDCs were recorded. The final results summary is shown in Table~\ref{tab:simstudy2_sensitivity}. 
It is noted that, even when the simulation  is run many times, the results are very consistent -- it is always the case that the Joe, Gumbel and survival Clayton have non-zero weights, with their mean weights being $0.338$, $0.159$ and $0.503$ respectively (with similar standard deviations).    
The distributions of the weights for the non-zero weighted copulas -- the Joe, Gumbel and survival Clayton -- are given in Figure~\ref{fig:simstudy2_sensitivity_weights_plots}. 
Figure~\ref{fig:simstudy2_sensitivity_weights_plots}(a) shows that the weights are volatile across simulations, ranging from 0 to 1, especially for the survival Clayton. Joe and Gumbel are more likely to have 0 or close to 0 weights.   
Figure~\ref{fig:simstudy2_sensitivity_weights_plots}(b) shows it is most likely that the survival Clayton and Joe together have the best fit (i.e. the sum of their weights equals 1) compared to survival Clayton and Gumbel, whereas Joe and Gumbel together seldom have the best fit. 
The survival Clayton shows a dominant goodness of fit overall in terms of individual copulas. 

\begin{figure}[!h]
	\centering
	\includegraphics[width=9cm]{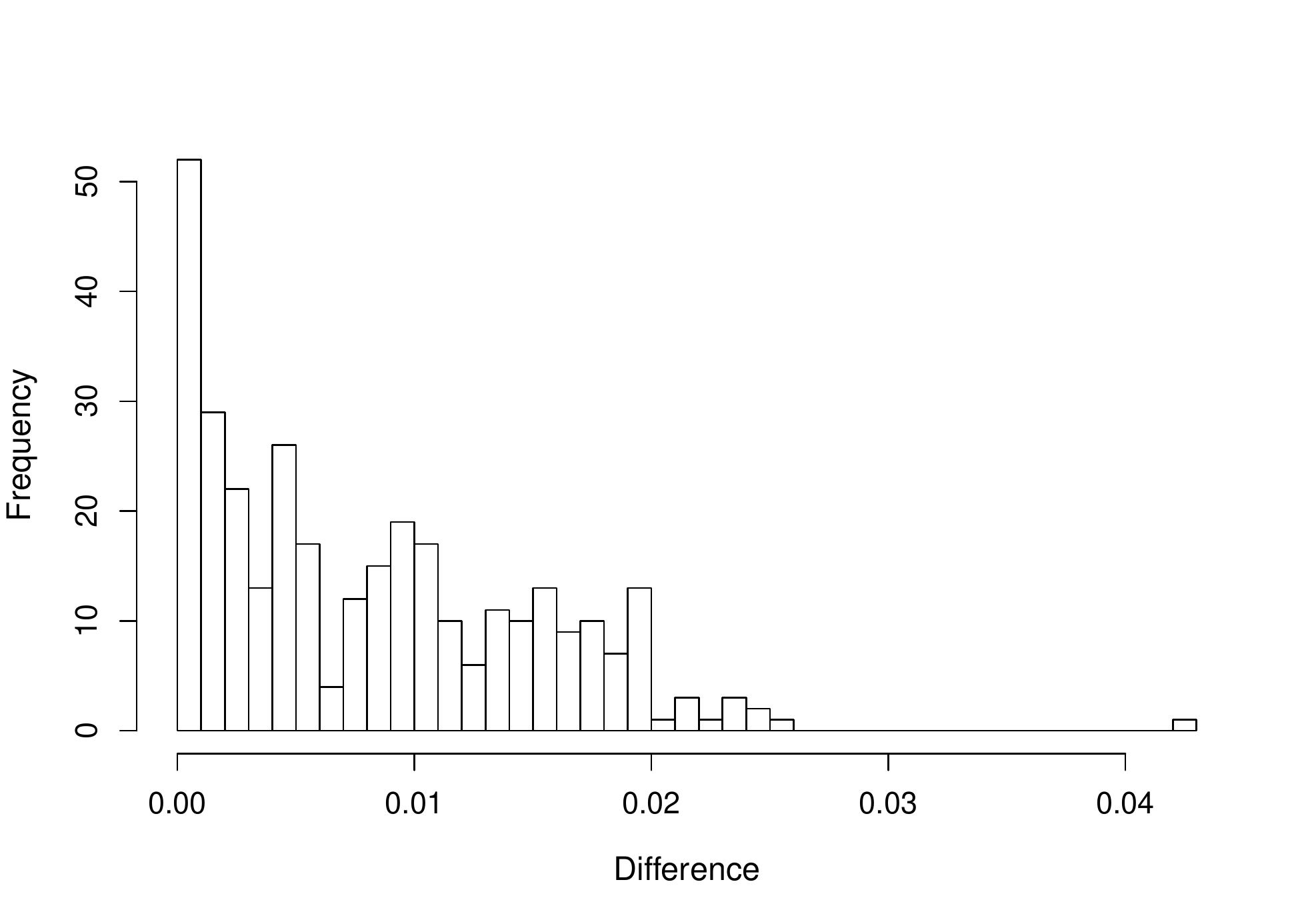}
	\caption{Histogram of the difference in estimated upper TDCs between the single ``best" fitting copula and the BMA method when the BMA TDC estimates are slightly worse in the sensitivity study. The differences are very small (mostly within 0.02).}
	\label{fig:simstudy2_sensitivity_diff}
\end{figure}

Table~\ref{tab:simstudy2_sensitivity} also shows the mean and standard deviation of the estimated $\widehat{\lambda}_U$ of each fitted copula from the 1000 iterations. 
Note that because the standard deviations are very small, the estimated upper TDCs are mostly consistent across iterations. 
The Joe copula always over-estimates $\lambda_U$, the Gumbel copula always under-estimates $\lambda_U$, and the survival Clayton's $\lambda_U$ is closest to the true value most of time with its mean of $0.498$ being very close to the true value of $0.499$.
Focusing on method 1, 
the estimated $\widehat{\lambda}_U$ from the best fitting copula in each iteration (i.e. the one with the highest BMA weight) and the BMA-induced upper TDCs are compared to the true value of $0.499$. 
In 67.3\% of cases the BMA-induced $\widehat{\lambda}_U$ are closest to the true value. 
For the remaining 32.7\% of instances it is typically the case that the dominating best copula (usually the survival Clayton) estimates upper TDC best but the small weights from other less well fitting copulas pull the TDC estimates slightly further from the true value. 
By observing the difference between the best estimated and BMA-estimated upper TDCs, their differences are mostly within only 0.02 or less, as seen in Figure~\ref{fig:simstudy2_sensitivity_diff}. 
In fact, in the 32.7\% of cases cited, when compared with the empirically estimated $\widehat{\lambda}_U$, most of the BMA-estimates are closer to the true value. 
Increasing the sample size to $3\times2000$, only 25\% of cases have BMA perform slightly worse and the differences are again very small. 
Therefore, BMA is mostly either comparable to the one ``best" copula or considerably improves the estimation of the TDCs beyond using a single copula, or alternatively it can again be said that it ``often helps and rarely hurts".

\begin{table}[H]
	\caption{Summary of BMA weights and the estimated upper TDCs of fitted copulas from 1000 iterations of simulation study II.}
	\label{tab:simstudy2_sensitivity}
	\centering
	\resizebox{.9\linewidth}{!}{
		\begin{tabular}{lrrrrrrrrrrrr}
			\toprule[.15 em]
			\ & t & Gaussian & Joe & Gumbel & Clayton & Frank & survival Clayton \\
			\midrule
			$\bar{W}_{C}$ & 0 (0) & 0 (0) & 0.338 (0.34) & 0.159 (0.31) & 0 (0) & 0 (0) & 0.503 (0.35) \\
			$\bar{\lambda}_{U}$ & 0.136 (0.04) & 0 (0) & 0.543 (0.02) & 0.430 (0.02) & 0 (0) & 0 (0) & 0.498 (0.02) \\
			\bottomrule[.15 em]
	\end{tabular} }
\end{table}	

\section{Irish GI insurer motor data}
\label{sec:irish}

A large motor insurance claims data set was obtained from an Irish general insurance company. The data represent the insurer's book of business written in Ireland over the period January 2013 to June 2014, containing 452,266 policies and their characteristics. 
In Ireland, motor insurance is required for all motorists and is enforced by Irish law, with third party (TP) cover being the minimum requirement. 
Across the portfolio there are five categories of claims: accidental damage (AD), third party property damage (PD), third party bodily injury (BI), windscreen (WS) and fire/theft (FT).
This insurer provides two different types of coverage: TP, fire and theft cover and comprehensive cover. 
Interest lies in the dependence relationship among these five categories, especially in the upper tails, which is an important quantity indicating the effects of potential catastrophic events such as a very serious car accident. 
Only comprehensive cover was selected since it includes all five categories, so that their overall and tail dependencies could be investigated.
From a claim severity perspective, i.e. excluding policies on which no claims have been made on each pair, among the five categories the pair AD and PD represent the highest overall correlation ($\rho_{(AD,PD)}=0.24$) and the highest empirical upper tail dependence ($\bar{\lambda}_{U, emp}=0.156$) compared to other pairs, despite overall correlation not necessarily indicating tail dependence (\citealp{Sweeting2013}). 
AD and BI have the highest empirical upper TDC estimated nonparametrically, slightly higher than that between AD and PD, mainly because the BI category has more claims with higher severity.
However, the number of policies having made both AD and BI claims is much smaller hence a smaller sample size.   
Therefore, AD and PD provide the main focus and only policies having made claims on both AD and PD are selected. 
In these two categories there exist a few extremely large claims that are much greater than \euro $20,000$. For upper tail dependence, it is naturally preferable to include such observations. However, they lead to numerical issues in analysis when using covariates in GLM regression settings on the marginals, so they are deleted. If copula parameter estimation is done via MPLE based on ranked observations on marginals as in Sections~\ref{sec:simstudy3} and~\ref{sec:simstudy2}, these extreme values can be kept.
This leads to 2,098 policies remaining in the data set. 
The logic of investigating their tail dependence is that, given a customer made claims on both categories, large claims in one risk would reflect possible large claims in the other, hence identifying high-risk customers as well as information to be used in future claim reserve calculation or joint product pricing. 
In the data set, there are about 100 covariates presented. Due to data confidentiality issues, only seven predictors are selected and used in this article, including policyholder's information (e.g. licence category), insured car information (e.g. fuel type, transmission) and policy information (e.g. no-claim discount). Table~\ref{tab:realdata_predictor_description} shows these predictors and their categorical levels. 

\begin{table}[ht]
	\centering
	\caption{Categorical levels of the selected variables in the Irish GI data set. Note that ``No claim discount" represents the number of years with no claim.}
	\label{tab:realdata_predictor_description}
	\begin{tabular}{ll}
		\toprule[.15 em]
		Variables  & Categories  \\
		\midrule 
		Vehicle fuel type 	  & Diesel; Petrol; Unknown  \\
		Vehicle transmission & Automatic; Manual; Unknown \\
		Annual mileage           & 0-5000; $\ldots$; 45001-50000; 50001+  \\
		Number of registered drivers & 1; 2; 3; 4; 5; 6; 7 \\
		No claim discount & 0; 0.1; 0.2; 0.3; 0.4; 0.5; 0.6\\
		No claim discount protection 	  & No; Yes; Unknown \\
		Main driver license category& B; C; D; F; I; N \\
		\bottomrule[.15 em]
	\end{tabular}
\end{table}

\begin{figure}[H]
	\centering
	\begin{minipage}{0.7\textwidth}
		\includegraphics[width=\linewidth, height=.8\linewidth]{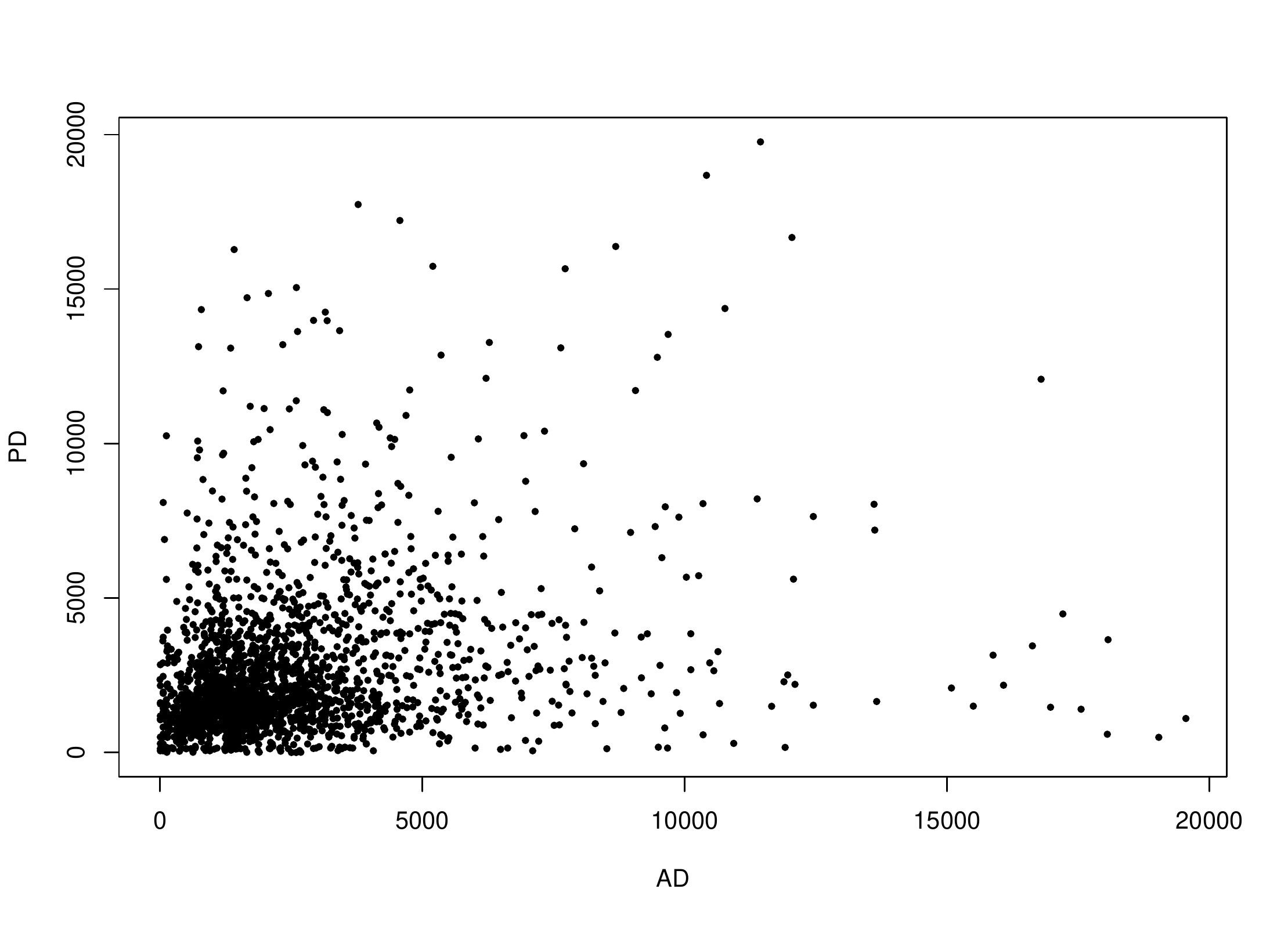}
		\caption*{(a)}
	\end{minipage}
	\begin{minipage}{0.7\textwidth}
		\includegraphics[width=\linewidth, height=.8\linewidth]{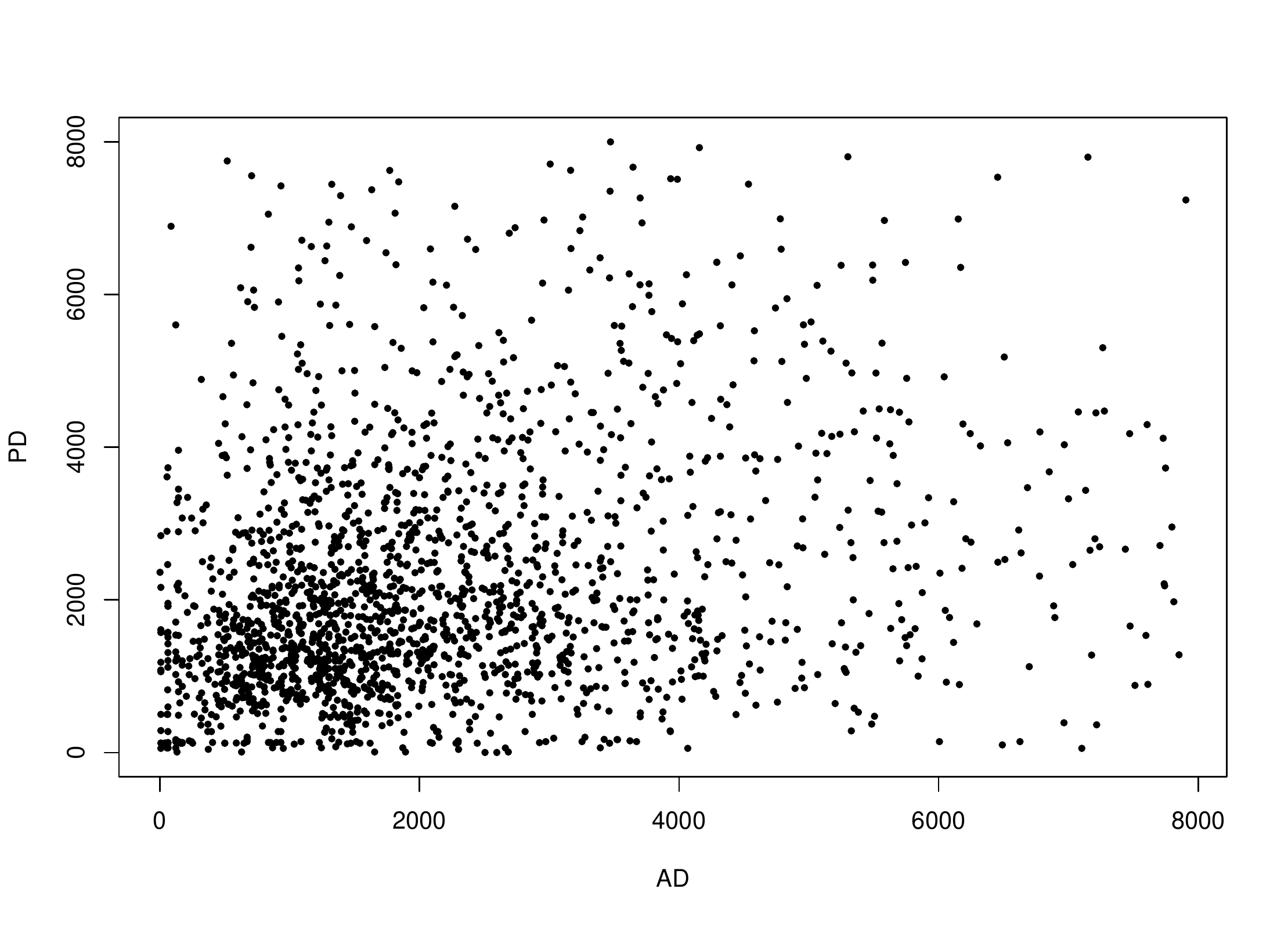}
		\caption*{(b)}
	\end{minipage}
	\caption{Plot of accidental damage (AD) and third party property damage (PD) claims of the Irish GI insurer data set: (a) shows all claims including the extreme values; (b) shows a truncated version for AD claims $<20,000$ and PD claims $<20,000$.}
	\label{fig:realdata_plot}
\end{figure}

This data set has been investigated in \cite{Hu2019}, which concludes that the data is very heterogeneous, with subgroups having different claim behaviour, and can be modelled as following a mixture of bivariate gamma distributions as defined in Section~\ref{sec:simstudy3}. 
This potential mixture is also reflected in the data plot -- 
Figure~\ref{fig:realdata_plot}(a) shows the scatter plot of the AD and PD claim sizes. 
The claims from the two risk perils are very dispersed: many have large claim sizes on individual risks and some made large claims on both risk perils simultaneously. 
This indicates that when there is a relatively high PD claim it is possible to also have a relatively high AD claim attached, and vice versa, although the possibility may not be very strong.
Overall, the spread of large claims suggests that there is at least some degree of upper tail dependence present.
Furthermore, when focusing on the range AD $\leq 8,000$ and PD $\leq 8,000$ as in Figure~\ref{fig:realdata_plot}(b), it is clear to observe that there is a dense cluster in the interval between 500 and 3000, and there are many claims close to both axes. 
These characteristics may explain why the overall dependence is weak, and there is heterogeneity of the dependence structure which can be targeted via a finite mixture of copulas.

When a copula and corresponding marginal distributions are given, this specifies a bivariate distribution; if covariates are available they can further assist model estimation. Therefore, in this analysis instead of using pseudo observations on marginal distributions, it is assumed that the marginals are also a mixture of univariate gamma distributions (as gamma is the typical distribution used for claim severity modelling), and covariates are incorporated into the marginals through GLM frameworks, following the finite mixture of copula regressions approach in Equation~\ref{eq:mixture_copulas_model}. 
It can also be considered as an extension to Section~\ref{sec:simstudy3}: the data can be seen as a finite mixture of bivariate gamma distributions from \cite{Hu2019}, and as there are no explicit copulas corresponding to such bivariate gamma distributions, multiple competing copulas can be averaged over; therefore, in this example it is expected that, instead of averaging standalone copulas, multiple mixtures of copulas can have competing goodness-of-fit and can be averaged over, similar to the use of BMA for Gaussian distribution mixtures in \cite{Wei2015} and \cite{Russell2015}.  

\subsection{Empirical estimation}

\begin{figure}[h!]
	\centering
	\begin{minipage}{0.7\textwidth}
		\includegraphics[width=\linewidth, height=.7\linewidth]{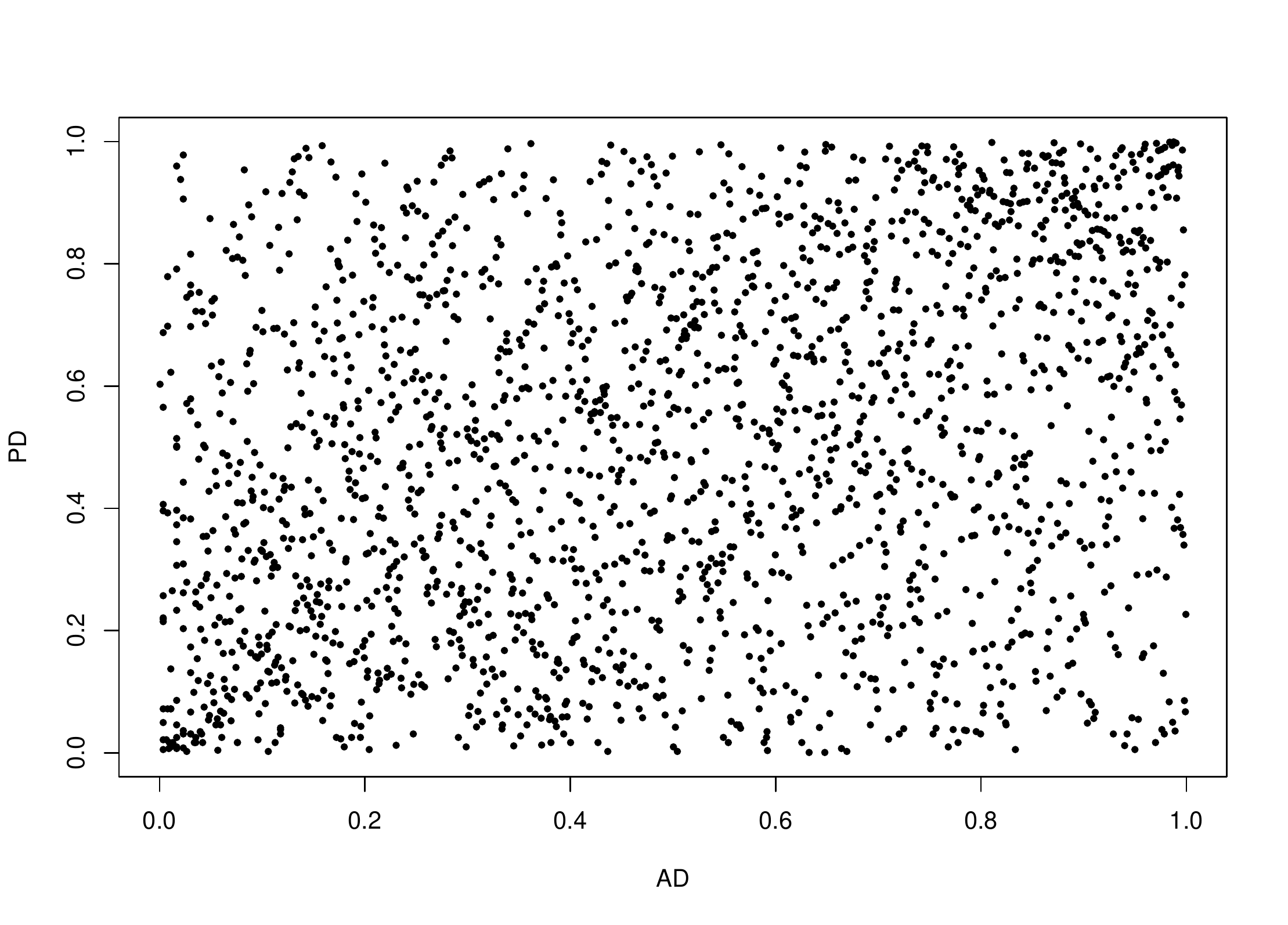}
		\caption*{(a)}
	\end{minipage}
	\begin{minipage}{0.7\textwidth}
		\includegraphics[width=\linewidth, height=.7\linewidth]{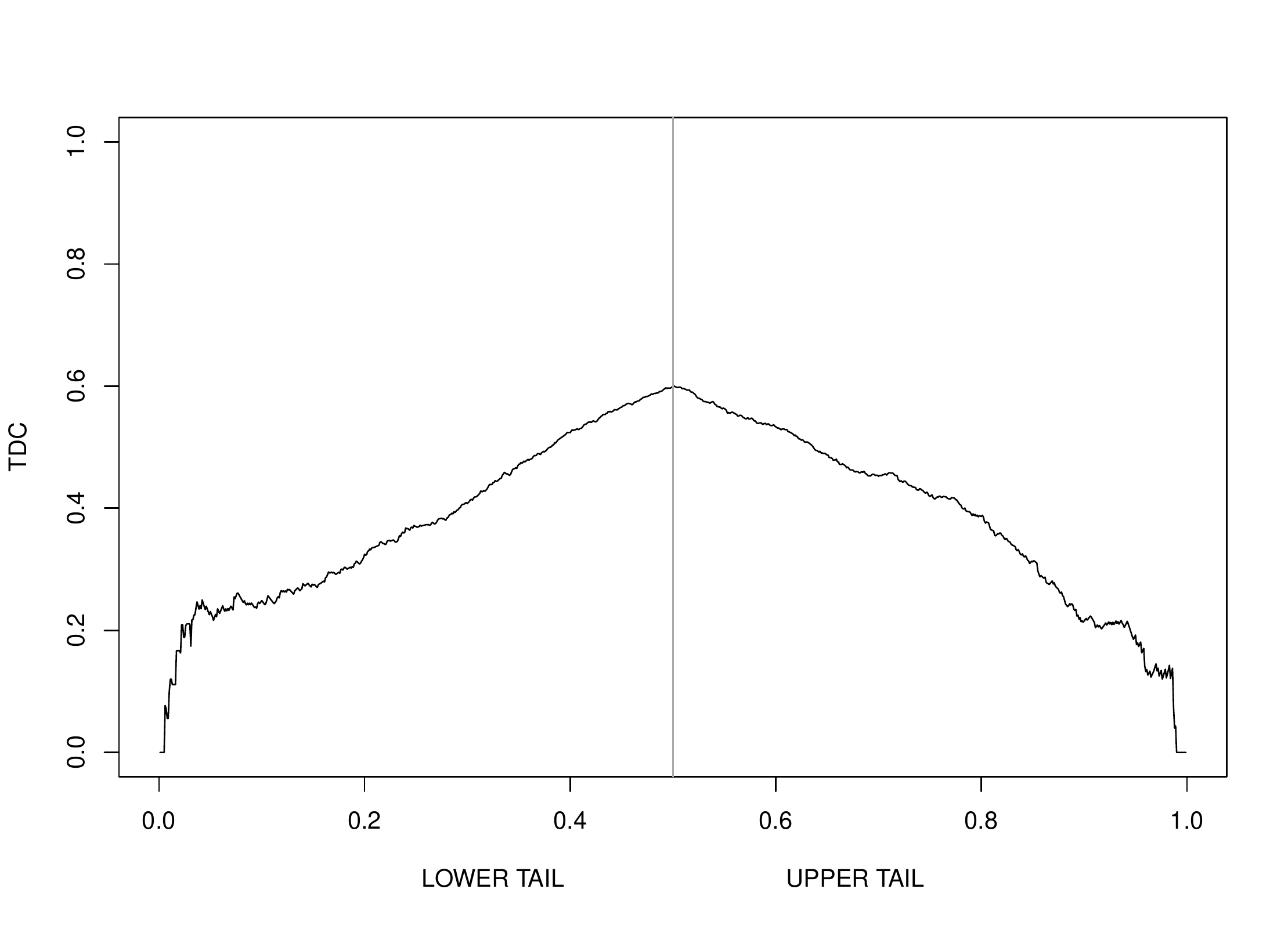}
		\caption*{(b)}
	\end{minipage}
	\caption{Pseudo observation plots: (a) shows the pseudo observation scatter plot of accidental damage (AD) and third party property damage (PD) claims of the Irish insurer, in copula space $[0,1]\times [0,1]$; (b)upper and lower TDC trajectories between accidental damage (AD) and third party property damage (PD) of the Irish insurer data, based on the pseudo data and empirical copula.}
	\label{fig:realdata_pseudo_data_plot}
\end{figure}

Due to the fact that the true TDCs are unknown for this data set, empirical methods are first implemented to estimate them, i.e. both copula and marginals are estimated empirically.
Figure~\ref{fig:realdata_pseudo_data_plot}(a) shows the pseudo-data based on their ranks. In this copula space there does not seem to exist obvious mixing behaviour;
and the top right corner (i.e. upper tail) does not exhibit a strong concentration of points, nor does the lower left corner (i.e. lower tail), which indicates relatively weak tail dependence in both upper and lower tails, if it exists.  
By using an empirical copula and plotting the trajectory of empirical upper and lower TDCs as in Equation~\ref{eq:tdc_trajectory},
the relatively smooth trends in Figure~\ref{fig:realdata_pseudo_data_plot} suggest both TDCs may exist but rather weakly, and the upper TDC should be weaker than the lower TDC (if any). 

Using all estimators described in Section~\ref{sec:utdc}, 
Table~\ref{tab:newrealdata_TDC_estimation} shows the approximated upper and lower TDCs.
As expected, the nonparametric empirical estimators give relatively unstable results, $\widehat{\lambda}_U$ ranging from $0.081$ to $0.164$ and $\widehat{\lambda}_L$ ranging from $0.103$ to $0.200$, suggesting the lower TDC is stronger than the upper TDC.
However, given the data plots it could be argued that the empirical estimators potentially over-estimate the TDCs. 

Estimating the TDCs is vital for the insurer because, for example, the upper TDC could be used when considering catastrophic events leading to claim sizes greater than $10,000$ or even greater than $100,000$, especially when it comes to bodily injuries indicated in the Irish insurer data. A difference of $0.1$ in upper TDC estimations could potentially lead to tens of thousands of Euros of difference in claim reserve calculations.  
The unstable empirical estimation may also be due to the real data set having more complicated dependence structure regarding its tails. 
Furthermore, it is worth noting again that the empirical estimators cannot take into account that TDCs may not exist (i.e. $\lambda =0$) -- they always over-estimate in this case and in the case where the TDCs are very small (if they exist). 
Therefore, parametric model-based estimation taking account of model uncertainty may provide a better solution to this problem. 

\begin{table}[h!]
	\centering
	\caption{Empirically estimated upper and lower TDCs $\widehat{\lambda}_U$ and $\widehat{\lambda}_L$ for the Irish insurer data based on estimators in Section~\ref{sec:utdc}.}
	\label{tab:newrealdata_TDC_estimation}
	\begin{tabular}{lrrrrrrc}
		\toprule[.15 em]
		\ & $\widehat{\lambda}^{(1)}$ & $\widehat{\lambda}^{(2)}$ & $\widehat{\lambda}^{(3)}$ & $\widehat{\lambda}^{(4)}$ & $\widehat{\lambda}^{(5)}$ & mean & range \\
		\midrule
		$\widehat{\lambda}_{U}$ & 0.164 & 0.147 & 0.108 & 0.081 & 0.133 & 0.127 & [0.081, 0.164] \\
		$\widehat{\lambda}_{L}$ & 0.153 & 0.135 & 0.103 & 0.110 & 0.200 & 0.141 & [0.103, 0.200] \\
		\bottomrule[.15 em]
	\end{tabular} 
\end{table}

\subsection{Full parametric estimation of finite mixtures of copulas}


Finite mixtures of copulas with covariates are fitted using a full parametric likelihood based approach:
the number of components $G$ and (combinations of) copulas need to be specified. The many modelling choices surrounding the use of different copulas for different components, the label switching issue for mixture models, and the stochastic nature of EM algorithm's dependence on the initial values make model selection very complex. The choice of component ordering issue is discussed in \cite{Kosmidis2016} -- one solution is to fit all models that result from all possible permutations of the component copulas, then for each unique permutation the same initializing classification is used to control the order of copula components in clustering.
Furthermore, covariates can be different for each marginal distribution of this finite mixture of copulas. However, this will add more complexity in model selection, and hence for simplicity all covariates in Table~\ref{tab:realdata_predictor_description} are used in the marginal GLMs.

The Akaike information criterion (AIC) (\citealp{Akaike1974}) is a suitable model selection criterion for this type of finite mixture of copulas with covariates, especially regarding selection of $G$. Based on the authors' experience it always selects a superior model compared to other criteria. The same general conclusions can also be found in other types of bivariate or multivariate regression settings, for example see \cite{Karlis2003}; \cite{Karlis2007}; \cite{Bermudez2009}; \cite{Bermudez2012}. 
Other information criteria available in the literature, for example the Bayesian information criterion (BIC) (\citealp{Schwarz1978}) or integrated complete-data likelihood (ICL) (\citealp{Biernacki2000}) can also be considered for model selection in such mixture model settings, although they may lead to different selection conclusions.  

For the Irish GI data, AIC is used to select the number of components $G$, resulting in $G=2$ (when $G=2$, most fitted models have better AIC values than those with other $G$ values). 
Then BIC is used to calculate the model probabilities. 
The 9 copula choices in Sections~\ref{sec:simstudy3} and~\ref{sec:simstudy2} are again used in this example; when $G=2$, there are $9\times9=81$ permutations of copulas to choose from, and the best 12 combinations of copulas with non-zero model probabilities are listed in Table~\ref{tab:realdata_copula_mixture_summary}. 
Note that the greedy search over all permutations of copulas has been repeated many times to check the stability of the results: due to the label switching issue for mixture models, and the stochastic nature of the EM algorithm's dependence on initial values, different runs may lead to very slightly different results regarding the ranking of the best few models. The EM algorithm is stable in finding the same optimal solutions given fixed initial values, see a simulation study example of this in Appendix~\ref{app:newsimstudy}. 

\begin{table}[ht!]
	\centering
	\caption{Summary of the best 12 fitted mixtures of copulas for $G=2$, including copula parameters, mixing probabilities, AIC and BIC values, BMA weights ($W_j$) and the estimated $\widehat{\lambda}_U$ and $\widehat{\lambda}_L$ based on Equation~\ref{eq:mixture_utdc}. }
	\label{tab:realdata_copula_mixture_summary}
	\resizebox{.9\textwidth}{!}{
		\begin{tabular}{lllccccrr}
			\toprule[.15em] 
			\rowcolor{lightgray!60} \ & Component $1$ &Component $2$ &$\tau_g$ &AIC &BIC &BMA $W_j$ &$\widehat{\lambda}_{L}$ &$\widehat{\lambda}_{U}$ \\
			\midrule
			1& Frank (2.02) & Clayton (0.50) & (0.49, 0.51) & 73566.60 &	74216.14 & 0.52 & 0.13 & 0 \\
			2& Frank (2.13) & Gumbel (1.22) & (0.32, 0.68) & 73568.33 &	74217.77 & 0.23 & 0	& 0.16 \\
			3& Frank (1.98) & survival Gumbel (1.26) & (0.33, 0.67) & 73570.40 & 74219.84 & 0.08 & 0.18 & 0 \\
			4& Gaussian (0.25) & survival Gumbel (1.32) & (0.32, 0.68) & 73571.36 & 74220.80 & 0.05 &	0.21 & 0 \\
			5& Gaussian (0.27) & Gumbel (1.27)	& (0.32, 0.68) & 73572.60 & 74222.04 & 0.03 & 0 & 0.19 \\
			6& Frank (2.05) & survival Joe (1.42) & (0.50, 0.50) &	73573.01 & 74222.45& 0.02 & 0.19 & 0 \\
			7& Gaussian (0.28) & Clayton (0.52) & (0.35, 0.65) & 73573.46 &	74222.90 & 0.02 & 0.17 & 0 \\
			8& Gaussian (0.30) & Frank (2.04) & (0.33, 0.67) & 73574.57 &	74224.01 & 0.01& 0 & 0 \\
			9& Gaussian (0.28) & survival Joe (1.44) &	(0.36, 0.64) &	73574.76 & 74224.20 &	0.01 & 0.24 & 0 \\
			10& Gaussian (0.27) & t (0.39, 9.06) & (0.52, 0.48) & 73569.13 & 74224.22 &	0.01 & 0.03 & 0.03 \\
			11& Gaussian (0.30) & survival Clayton (0.37) &	(0.33, 0.67) &	73575.15 & 74224.59 & 0.01 & 0 &	0.10
			\\
			12& Frank (2.07) & t (0.29, 11.30) &	(0.32, 0.68) & 73570.07 &	74225.16 & 0.01 & 0.02 & 0.02 \\
			\bottomrule[.15 em]
	\end{tabular} }
\end{table}

The dominating copula mixture is Clayton and Frank with similar mixing probabilities, which accounts for 52\% model weight. This mixture does not have an upper TDC but does have relatively strong lower tail dependence, which is consistent with the findings through empirical estimation. Furthermore, the best 12 models have relatively close BIC values to each other, suggesting model uncertainty is an important factor to consider in modelling. Among all 12 models, 11 have either lower or upper TDC or both, which shows that the tail dependence phenomenon indeed exists, contrasting to the empirical estimators which cannot verify its existence.  

\begin{figure}[p!]
	\centering
	\begin{minipage}{0.49\textwidth}
		\includegraphics[width=\linewidth, height=.8\linewidth]{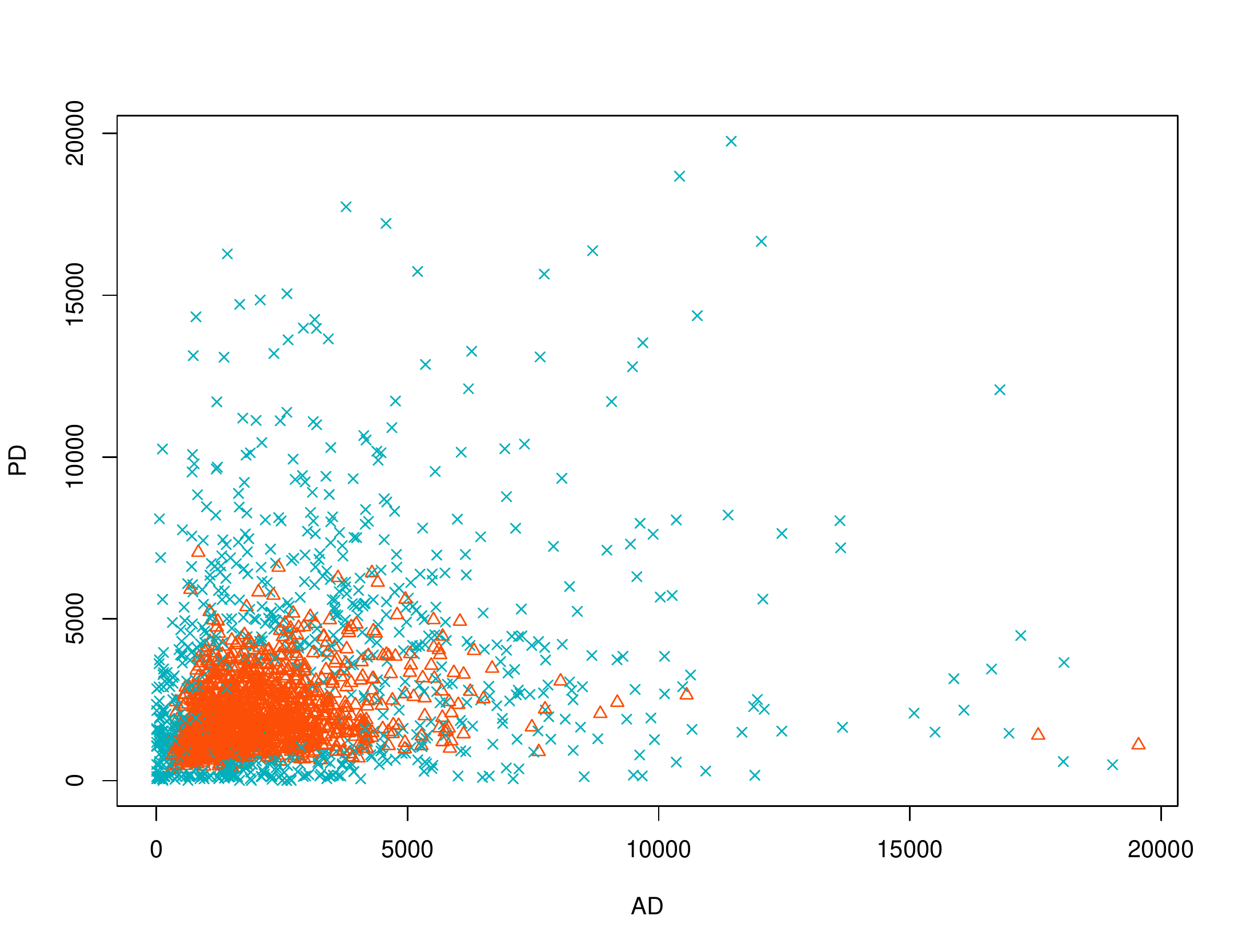}
		\caption*{(1) Frank + Clayton}
	\end{minipage}
	\begin{minipage}{0.49\textwidth}
		\includegraphics[width=\linewidth, height=.8\linewidth]{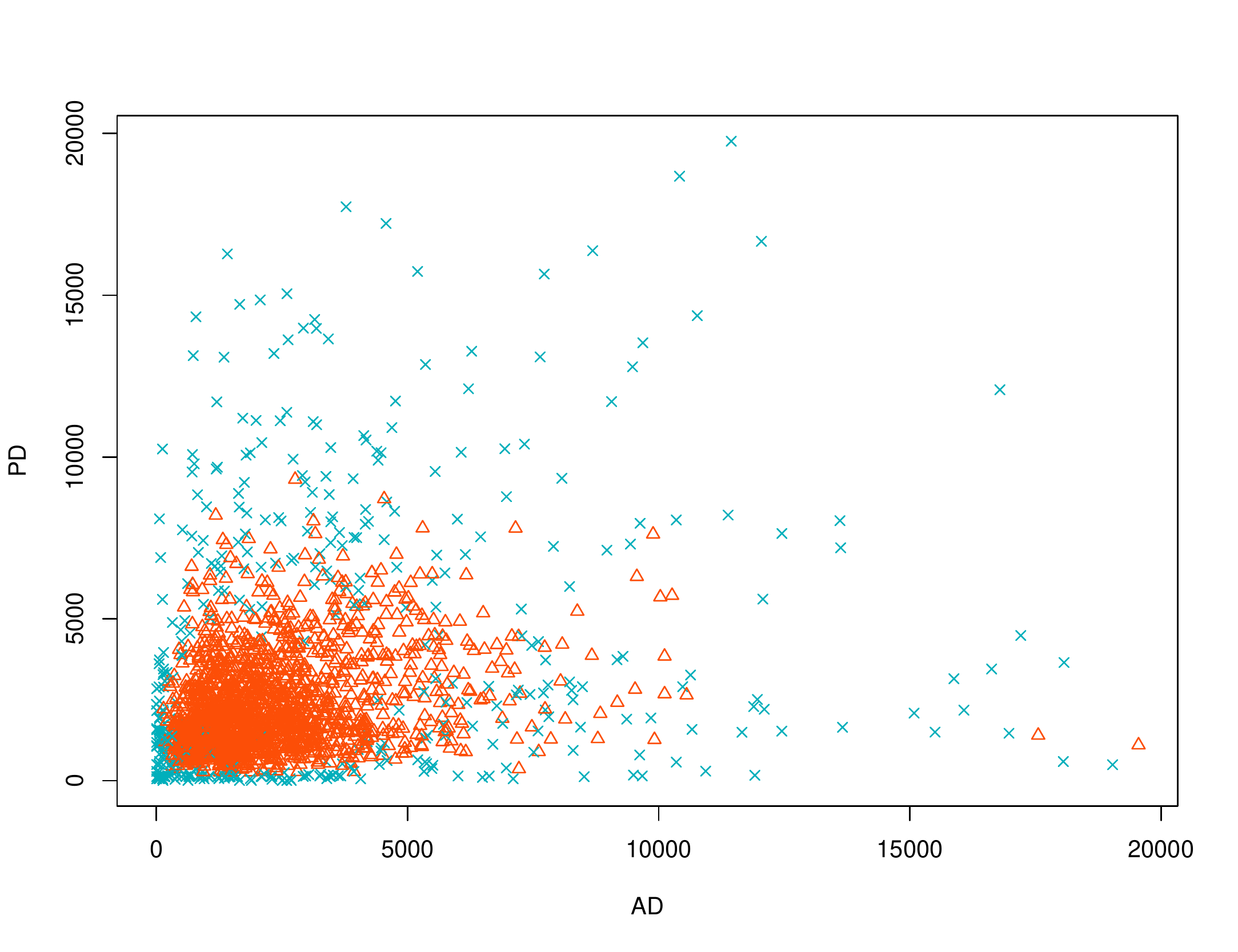}
		\caption*{(2) Frank + Gumbel}
	\end{minipage}
	\begin{minipage}{0.49\textwidth}
		\includegraphics[width=\linewidth, height=.8\linewidth]{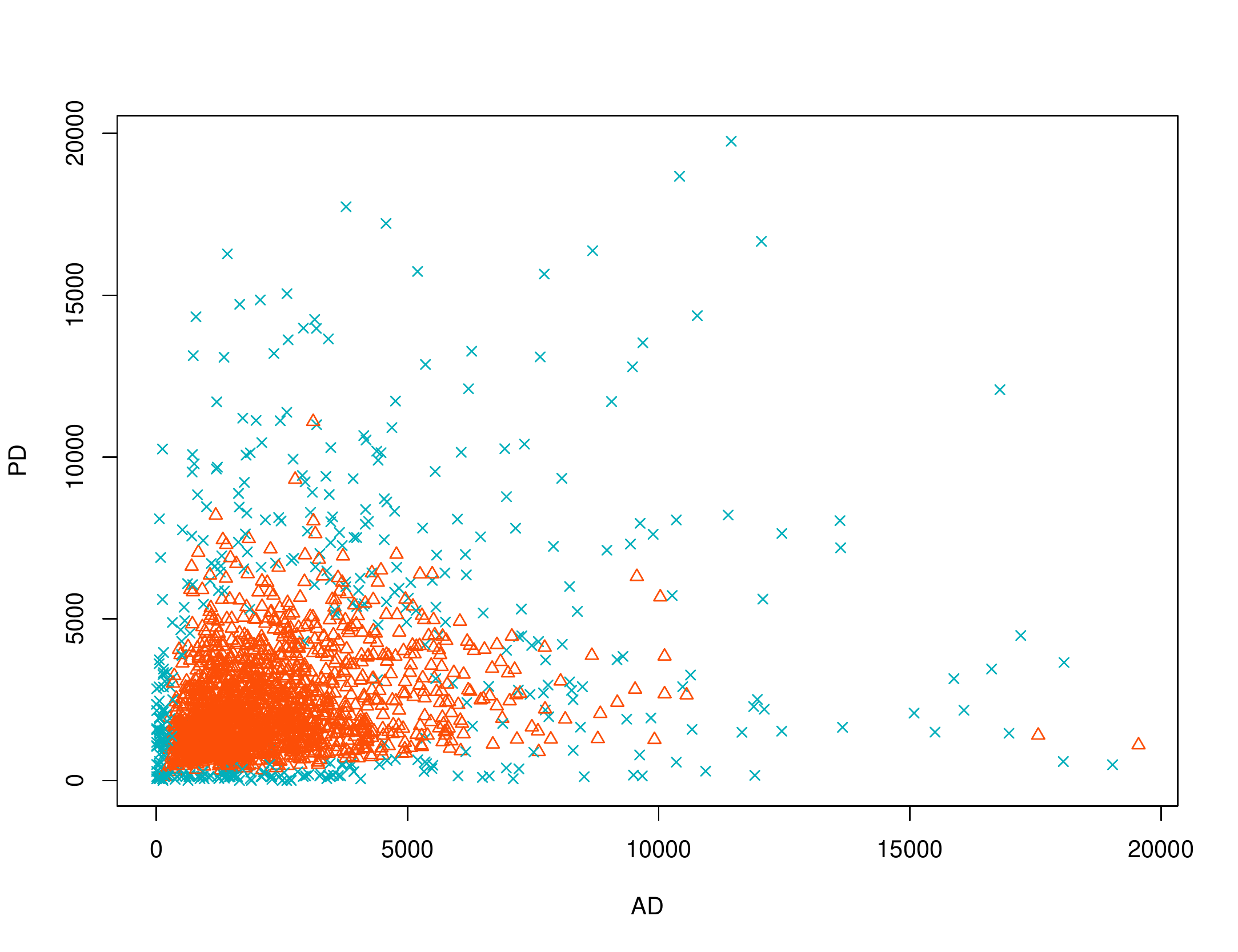}
		\caption*{(3) Frank + survival Gumbel}
	\end{minipage}
	\begin{minipage}{0.49\textwidth}
		\includegraphics[width=\linewidth, height=.8\linewidth]{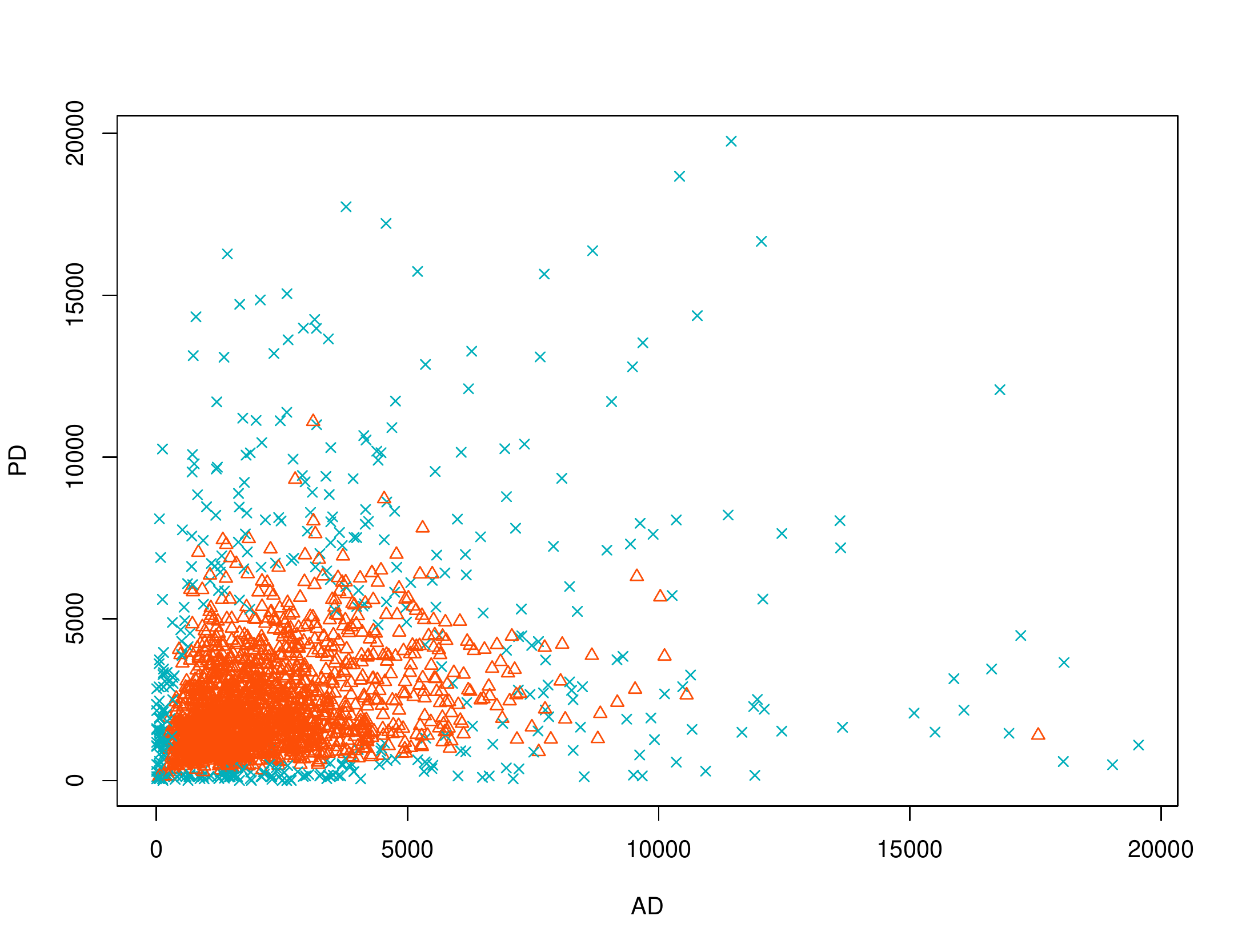}
		\caption*{(4) Gaussian + survival Gumbel}
	\end{minipage}
	\begin{minipage}{0.49\textwidth}
		\includegraphics[width=\linewidth, height=.8\linewidth]{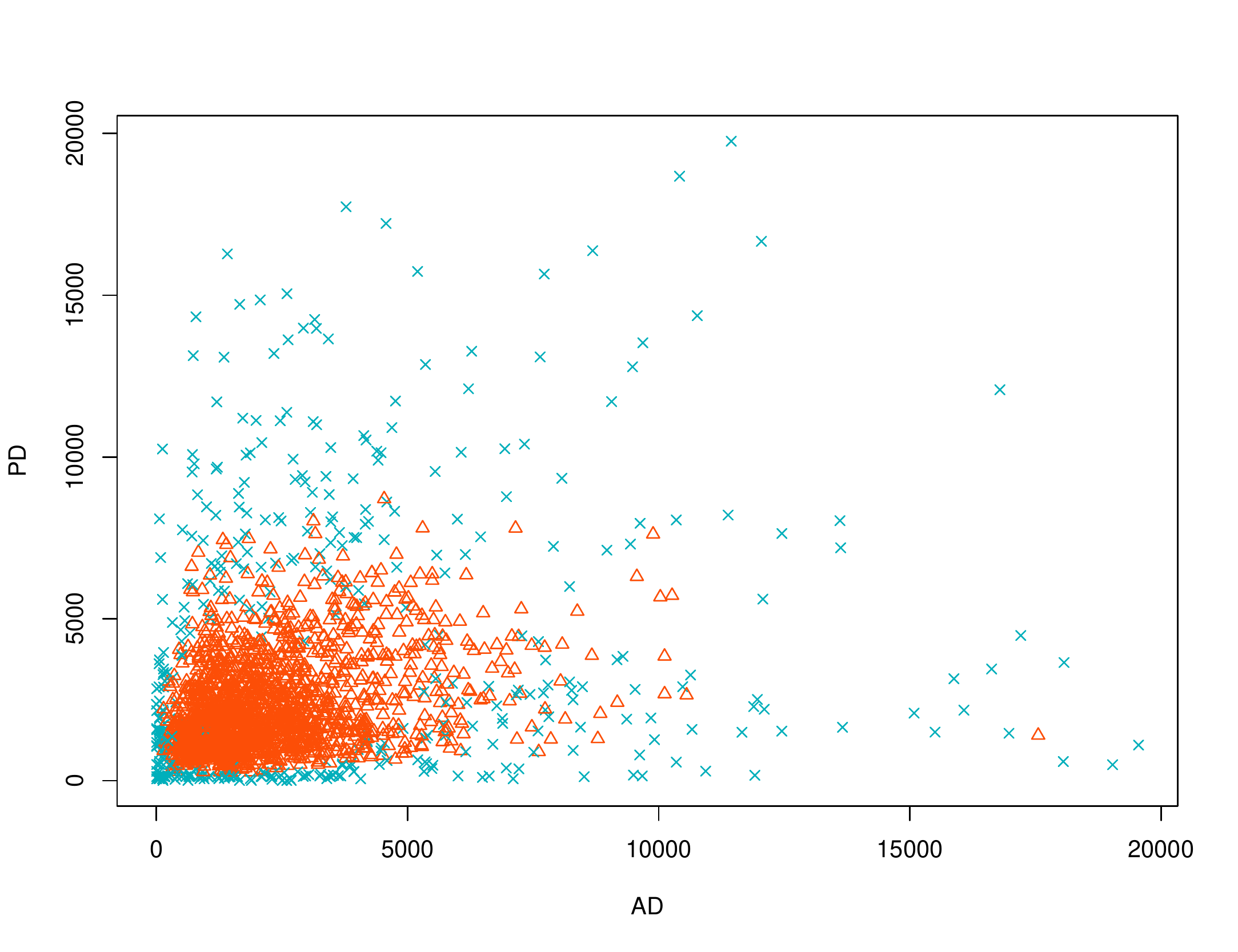}
		\caption*{(5) Gaussian + Gumbel}
	\end{minipage}
	\begin{minipage}{0.49\textwidth}
		\includegraphics[width=\linewidth, height=.8\linewidth]{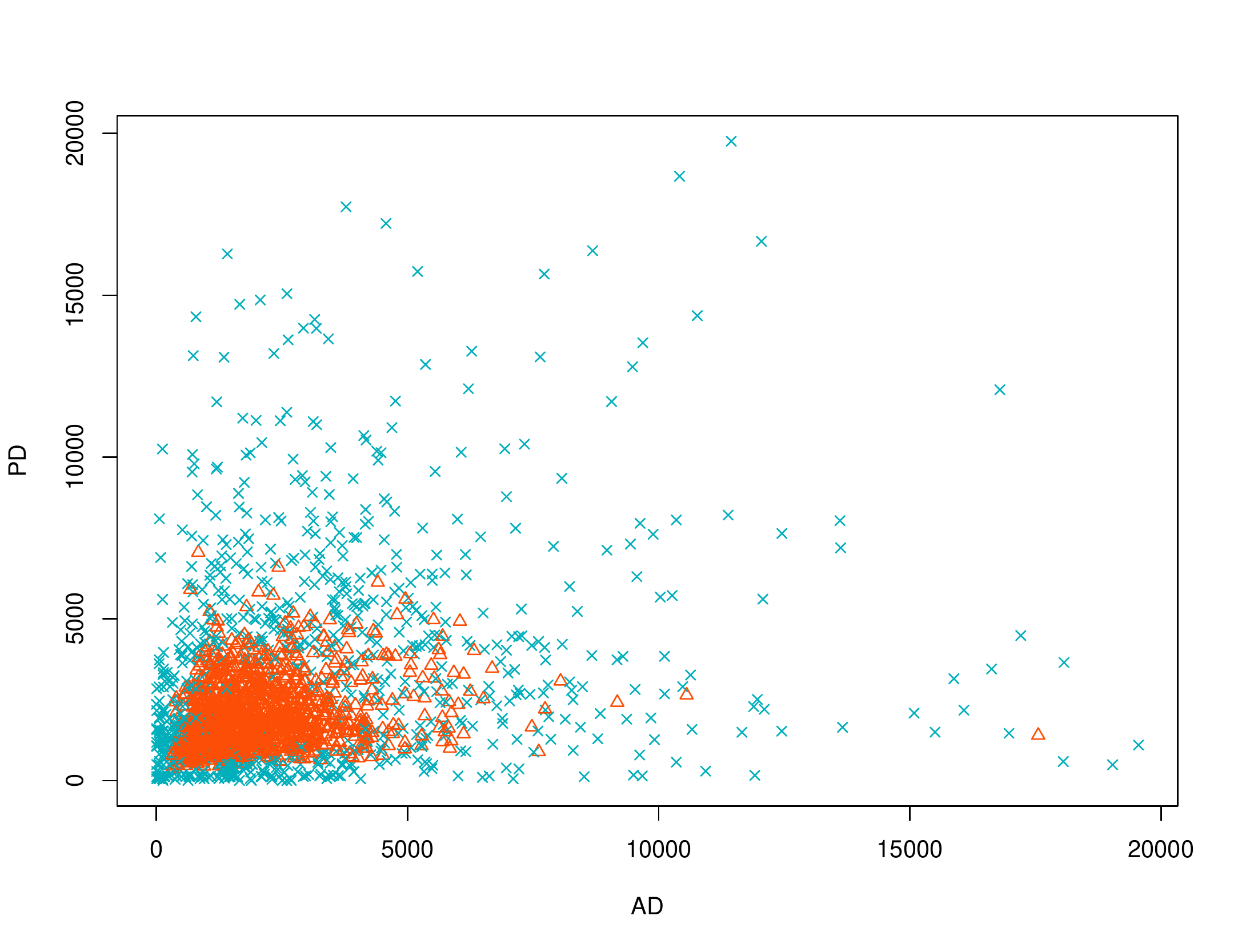}
		\caption*{(6) Frank + survival Joe}
	\end{minipage}
	\caption{Classification of accidental damage (AD) and third party property damage (PD) bivariate claims by different finite mixtures of copulas, coloured by membership, corresponding to Table~\ref{tab:realdata_copula_mixture_summary}. Only the best 6 models are shown here due to space limit; the remaining models have similar clustering results as those presented here. The teal ``$\times$" denotes component 1, the red ``$\triangle$" denotes component 2. These optimal models all have $G=2$.}
	\label{fig:realdata_copula_mixture_clustering}
\end{figure}

Finite mixtures of copulas can be regarded as a clustering method, to segment the motor policy customers into homogeneous groups; the clustering results (only the best 6 models due to space limitations) are shown in Figure~\ref{fig:realdata_copula_mixture_clustering}. 
All models have similar clustering results: component 1 (either Frank or Gaussian copula) consists of the very large claims, as well as those who made very small AD claims but relatively large PD claims and who made very small PD claims but relatively large AD claims. The second component consists of the medium claims of both AD and PD, i.e. the dense cluster in the middle of the data. This clustering result makes sense, because component 1 can be interpreted as a noise component, which is always modelled by a Frank or Gaussian copula that does not have any tail dependence; whereas component 2 is a more flexible component that can be modelled by a copula with either upper or lower tail dependence or both.     
\\

\noindent\textbf{Method 1:}\\

The fitted TDCs are directly averaged based on the weights, giving the results in Table~\ref{tab:newrealdata_method1_result}.
When comparing the BMA-weighted estimation to the single best fitting model with highest model weight (i.e. mixture of Frank and Clayton in Table~\ref{tab:realdata_copula_mixture_summary}), it shows that: (1) the copula-averaged upper TDC is better and potentially closer to the true value (approximated by the empirical estimations), whereas the Frank+Clayton mixture under-estimates the upper TDC and the upper TDC indeed exists but is rather weak, as concluded from Table~\ref{tab:realdata_copula_mixture_summary};  
(2) the copula-averaged lower TDC is closer to the potential true value and is at least similar to the Frank+Clayton mixture if not better. As concluded earlier, the empirical results may have over-estimated the TDCs. 

\begin{table}[h!]
	\centering
	\caption{Results of method 1: comparison of upper and lower TDCs estimated for the Irish GI insurer data.}
	\label{tab:newrealdata_method1_result}
	\begin{tabular}{l|ccc}
		\toprule[.15 em]
		\ & Empirical & Frank+Clayton & BMA \\
		\midrule
		$\lambda_U$ & [0.081, 0.164] & 0    & 0.05 \\
		$\lambda_L$ & [0.103, 0.200] & 0.13 & 0.10 \\ 
		\bottomrule[.15 em]
	\end{tabular}
\end{table}

\noindent\textbf{Method 2} \\

The second method invokes the use of empirical estimators --
because the true TDCs are unknown it is also unknown which nonparametric estimator provides the best estimation, hence all five estimators are tested, with the results shown in Table~\ref{tab:newrealdata_method2_result}. 
Note that this process has been repeated 1000 times to ensure stable results, and the mean and standard deviation of the estimated $\widehat{\lambda}_U$ and $\widehat{\lambda}_L$ are calculated and compared.
Regardless of the drawbacks of the empirical estimators, the results show that no matter which nonparametric estimator is used, the TDCs estimated from the BMA simulated data set are always closer to the (assumed) truth (which likely lies in the range $[0.081, 0.164]$ for $\lambda_U$ and $[0.103, 0.200]$ for $\lambda_L$ from Table~\ref{tab:newrealdata_TDC_estimation}) than those estimated from the simulated data from the best fitting individual copula (Frank+Clayton mixture). Therefore, it can be concluded that the BMA approach better reflects the true nature of the data, i.e. is the most similar to the original data set, which is the motivation of this second method.  
The underlying assumption can be verified using Wasserstein and $L^2$ distances, which shows that the BMA method outperforms the best fitting Frank+Clayton mixture in Table~\ref{tab:newrealdata_method2_distance}.

\begin{table}[h!]
	\centering
	\caption{Mean of the empirical nonparametric estimates of TDC over 1000 repetitions, using proposed method 2. In each iteration empirical TDCs are estimated using all five estimators on simulations from the best fitting individual copula and the BMA simulated set. Standard deviations are shown in brackets.}
	\label{tab:newrealdata_method2_result}
	\begin{tabular}{lcc||ccc}
		\toprule[.15 em]
		\ & Frank+Clayton & BMA & \ & Frank+Clayton & BMA \\
		\midrule
		$\overline{\lambda}_U^{(1)}$ & 0.074 (0.028) & 0.119 (0.049) & $\overline{\lambda}_L^{(1)}$ & 0.167 (0.052) & 0.145 (0.049) \\
		$\overline{\lambda}_U^{(2)}$ & 0.052 (0.029) & 0.096 (0.040) & $\overline{\lambda}_L^{(2)}$ & 0.150 (0.056) & 0.125 (0.052) \\
		$\overline{\lambda}_U^{(3)}$ & 0.031 (0.025) & 0.076 (0.040) & $\overline{\lambda}_L^{(3)}$ & 0.161 (0.055) & 0.138 (0.052) \\
		$\overline{\lambda}_U^{(4)}$ & 0.049 (0.023) & 0.092 (0.032) & $\overline{\lambda}_L^{(4)}$ & 0.147 (0.042) & 0.089 (0.048) \\ 
		$\overline{\lambda}_U^{(5)}$ & 0.040 (0.028) & 0.085 (0.040) & $\overline{\lambda}_L^{(5)}$ & 0.169 (0.053) & 0.145 (0.050) \\
		\bottomrule[.15 em]
	\end{tabular} 
\end{table}

\begin{table}[h!]
	\centering
	\caption{Comparison of distances between simulations from the best fitting individual copula and the BMA simulated data set to the Irish insurer data set. The underlined values indicate that BMA provides the best fit to the original data.}
	\label{tab:newrealdata_method2_distance}
	\begin{tabular}{l|rr}
		\toprule[.15 em]
		Distance & Frank+Clayton & BMA \\
		\midrule 
		Wasserstein & 57.55 & \underline{54.23} \\
		$L^2$ distance & 0.45 & \underline{0.34} \\
		\bottomrule[.15 em]
	\end{tabular} 
\end{table}

\section{Conclusion and further work}
\label{sec:conclusion}

This work investigated the use of Bayesian model averaging in conjunction with copulas, with a particular focus on more accurately estimating the tail dependence coefficient, which is a prime quantity of interest with respect to the dependence structure among different risks especially in terms of simulataneous pricing process of multiple products for insurance companies.
It is typically the case that when fitting copulas to insurance claims data, the best fitting copula is used to extract the quantities of interest, such as the TDCs, which provides a model-based approach for TDC estimation.
However, this approach does not take model uncertainty into account, and if another copula or copulas also give reasonable fit it may be sensible to blend multiple results.
It has been shown that BMA can always match or improve the estimation of TDCs compared to using individual fitted copulas, in both simulated studies and a real-world example. This estimation can be validated via either known true values or nonparametric estimates based on the empirical copula. 

To date, empirical estimation of the TDC is a common and popular approach. 
Although sometimes nonparametric estimators can give an accurate approximation, most of the time different nonparametric empirical copula estimators provide very different and unstable results.  
Hence only a range of approximating values or the mean of the empirical estimates can be considered.
More importantly, when the tail dependence is weak or non-existent, they often over-estimate the coefficient. 
Model-based estimation with copulas, especially invoking BMA, produces a much more stable estimation based on well-fitting copulas to the data, taking model uncertainty and goodness-of-fit into account. In particular, weak or non-existent tail dependence can be more accurately detected.
Therefore, BMA can be regarded as a robust parametric estimation procedure for TDCs, providing valuable insight as to the degree of tail dependence present in insurance claims data.

In one instance, if the data originate from one copula, then fitting that copula is expected to out-perform any other competing copulas in terms of log-likelihood, as well as produce a very accurate estimate of the true TDC. 
In this case, BMA does not negatively impact the TDC estimate.
The simulation studies in the article also confirm that including BMA is either comparable to the ``best" copula or considerably improves the estimation.   
Furthermore, it can be argued that model averaging works best where data exhibit a more complex dependence structure that cannot be described by one copula with high model certainty, which is typically the case for real-world insurance claims data.
BMA then permits multiple estimates of TDCs to be combined to produce a more accurate overall estimate, outperforming any copula individually.
This article investigated different ways of creating non-standard dependence structures, through mixtures of copulas and bivariate gamma distributions in two simulation studies.
It is noted that TDCs can be represented neatly by weighted TDCs of individual copulas in a mixture, including when multiple copulas are mixed whereby some account for the upper TDC and some account for the lower TDC (\citealp{Arakelian2014}). Therefore, the use of BMA can be extended to the context of mixtures of copulas, which is investigated via the Irish GI insurer data. 

By applying BMA to finite mixtures of copulas, the focus in this article is to better estimate tail dependence that can be disguised by heterogeneity in the data. From a clustering perspective this bears similarity to \cite{Wei2015} and \cite{Russell2015}, and the BMA approach can be used for improving clustering membership. Because covariates are used in our investigation to assist with marginal distribution estimation, from a prediction perspective this approach produces predicted values and can be regarded as future claims predictions via mixture of regressions while taking dependence into account.
However, due to the TDC estimation focus, these research strands are not investigated in this article. 

The main focus of this article is to explore the possibilities afforded by BMA in combining copulas, contrasting the goodness of fit of the selected copulas and BMA in producing accurate TDC estimates for the data. The investigation focused on bivariate copulas. 
A potential further advance could be the evaluation of the method in three dimensional or higher cases. 
When the dimension is three or more, more interesting and complex dependence structures can be constructed, for example via nested Archimedean copulas, vine copulas or extreme copulas.
For vine copulas, the dependence structure consists of a nested set of trees, which is typically not unique, and the bivariate copulas for each pair also need to be chosen. 
It seems natural to envisage that a process such as BMA would work well in those cases and can be explored in future work. 

This work investigated a limited number of copulas in the BMA context, selected due to their popularity. 
However, in the literature there are many varieties of copulas, for example the BB1 copula (\citealp{Joe1997}), the BB7 copula (\citealp{Joe1997}), the skewed-t copula (\citealp{Demarta2005}), and their rotated (including survival) versions. For more choices, see \cite{Nelsen2007}.
Analysis of a wider pool of options could be considered to further validate the capability of the BMA process for blending copula results.

\section*{Acknowledgements}
This work was supported by the Science Foundation Ireland funded Insight Research Centre (SFI/12/RC/2289\_P2).

\bibliographystyle{wb_stat}
\bibliography{Ref_copulaBMA}

\begin{thebibliography}{75}
\newcommand{\enquote}[1]{`#1'}
\providecommand{\natexlab}[1]{#1}
\expandafter\ifx\csname urlstyle\endcsname\relax
  \providecommand{\doi}[1]{doi:\discretionary{}{}{}#1}\else
  \providecommand{\doi}{doi:\discretionary{}{}{}\begingroup
  \urlstyle{rm}\Url}\fi

\bibitem[{Aitken(1927)}]{Aitken1927}
Aitken, AC (1927), \enquote{{On Bernoulli's numerical solution of algebraic
  equations},} \emph{Proceedings of the Royal Society of Edinburgh},
  \textbf{46}, pp. 289--305.

\bibitem[{Akaike(1974)}]{Akaike1974}
Akaike, H (1974), \enquote{A new look at the statistical model identification,}
  \emph{IEEE Transactions on Automatic Control}, \textbf{19}(6), pp. 716--723.

\bibitem[{Apputhurai \& Stephenson(2011)}]{Apputhurai2011}
Apputhurai, P \& Stephenson, A (2011), \enquote{{Accounting for uncertainty in
  extremal dependence modeling using Bayesian model averaging techniques},}
  \emph{Journal of Statistical Planning and Inference}, \textbf{141}(5), pp.
  1800--1807.

\bibitem[{Arakelian \& Karlis(2014)}]{Arakelian2014}
Arakelian, V \& Karlis, D (2014), \enquote{{Clustering dependencies via
  mixtures of copulas},} \emph{Communications in Statistics - Simulation and
  Computation}, \textbf{43}(7), pp. 1644--1661.

\bibitem[{Bedford \& Cooke(2002)}]{Bedford2002}
Bedford, T \& Cooke, RM (2002), \enquote{Vines -- a new graphical model for
  dependent random variables,} \emph{The Annals of Statistics}, \textbf{30}(4),
  pp. 1031--1068.

\bibitem[{Berm{\'{u}}dez(2009)}]{Bermudez2009}
Berm{\'{u}}dez, L (2009), \enquote{{A priori ratemaking using bivariate Poisson
  regression models},} \emph{Insurance: Mathematics and Economics},
  \textbf{44}(1), pp. 135--141.

\bibitem[{Berm{\'{u}}dez \& Karlis(2012)}]{Bermudez2012}
Berm{\'{u}}dez, L \& Karlis, D (2012), \enquote{{A finite mixture of bivariate
  Poisson regression models with an application to insurance ratemaking},}
  \emph{Computational Statistics {\&} Data Analysis}, \textbf{56}(12), pp.
  3988--3999.

\bibitem[{Biernacki et~al.(2000)Biernacki, Celeux \& Govaert}]{Biernacki2000}
Biernacki, C, Celeux, G \& Govaert, G (2000), \enquote{{Assessing a mixture
  model for clustering with the integrated completed likelihood},} \emph{IEEE
  Transactions on Pattern Analysis and Machine Intelligence}, \textbf{22}(7),
  pp. 719--725.

\bibitem[{B{\"{o}}hning et~al.(1994)B{\"{o}}hning, Dietz, Schaub, Schlattmann
  \& Lindsay}]{Bohning1994}
B{\"{o}}hning, D, Dietz, E, Schaub, R, Schlattmann, P \& Lindsay, BG (1994),
  \enquote{The distribution of the likelihood ratio for mixtures of densities
  from the one-parameter exponential family,} \emph{Annals of the Institute of
  Statistical Mathematics}, \textbf{46}(2), pp. 373--388.

\bibitem[{Caillault \& Gu{\'{e}}gan(2005)}]{Caillault2005}
Caillault, C \& Gu{\'{e}}gan, D (2005), \enquote{{Empirical estimation of tail
  dependence using copulas: application to Asian markets},} \emph{Quantitative
  Finance}, \textbf{5}(5), pp. 489--501.

\bibitem[{Cheriyan(1941)}]{Cheriyan1941}
Cheriyan, K (1941), \enquote{{A bivariate correlated gamma-type distribution
  function},} \emph{Journal of the Indian Mathematical Society}, \textbf{5},
  pp. 133--144.

\bibitem[{Cherubini et~al.(2004)Cherubini, Luciano \&
  Vecchiato}]{Cherubini2004}
Cherubini, U, Luciano, E \& Vecchiato, W (2004), \emph{Copula Methods in
  Finance}, Wiley, West Sussex.

\bibitem[{Clayton(1978)}]{Clayton1978}
Clayton, DG (1978), \enquote{{A model for association in bivariate life tables
  and its application in epidemiological studies of familial tendency in
  chronic disease incidence},} \emph{Biometrika}, \textbf{65}(1), pp. 141--151.

\bibitem[{Coles et~al.(1999)Coles, Heffernan \& Tawn}]{Coles1999}
Coles, S, Heffernan, J \& Tawn, J (1999), \enquote{Dependence measures for
  extreme value analyses,} \emph{Extremes}, \textbf{2}(4), pp. 339--365.

\bibitem[{{CRO Forum}(2019)}]{CROForum2019}
{CRO Forum} (2019), \enquote{The heat is on -- insurability and resilience in a
  changing climate,} {Emerging Risk Initiative -- Position Paper}, CRO Forum,
  Amsterdam.

\bibitem[{Demarta \& McNeil(2005)}]{Demarta2005}
Demarta, S \& McNeil, AJ (2005), \enquote{The t copula and related copulas,}
  \emph{International Statistical Review}, \textbf{73}(1), pp. 111--129.

\bibitem[{Dempster et~al.(1977)Dempster, Laird \& Rubin}]{Dempster1977}
Dempster, AP, Laird, NM \& Rubin, DB (1977), \enquote{{Maximum likelihood from
  incomplete data via the EM algorithm},} \emph{Journal of the Royal
  Statistical Society. Series B (Methodological)}, \textbf{39}, pp. 1--38.

\bibitem[{Denuit et~al.(2006)Denuit, Dhaene, Goovaerts \& Kaas}]{Denuit2005}
Denuit, M, Dhaene, J, Goovaerts, M \& Kaas, R (2006), \emph{Actuarial Theory
  for Dependent Risks: Measures, Orders and Models}, Wiley, West Sussex.

\bibitem[{Dobric \& Schmid(2005)}]{Dobric2005}
Dobric, J \& Schmid, F (2005), \enquote{Nonparametric estimation of the lower
  tail dependence {$\lambda_L$} in bivariate copulas,} \emph{Journal of Applied
  Statistics}, \textbf{32}(4), pp. 387--407.

\bibitem[{Draper(1995)}]{Draper1995}
Draper, D (1995), \enquote{{Assessment and propagation of model uncertainty},}
  \emph{Journal of the Royal Statistical Society. Series B (Methodological)},
  \textbf{57}(1), pp. 45--97.

\bibitem[{Embrechts et~al.(2003)Embrechts, Lindskog \& McNeil}]{Embrechts2003}
Embrechts, P, Lindskog, F \& McNeil, A (2003), \enquote{{Modelling dependence
  with copulas and applications to risk management},} in {Svetlozar T. Rachev}
  (ed.), \emph{Handbook of Heavy Tailed Distributions in Finance},
  North-Holland, Amsterdam, pp. 329--384.

\bibitem[{Engelke et~al.(2019)Engelke, Opitz \& Wadsworth}]{Engelke2019}
Engelke, S, Opitz, T \& Wadsworth, J (2019), \enquote{Extremal dependence of
  random scale constructions,} \emph{arXiv preprint: 1803.04221v2}.

\bibitem[{{European Insurance annd Occupational Pensions
  Authority}(2010)}]{SolvencyII}
{European Insurance annd Occupational Pensions Authority} (2010),
  \enquote{{CEIOPS}'s advice for level 2 implementing measures on {Solvency
  II}: {SCR} standrd formula - correlations,}
  \url{https://eiopa.europa.eu/publications/solvency-ii-final-l2-advice},
  accessed 29 April 2019.

\bibitem[{Everitt et~al.(2011)Everitt, Landau, Leese \& Stahl}]{Everitt2011}
Everitt, B, Landau, S, Leese, M \& Stahl, D (2011), \emph{Cluster Analysis},
  Wiley, West Sussex, 5 edn.

\bibitem[{Fischer \& D{\"{o}}rflinger(2006)}]{Fischer2006}
Fischer, MJ \& D{\"{o}}rflinger, M (2006), \emph{A note on a non-parametric
  tail dependence estimator}, no. 76/2006 in Diskussionspapiere --
  Friedrich-Alexander-Universit{\"{a}}t Erlangen-N{\"{u}}rnberg, Lehrstuhl
  f{\"{u}}r Statistik und {\"{O}}konometrie, Universit{\"{a}}t
  Erlangen-N{\"{u}}rnberg, Lehrstuhl f{\"{u}}r Statistik und empirische
  Wirtschaftsforschung, N{\"{u}}rnberg.

\bibitem[{Frahm et~al.(2005)Frahm, Junker \& Schmidt}]{Frahm2005}
Frahm, G, Junker, M \& Schmidt, R (2005), \enquote{Estimating the
  tail-dependence coefficient: properties and pitfalls,} \emph{Insurance:
  Mathematics and Economics}, \textbf{37}(1), pp. 80--100.

\bibitem[{Frank(1979)}]{Frank1979}
Frank, M (1979), \enquote{On the simultaneous associativity of f(x, y) and x +
  y - f(x, y).} \emph{Aequationes Mathematicae}, \textbf{19}(1), pp. 194--226.

\bibitem[{Frees \& Valdez(1998)}]{Frees1998}
Frees, EW \& Valdez, Ea (1998), \enquote{{Understanding relationships using
  copulas},} \emph{North American Actuarial Journal}, \textbf{2}(1), pp. 1--25.

\bibitem[{Genest \& Favre(2007)}]{Genest2007}
Genest, C \& Favre, AC (2007), \enquote{{Everything you always wanted to know
  about copula modeling but were afraid to ask},} \emph{Journal of Hydrologic
  Engineering}, \textbf{12}(4), pp. 347--368.

\bibitem[{Genest \& Rivest(1993)}]{Genest1993}
Genest, C \& Rivest, LP (1993), \enquote{Statistical inference procedures for
  bivariate archimedean copulas,} \emph{Journal of the American Statistical
  Association}, \textbf{88}(423), pp. 1034--1043.

\bibitem[{Gormley \& Murphy(2011)}]{Gormley2011}
Gormley, IC \& Murphy, TB (2011), \enquote{{Mixture of experts modelling with
  social science applications},} in Mengersen, K, Robert, C \& Titterington, M
  (eds.), \emph{Mixture: Estimation and Applications}, Wiley, West Sussex.

\bibitem[{Gr{\o}nneberg \& Hjort(2014)}]{Gronneberg2014}
Gr{\o}nneberg, S \& Hjort, NL (2014), \enquote{The copula information
  criteria,} \emph{Scandinavian Journal of Statistics}, \textbf{41}(2), pp.
  436--459.

\bibitem[{Gumbel(1960)}]{Gumbel1960}
Gumbel, E (1960), \enquote{{Distributions de valeurs extr{\^{e}}mes en
  plusieurs dimensions},} \emph{Publications de l'Institut de Statistique de
  l'Universit{\'{e}} de Paris}, \textbf{9}, pp. 171--173.

\bibitem[{Hoeting et~al.(1999)Hoeting, Madigan, Raftery \&
  Volinsky}]{Hoeting1999}
Hoeting, JA, Madigan, D, Raftery, AE \& Volinsky, CT (1999), \enquote{{Bayesian
  model averaging: a tutorial},} \emph{Statistical Science}, \textbf{14}(4),
  pp. 382--417.

\bibitem[{Hofert et~al.(2018{\natexlab{a}})Hofert, Kojadinovic, M{\"a}chler \&
  Yan}]{Hofert2018}
Hofert, M, Kojadinovic, I, M{\"a}chler, M \& Yan, J (2018{\natexlab{a}}),
  \emph{copula: multivariate dependence with copulas}, {R} package version
  0.999-19.

\bibitem[{Hofert et~al.(2018{\natexlab{b}})Hofert, Kojadinovic, M{\"a}chler \&
  Yan}]{Hofert2018book}
Hofert, M, Kojadinovic, I, M{\"a}chler, M \& Yan, J (2018{\natexlab{b}}),
  \emph{{Elements of Copula Modeling with R}}, Springer, Switzerland.

\bibitem[{Hu et~al.(2019{\natexlab{a}})Hu, Murphy \& O'Hagan}]{Hu2019}
Hu, S, Murphy, TB \& O'Hagan, A (2019{\natexlab{a}}), \enquote{Bivariate gamma
  mixture of experts models for joint insurance claims modelling,} \emph{arXiv
  preprint:1904.04699}.

\bibitem[{Hu et~al.(2019{\natexlab{b}})Hu, Murphy \& O'Hagan}]{mvClaim2019}
Hu, S, Murphy, TB \& O'Hagan, A (2019{\natexlab{b}}), \emph{{mvClaim:
  multivariate general insurance claims modelling}}, {R} package version 0.1.0,
  \url{https://github.com/senhu/mvClaim}.

\bibitem[{Hu et~al.(2018)Hu, O'Hagan \& Murphy}]{Hu2018}
Hu, S, O'Hagan, A \& Murphy, TB (2018), \enquote{Motor insurance claim
  modelling with factor collapsing and {B}ayesian model averaging,}
  \emph{Stat}, \textbf{7}(1), p. e180.

\bibitem[{Huser et~al.(2016)Huser, Davison \& Genton}]{Huser2016}
Huser, R, Davison, AC \& Genton, MG (2016), \enquote{Likelihood estimators for
  multivariate extremes,} \emph{Extremes}, \textbf{19}(1), pp. 79--103.

\bibitem[{Joe(1997)}]{Joe1997}
Joe, H (1997), \emph{Multivariate models and dependence concepts}, Chapman {\&}
  Hall/CRC, Boca Raton.

\bibitem[{Joe(2005)}]{Joe2005}
Joe, H (2005), \enquote{{Asymptotic efficiency of the two-stage estimation
  method for copula-based models},} \emph{Journal of Multivariate Analysis},
  \textbf{94}(2), pp. 401--419.

\bibitem[{Joe(2014)}]{Joe2014}
Joe, H (2014), \emph{Dependence modeling with copulas}, Chapman {\&} Hall/CRC,
  Boca Raton.

\bibitem[{Joe \& Xu(1996)}]{Joe1996}
Joe, H \& Xu, JJ (1996), \enquote{The estimation method of inference functions
  for margins for multivariate models,} Technical Report 166, Department of
  Statistics, University of British Columbia.

\bibitem[{Karlis \& Meligkotsidou(2007)}]{Karlis2007}
Karlis, D \& Meligkotsidou, L (2007), \enquote{{Finite mixtures of multivariate
  Poisson distributions with application},} \emph{Journal of Statistical
  Planning and Inference}, \textbf{137}(6), pp. 1942--1960.

\bibitem[{Karlis \& Ntzoufras(2003)}]{Karlis2003}
Karlis, D \& Ntzoufras, I (2003), \enquote{{Analysis of sports data by using
  bivariate Poisson models},} \emph{Journal of the Royal Statistical Society:
  Series D (The Statistician)}, \textbf{52}(3), pp. 381--393.

\bibitem[{Ko \& Hjort(2019)}]{Ko2019cic}
Ko, V \& Hjort, NL (2019), \enquote{Copula information criterion for model
  selection with two-stage maximum likelihood estimation,} \emph{Econometrics
  and Statistics}, {In press}.

\bibitem[{Ko et~al.(2019)Ko, Hjort \& Hob{\ae}k~Haff}]{Ko2019fic}
Ko, V, Hjort, NL \& Hob{\ae}k~Haff, I (2019), \enquote{Focused information
  criteria for copulas,} \emph{Scandinavian Journal of Statistics},
  \textbf{46}(2), pp. 1--25.

\bibitem[{Kosmidis \& Karlis(2016)}]{Kosmidis2016}
Kosmidis, I \& Karlis, D (2016), \enquote{{Model-based clustering using copulas
  with applications},} \emph{Statistics and Computing}, \textbf{26}(5), pp.
  1079--1099.

\bibitem[{Kr{\"{a}}mer et~al.(2013)Kr{\"{a}}mer, Brechmann, Silvestrini \&
  Czado}]{Kramer2013}
Kr{\"{a}}mer, N, Brechmann, EC, Silvestrini, D \& Czado, C (2013),
  \enquote{Total loss estimation using copula-based regression models,}
  \emph{Insurance: Mathematics and Economics}, \textbf{53}(3), pp. 829--839.

\bibitem[{Ledford \& Tawn(1996)}]{Ledford1996}
Ledford, AW \& Tawn, JA (1996), \enquote{Statistics for near independence in
  multivariate extreme values,} \emph{Biometrika}, \textbf{83}(1), pp.
  169--187.

\bibitem[{Madadgar \& Moradkhani(2014)}]{Madadgar2014}
Madadgar, S \& Moradkhani, H (2014), \enquote{Improved {B}ayesian
  multimodeling: integration of copulas and {B}ayesian model averaging,}
  \emph{Water Resources Research}, \textbf{50}(12), pp. 9586--9603.

\bibitem[{Madigan \& Raftery(1994)}]{Madigan1994}
Madigan, D \& Raftery, AE (1994), \enquote{{Model selection and accounting in
  graphical models for model uncertainty using Occam's window},} \emph{Journal
  of the American Statistical Association}, \textbf{89}(428), pp. 1535--1546.

\bibitem[{Masarotto \& Varin(2017)}]{Masarotto2017}
Masarotto, G \& Varin, C (2017), \enquote{Gaussian copula regression in {R},}
  \emph{Journal of Statistical Software}, \textbf{77}(8), pp. 1--26.

\bibitem[{M{\"o}ller et~al.(2013)M{\"o}ller, Lenkoski \&
  Thorarinsdottir}]{Moller2013}
M{\"o}ller, A, Lenkoski, A \& Thorarinsdottir, TL (2013), \enquote{Multivariate
  probabilistic forecasting using ensemble {B}ayesian model averaging and
  copulas,} \emph{Quarterly Journal of the Royal Meteorological Society},
  \textbf{139}(673), pp. 982--991.

\bibitem[{Nelsen(2007)}]{Nelsen2007}
Nelsen, RB (2007), \emph{An introduction to copulas}, Springer, New York.

\bibitem[{Pelster \& Vilsmeier(2016)}]{Pelster2016}
Pelster, M \& Vilsmeier, J (2016), \enquote{The determinants of {CDS} spreads:
  evidence from the model space,} Bundesbank Discussion Paper 43/2016, Deutsche
  Bundesbank, Frankfurt.

\bibitem[{Pfaff(2016)}]{Pfaff2016}
Pfaff, B (2016), \emph{{Financial Risk Modelling and Portfolio Optimisation
  with R}}, Wiley, London, 2nd edn.

\bibitem[{{R Core Team}(2018)}]{R2018}
{R Core Team} (2018), \emph{R: a language and environment for statistical
  computing}, R Foundation for Statistical Computing, Vienna, Austria.

\bibitem[{Raftery(1984)}]{Raftery1984}
Raftery, AE (1984), \enquote{A continuous multivariate exponential
  distribution,} \emph{Communications in Statistics - Theory and Methods},
  \textbf{13}(8), pp. 947--965.

\bibitem[{Ramabhadran(1951)}]{Ramabhadran1951}
Ramabhadran, V (1951), \enquote{{A multivariate gamma-type distribution},}
  \emph{Sankhya}, \textbf{11}, pp. 45--46.

\bibitem[{Ramsey et~al.(2018)Ramsey, Park \& Li}]{Ramsey2018}
Ramsey, F, Park, E \& Li, X (2018), \enquote{{Bayesian model averaging for
  copulas: an application to dependence between crop yields and futures
  prices},} SCC-76 Meeting, 2018, April 5-7, Kansas City, Missouri 276153,
  SCC-76: Economics and Management of Risk in Agriculture and Natural
  Resources.

\bibitem[{Russell et~al.(2015)Russell, Murphy \& Raftery}]{Russell2015}
Russell, N, Murphy, TB \& Raftery, AE (2015), \enquote{Bayesian model averaging
  in model-based clustering and density estimation,} \emph{arXiv
  preprint:1506.09035}.

\bibitem[{Sabourin et~al.(2013)Sabourin, Naveau \& Foug{\`e}res}]{Sabourin2013}
Sabourin, A, Naveau, P \& Foug{\`e}res, AL (2013), \enquote{Bayesian model
  averaging for multivariate extremes,} \emph{Extremes}, \textbf{16}(3), pp.
  325--350.

\bibitem[{Salmon(2012)}]{Salmon2012}
Salmon, F (2012), \enquote{{The formula that killed Wall Street},}
  \emph{Significance}, \textbf{9}(1), pp. 16--20.

\bibitem[{Schmidt \& Stadtm{\"u}ller(2006)}]{Schmidt2006}
Schmidt, R \& Stadtm{\"u}ller, U (2006), \enquote{Non-parametric estimation of
  tail dependence,} \emph{Scandinavian Journal of Statistics}, \textbf{33}(2),
  pp. 307--335.

\bibitem[{Schuhmacher et~al.(2018)Schuhmacher, B{\"{a}}hre, Gottschlich,
  Hartmann, Heinemann \& Schmitzer}]{Schuhmacher2018}
Schuhmacher, D, B{\"{a}}hre, B, Gottschlich, C, Hartmann, V, Heinemann, F \&
  Schmitzer, B (2018), \emph{{transport}: computation of optimal transport
  plans and {W}asserstein distances}, {R package version 0.10-0}.

\bibitem[{Schwarz(1978)}]{Schwarz1978}
Schwarz, G (1978), \enquote{Estimating the dimension of a model,} \emph{The
  Annals of Statistics}, \textbf{6}(2), pp. 461--464.

\bibitem[{Scrucca et~al.(2017)Scrucca, Fop, Murphy \& Raftery}]{Scrucca2017}
Scrucca, L, Fop, M, Murphy, TB \& Raftery, AE (2017), \enquote{{mclust} 5:
  clustering, classification and density estimation using {G}aussian finite
  mixture models,} \emph{{The R Journal}}, \textbf{8}(1), pp. 205--233.

\bibitem[{Sibuya(1960)}]{Sibuya1960}
Sibuya, M (1960), \enquote{Bivariate extreme statistics,} \emph{Annals of the
  Institute of Statistical Mathematics}, \textbf{11}(3), pp. 195--210.

\bibitem[{Sklar(1959)}]{Sklar1959}
Sklar, A (1959), \enquote{{Fonctions de r{\'{e}}partition {\`{a}} n dimensions
  et leurs marges},} \emph{Publications de l'Institut Statistique de
  l'Universit{\'{e}} de Paris}, \textbf{8}, pp. 229--231.

\bibitem[{Sweeting \& Fotiou(2013)}]{Sweeting2013}
Sweeting, P \& Fotiou, F (2013), \enquote{{Calculating and communicating tail
  association and the risk of extreme loss},} \emph{British Actuarial Journal},
  \textbf{18}(1), pp. 13--72.

\bibitem[{Vallender(1974)}]{Vallender1974}
Vallender, S (1974), \enquote{{Calculation of the Wasserstein distance between
  probability distributions on the line},} \emph{Theory of Probability {\&} Its
  Applications}, \textbf{18}(4), pp. 784--786.

\bibitem[{Wei \& McNicholas(2015)}]{Wei2015}
Wei, Y \& McNicholas, PD (2015), \enquote{{Mixture model averaging for
  clustering},} \emph{Advances in Data Analysis and Classification},
  \textbf{9}(2), pp. 197--217.

\bibitem[{Yan(2007)}]{Yan2007}
Yan, J (2007), \enquote{Enjoy the joy of copulas: with a package copula,}
  \emph{Journal of Statistical Software}, \textbf{21}(4), pp. 1--21.

\end{thebibliography}

\newpage
\appendix
\section*{Appendix}

\section{Simulation study III}
\label{app:simstudy4}

\begin{figure}[H]
	\centering
	\begin{minipage}{.45\linewidth}
		\includegraphics[width=\linewidth]{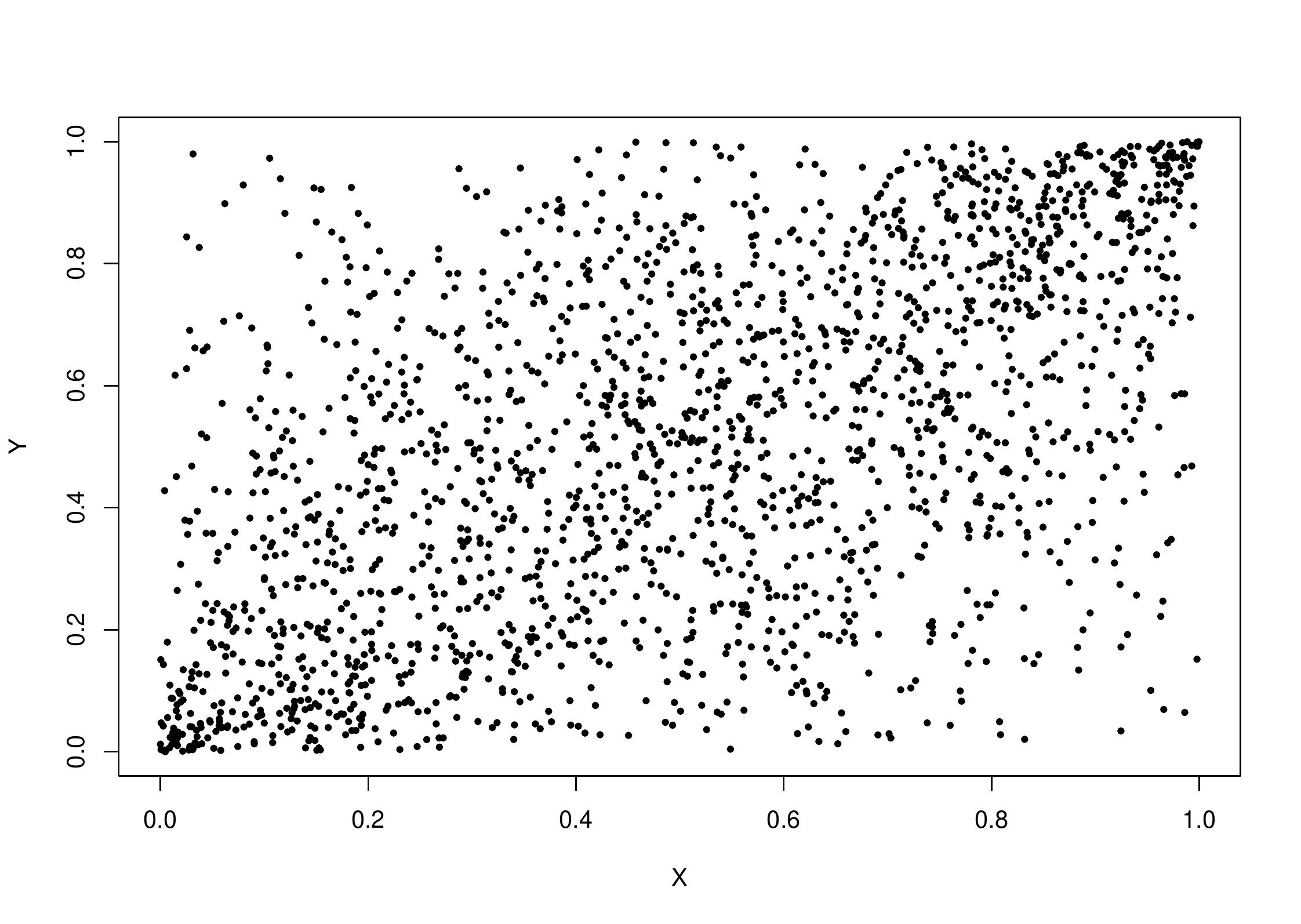}
		\caption*{(a)}	
	\end{minipage}
	\begin{minipage}{.45\linewidth}
		\includegraphics[width=\linewidth]{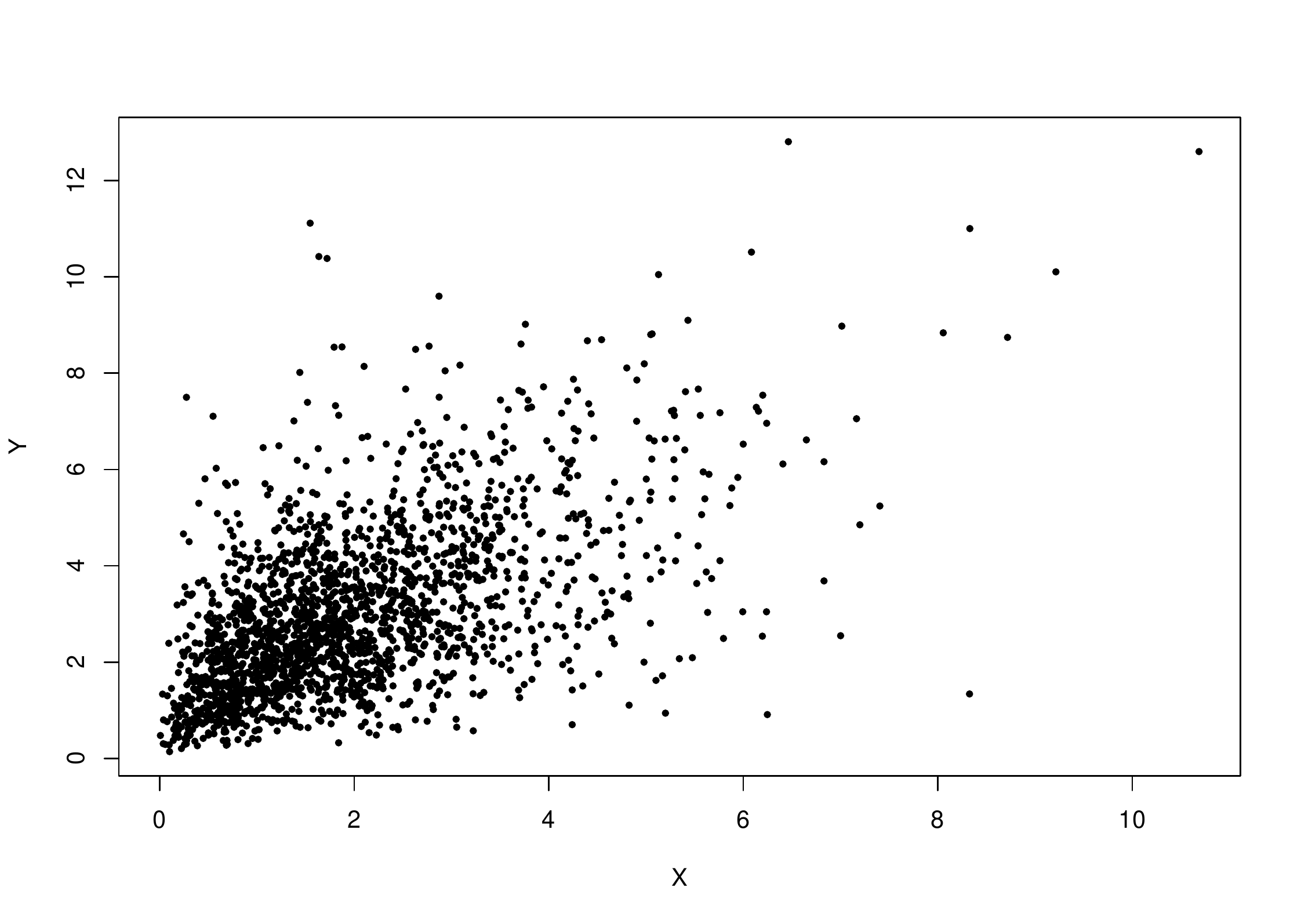}		
		\caption*{(b)}
	\end{minipage}
	\caption{Plot of the simulation data: (a) shows the data in the copula $[0,1]\times [0,1]$ space; (b) shows the data in the data space after taking marginals into account.}
	\label{fig:simstudy4_plot}
\end{figure}

In this section a simulated data set of size 2000 is investigated, in which the dependence structure is elicited from a t-copula parameterised to give weak lower and upper tail dependence ($\lambda_L = \lambda_U = 0.23$) and the two marginal distributions are gamma$(2, 1)$ and gamma$(3, 1)$ respectively.  
It is expected that, from the set of all bivariate copulas trialled, only the t-copula will produce an accurate estimate of the true upper and lower TDCs. 
The simulated data in both the copula space and the data space are shown in Figure~\ref{fig:simstudy4_plot}(a) and Figure~\ref{fig:simstudy4_plot}(b) respectively. 

For this simulated data set with t-copula induced weak tail dependence structure, the t-copula achieved a significantly greater log-likelihood and hence superior BIC result compared with all other fitted copulas, as seen in Table~\ref{tab:simstudy4_copula_summary}. This is to be expected since the simulated data came directly from a t-copula that could not be well approximated by other alternatives. Consequently, the t-copula gives the most accurate estimate of the TDCs. Any BMA implementation with other copulas will pull the TDC estimations further away from the true values. 
If BMA in this case featured positive weights on any other individual copulas it would pull the TDC estimate further from its true value -- best estimated from the t-copula. However, the nature of the BIC values is such that all BMA weight goes to the t-copula, which is reassuring, meaning the BMA estimate of the TDC is equal to that of the t-copula (0.22) and very close to the true value (0.23). 

\begin{table}[H]
	\centering
	\caption{Summary of the fitted copulas, including fitted BIC values, BMA weights ($W_j$) calculated based on BICs and the estimated $\widehat{\lambda}_U$ and $\widehat{\lambda}_L$ based on the fitted copula parameters.}
	\label{tab:simstudy4_copula_summary}
	\resizebox{.7\textwidth}{!}{
		\begin{tabular}{lrrrrrrrrrrrr}
			\toprule[.15 em]
			\ & t & Gaussian & Joe & survival Joe& Gumbel \\
			\midrule
			BIC & -856.9& -809.4& -621.7& -607.2& -796.6  \\
			BMA $W_j$ & 1 & 0 & 0 & 0 & 0  \\
			$\widehat{\lambda}_{U}$ & 0.22 & 0 & 0.53 & 0 & 0.46  \\
			$\widehat{\lambda}_{L}$ & 0.22 & 0 & 0 & 0.52 & 0  \\
			\midrule[.15 em]
			\ & survival Gumbel & Clayton & survival Clayton & Frank  \\
			\midrule
			BIC & -783.5 & -646.3 & -653.8  & -747.9 \\
			BMA $W_j$ & 0 & 0 & 0  & 0  \\
			$\widehat{\lambda}_{U}$ & 0 & 0 & 0.47 & 0  \\
			$\widehat{\lambda}_{L}$ & 0.46 & 0.47 & 0 & 0  \\
			\bottomrule[.15 em]
	\end{tabular} }
\end{table}

\section{EM algorithm for finite mixture of copula regression}
\label{app:EM_mixture_copula_regression}

In this section, the EM algorithm for the finite mixture of copulas (regressions) is illustrated in detail, similar to the work in \cite{Kosmidis2016} and \cite{Arakelian2014}; the methodological difference in this work is that the marginals are of GLM frameworks in each component $g$. 
Given the data $\bm{y}_1, \ldots, \bm{y}_N$, the observed likelihood is 
\begin{equation*}
\begin{split}
\mathcal{L}(\bm{\theta}) &= \prod_{i=1}^{N} \sum_{g=1}^{G} \tau_g h_g(\bm{y}_i;\bm{\theta}_g) \\
&= \prod_{i=1}^{N} \sum_{g=1}^{G} \tau_g c_g\left(F_1(y_{1i};\bm{\beta}_{1g},\bm{\gamma}_{1g}), F_2(y_{2i};\bm{\beta}_{2g},\bm{\gamma}_{2g});\bm{\alpha}_g\right) f_1(y_{1i};\bm{\beta}_{1g},\bm{\gamma}_{1g}) f_2(y_{2i};\bm{\beta}_{2g},\bm{\gamma}_{2g})
\end{split}
\end{equation*}
Maximum likelihood estimation approach via the EM algorithm (\citealp{Dempster1977}) is used for model fitting and inference, where there is one latent variable to be estimated -- the missing group membership $\bm{z}_i = \{z_{i1}, \ldots, z_{iG}\}$ where $z_{ig} =1$ if observation $i$ belongs to cluster $g$ and $z_{ig} = 0$ otherwise. 
The complete data likelihood is then:
\begin{equation*}
\mathcal{L}_c = \prod_{i=1}^{N}\prod_{g=1}^{G}\left[\tau_g h_g(\bm{y}_i;\bm{\theta}_g)\right] ^{z_{ig}} , 
\end{equation*}
and the complete data log-likelihood is:
\begin{equation*}
\begin{split}
\ell_c =& \sum_{i=1}^{N} \sum_{g=1}^{G} z_{ig} \log [ \tau_g h_g(\bm{y}_i; \bm{\theta}_g) ] \\
=& \sum_{i=1}^{N} \sum_{g=1}^{G} z_{ig} \log \tau_g + \sum_{i=1}^{N} \sum_{g=1}^{G} z_{ig} \log h_g(\bm{y}_i; \bm{\theta}_g) \\
=& \sum_{i=1}^{N} \sum_{g=1}^{G} z_{ig} \log \tau_g + \sum_{i=1}^{N} \sum_{g=1}^{G} z_{ig} \log c_g \left(F_1(y_{1i};\bm{\beta}_{1g},\bm{\gamma}_{1g}), F_2(y_{2i};\bm{\beta}_{2g},\bm{\gamma}_{2g});\bm{\alpha}_g\right) \\ 
& + \sum_{i=1}^{N} \sum_{g=1}^{G} z_{ig} \log f_1(y_{1i};\bm{\beta}_{1g},\bm{\gamma}_{1g}) + \sum_{i=1}^{N} \sum_{g=1}^{G} z_{ig} \log f_2(y_{2i};\bm{\beta}_{2g},\bm{\gamma}_{2g}) \\
\end{split}
\label{eq:mixture_copulas_complete_loglike}
\end{equation*}

The expectation of the complete data log-likelihood can be obtained in the E-step of the EM algorithm, followed by an M-step that maximises the expectation of the complete data log-likelihood. The estimated parameters, on convergence, achieve at least local maxima of the observed likelihood of the data. At the t$^{th}$ iteration of the algorithm:

\vskip .5 cm
\noindent\textbf{\underline{E-step}:}
\begin{equation*}
\hat{z}_{ig}^{(t+1)} = \frac{\hat{\tau}_g^{(t)}h_g(\bm{y}_i;\hat{\bm{\theta}}_g^{(t)}) }{\sum_{g^{\prime}=1}^{G} \hat{\tau}_{g^{\prime}}^{(t)}h_{g^{\prime}}(\bm{y}_i;\hat{\bm{\theta}}_{g^{\prime}}^{(t)}) } .
\end{equation*}
\vskip .5 cm
\noindent\textbf{\underline{M-step}:}
\begin{equation*}
\begin{split}
\hat{\tau}_g^{(t+1)} &= \frac{\sum_{i=1}^{N}\hat{z}_{ig}^{(t+1)} }{n } , \\
\hat{\bm{\theta}}_g^{(t+1)} &= \underset{\bm{\theta}_g}{\arg\max} \left( \sum_{i=1}^{N} \sum_{g=1}^{G} z_{ig}^{(t+1)} \log h_g(\bm{y}_i; \bm{\theta}_g) \right) .
\end{split}
\end{equation*}

Note that the update of $\theta_g^{(t+1)}$ in the M-step can be implemented either by updating all parameters simultaneously, or by availing of the separation of the log of the copula likelihood and the marginal distributions in the complete data log-likelihood.  
Following the mixture of experts framework in \cite{Gormley2011} and \cite{Hu2019}, covariates can be incorporated into the mixing probabilities, which results in a multinomial logistic regression (when $G=2$ it becomes a logistic regression). 

Initialisation can be done via a standard clustering algorithm such as \textsf{mclust} (\citealp{Scrucca2017}) or agglomerative hierarchical clustering (\citealp{Everitt2011}) for initial clustering classification and mixing proportions. Within each component, the IMF method could be used - standard GLMs used for marginal distributions followed by estimation of copula parameters.  
The algorithm can be stopped when the increase in the observed log-likelihood is sufficiently small:
$\frac{\ell(\hat{\boldsymbol{\theta}}^{(t+1)}, \hat{\boldsymbol{\tau}}^{(t+1)}) - \ell(\hat{\boldsymbol{\theta}}^{(t)}, \hat{\boldsymbol{\tau}}^{(t)}) }{\ell(\hat{\boldsymbol{\theta}}^{(t+1)}, \hat{\boldsymbol{\tau}}^{(t+1)})} < \epsilon .
$
Typically $\epsilon$ is set as $1\times 10^{-5}$. 
Other suitable terminating conditions can also be considered, such as Aitkens acceleration criterion (\citealp{Aitken1927}; \citealp{Bohning1994}).

\newpage
\section{Simulation study IV}
\label{app:newsimstudy}

To compensate for the fact that the Irish insurer data set is not publicly available, in this simulation study artificially simulated data set is examined using a mixture of copula regressions, primarily to show that using the EM algorithm for such mixtures results in stable and accurate estimates of the true parameter values. 
Given the true TDCs are known, this example also shows that copula based models are effective in TDC estimation. 

\begin{figure}[ht!]
	\centering
	\includegraphics[width=1\textwidth, height=.9\textwidth]{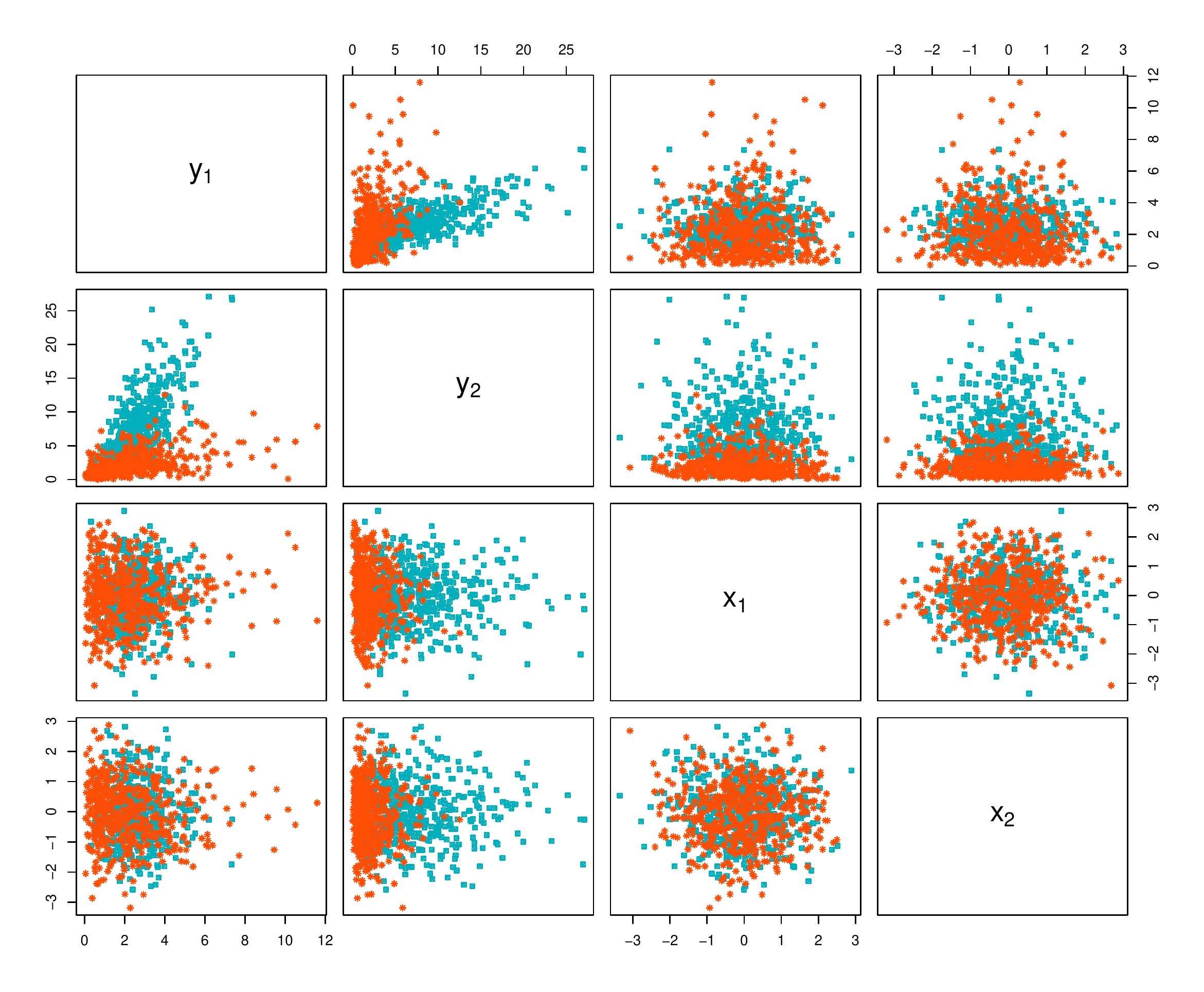}
	\caption{Scatter plot matrix of the simulated data set, colored by component membership: the teal ``$\boxtimes$" is component 1 and the read ``$\ast$" is component 2.}
	\label{fig:newsimstudy_data_plot}
\end{figure}

A random sample of $N=1000$ values were generated from the Gaussian distributions as covariates such that $\bm{x}_1 \sim \bm{x}_2 \sim \mathcal{N}(0,1)$. 
The true number of components is $G=2$, and the two components are equally mixed, i.e. $\bm{\tau} = (0.5, 0.5)$.   
Component 1 follows a Gumbel(2) copula, while component 2 follows a Frank(3) copula. For the marginal distributions, they all have gamma distributions, with shape parameters as $\gamma_{y_1}^{(g=1)}=6, \gamma_{y_2}^{(g=1)}=3, \gamma_{y_1}^{(g=2)}=2, \gamma_{y_2}^{(g=2)}=2$:
\begin{equation*}
\begin{split}
y_1^{(g=1)} &= 1 + 0.1 \bm{x}_1 + 0.1 \bm{x}_2 \\ 
y_2^{(g=1)} &= 2 + 0.2 \bm{x}_1 + 0.2 \bm{x}_2 \\
y_1^{(g=2)} &= 0.8 + 0.2 \bm{x}_1 + 0.2 \bm{x}_2 \\
y_2^{(g=2)} &= 0.8 + 0.2 \bm{x}_1 + 0.2 \bm{x}_2 \\
\end{split}
\end{equation*}
The final simulated data consists of 500 samples from component 1, 500 samples from component 2, covariates $\bm{x}_1$ and $\bm{x}_2$, and the true labels. 
Figure~\ref{fig:newsimstudy_data_plot} shows the scatterplot matrix of the generated data. In this set-up, covariates ${x}$ are not strongly correlated with $\bm{y}$ in a universal sense, but component-wise they are much more strongly correlated. Both $\bm{x}_1$ and $\bm{x}_2$ are expected to be significant covariates in the copula regressions. More noticeably for $\bm{y}_1$ and $\bm{y}_2$ the two components are slightly overlapped.  

The true TDCs are known as $\lambda_U = 0.29$ and $\lambda_L=0$, based on Equation~\ref{eq:mixture_utdc}. First, using the nonparametric estimators TDCs are estimated and the results are shown in Table~\ref{tab:newsimstudy_TDC_estimation}. It shows that they all under-estimate the upper TDC and over-estimate the lower TDC.

\begin{table}[hb!]
	\centering
	\caption{Empirically estimated upper and lower TDCs $\widehat{\lambda}_U$ and $\widehat{\lambda}_L$ for the Irish insurer data based on estimators in Section~\ref{sec:utdc}.}
	\label{tab:newsimstudy_TDC_estimation}
	\resizebox{.6\textwidth}{!}{
		\begin{tabular}{lrrrrrrc}
			\toprule[.15 em]
			\ & $\widehat{\lambda}^{(1)}$ & $\widehat{\lambda}^{(2)}$ & $\widehat{\lambda}^{(3)}$ & $\widehat{\lambda}^{(4)}$ & $\widehat{\lambda}^{(5)}$ & mean & range \\
			\midrule
			$\widehat{\lambda}_{U}$ & 0.166 & 0.141 & 0.144 & 0.148 & 0.129 & 0.145 & [0.129, 0.166] \\
			$\widehat{\lambda}_{L}$ & 0.158 & 0.132 & 0.122 & 0.048 & 0.161 & 0.124 & [0.048, 0.158] \\
			\bottomrule[.15 em]
	\end{tabular} }
\end{table}

\begin{figure}[H]
	\centering
	\begin{minipage}{.48\textwidth}
		\includegraphics[width=\linewidth]{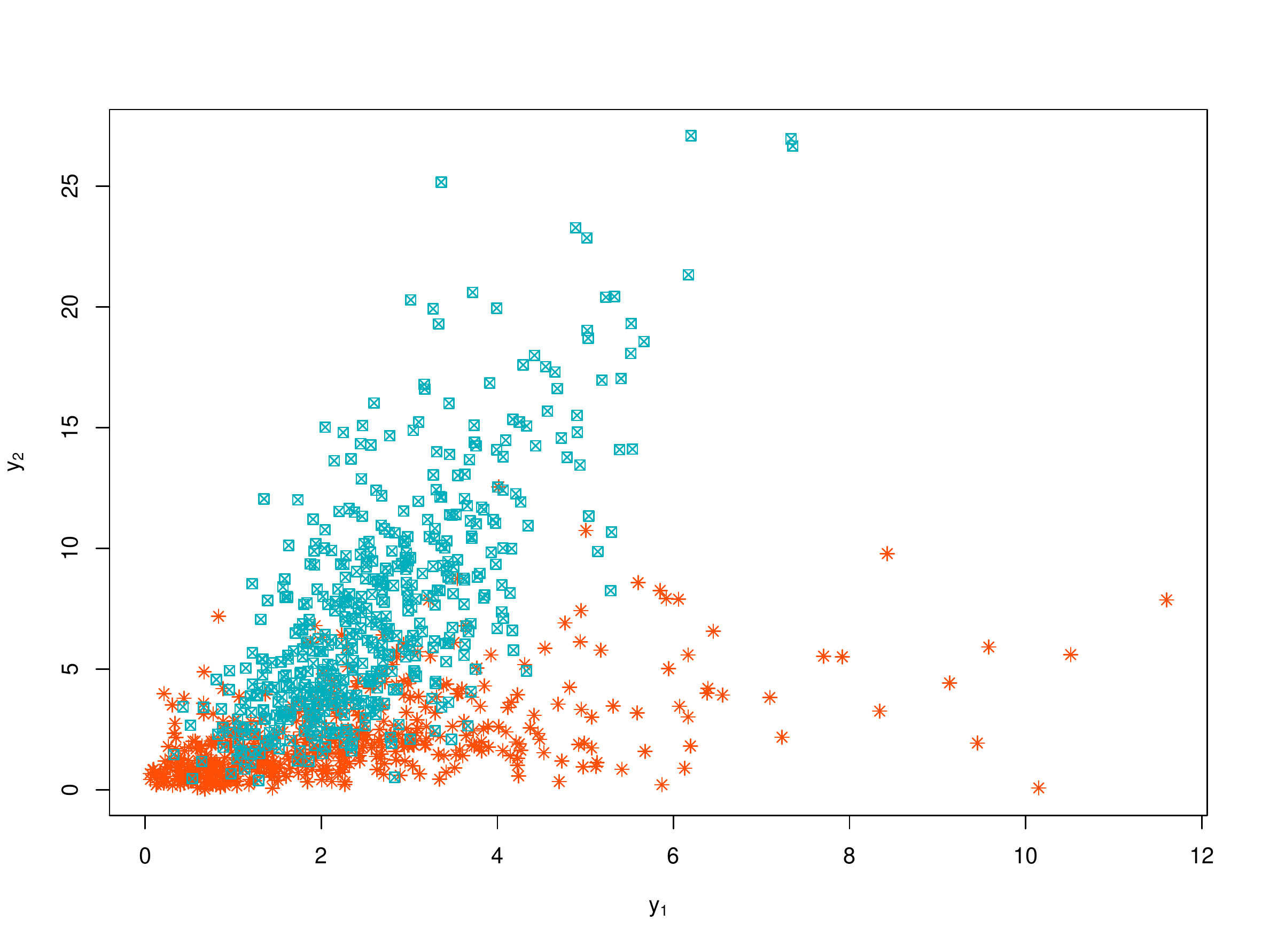}
		\caption*{(a)}
	\end{minipage}
	\begin{minipage}{.48\textwidth}
		\includegraphics[width=\linewidth]{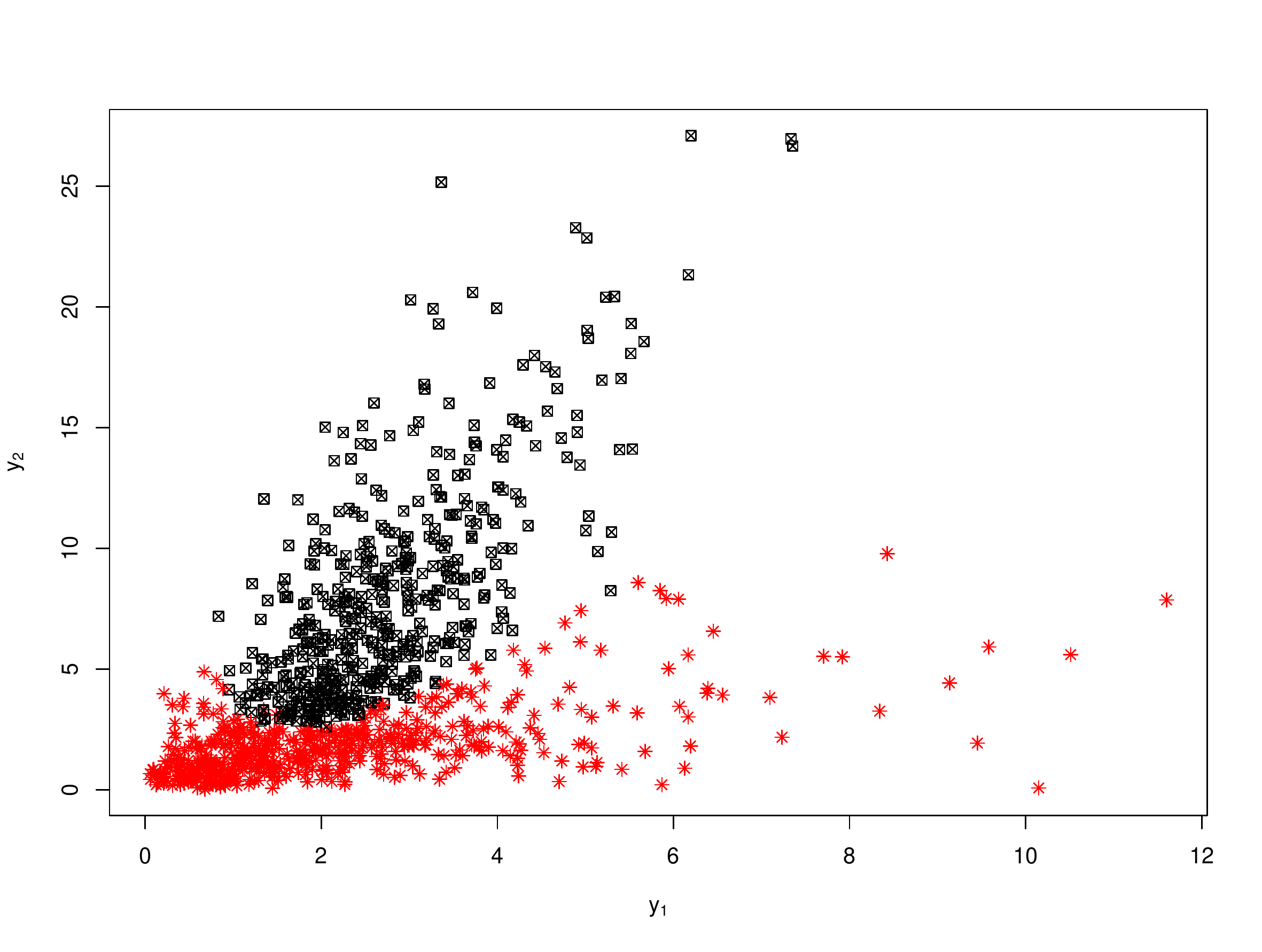}
		\caption*{(b)}
	\end{minipage}
	\caption{Plot of the response variable $\bm{y}$: (a) shows the original classification in the data; (b) shows the classification via the optimal model with Gumbel+Frank mixture of copulas.}
	\label{fig:newsimstudy_model_classification}
\end{figure}

When fitting a mixture of copula regressions to the data, different permutations of copulas again need to be considered for copula selection based on a combination of AIC and BIC as in Section~\ref{sec:irish}. It is expected, and verified by a greedy search over all copula permutations, that the optimal copula combination is Gumbel+Frank and $G=2$. The model selection process is repeated 20 times to make sure that the estimation process is stable and consistent. 
The estimated model parameters, together with comparisons to their true values, are shown in Table~\ref{tab:newsimstudy_model_estimation}. The classification plot is shown in Figure~\ref{fig:newsimstudy_model_classification}. The misclassification rate is 14\%, which is to be expected considering the overlapping area between the two components. Furthermore, the optimal model has estimates of the TDCs as $\widehat{\lambda}_U=0.27$ and $\widehat{\lambda}_L=0$, which are the closest to the true values among all competing models.  

\ \\

\ \\

\ \\

\begin{table}[ht!]
	\centering
	\caption{Estimated parameters of the optimal mixture of copula regression model: for each part of the model, i.e. margin $y_1$, margin $y_2$ and copula, both the true values (e.g. $\bm{\theta}, \alpha, \tau$) and the estimated values (e.g. $\widehat{\bm{\theta}}, \widehat{\alpha}, \widehat{\tau}$) are shown.}
	\label{tab:newsimstudy_model_estimation}
	\resizebox{.9\textwidth}{!}{
		\begin{tabular}{l||l|cc|cc|cc|cc}
			\toprule[.15 em]
			\multicolumn{2}{c}{ \ } & \multicolumn{2}{c}{margin $y_1$} & \multicolumn{2}{c}{margin $y_2$} & \multicolumn{2}{c}{copula} & \multicolumn{2}{c}{mixing} \\
			\midrule[.14 em]
			\ & \ & $\bm{\theta}_{1g}$ & $\widehat{\bm{\theta}}_{1g}$ & $\bm{\theta}_{2g}$ & $\widehat{\bm{\theta}}_{2g}$ & $\alpha_g$ & $\widehat{\alpha}_g$ & $\tau_g$ & $\widehat{\tau}_g$ \\ 
			\multirow{4}{*}{Component $g=1$} & intercept & 1 & 1.00 & 2 & 2.03 & \ & \ & \multirow{4}{*}{0.50} & \multirow{4}{*}{0.49} \\
			& $x_1$ & 0.1 & 0.02 & 0.2 & 0.02 & Gumbel & Gumbel & \\
			& $x_2$ & 0.1 & 0.01 & 0.2 & 0.03 & 2      & 1.85 & \\
			& shape & 6   & 6.23 & 3   & 2.50 & \  & \  & \\
			\midrule[.1 em]
			\multirow{4}{*}{Component $g=2$} & intercept & 0.8 & 0.78 & 0.8 & 0.79 & \ & \ & \multirow{4}{*}{0.50} & \multirow{4}{*}{0.51} \\
			& $x_1$ & 0.2 & 0.03 & 0.2 & -0.03 & Frank & Frank & \\
			& $x_2$ & 0.2 & -0.05 & 0.2 & 0.03 & 3 & 3.75 &\\
			& shape & 2   & 1.63 & 2   & 1.65  & \ & \  & \\
			\bottomrule[.15 em]
	\end{tabular} }
\end{table}

%

\end{document}